\documentclass[11pt,oneside,letterpaper]{article}
\usepackage{amssymb}
\usepackage{amsmath}
\usepackage{amsthm}
\usepackage[dvips]{graphicx}
\usepackage{setspace}
\usepackage{amsfonts}
\usepackage{fancyhdr}
\usepackage{xcolor}
\usepackage{graphicx}
\usepackage{rotating}
\usepackage{comment}
\usepackage{color}
\usepackage{cite}
\usepackage{braket}
\usepackage{caption}
\usepackage[utf8]{inputenc}
\usepackage{mathtools}
\usepackage{tikz}
\usepackage{overpic}
\usepackage{stackengine}
\usepackage[normalem]{ulem}
\usepackage{bbold}

\definecolor{darkgreen}{rgb}{0,0.5,0}
\definecolor{darkblue}{rgb}{0,0,0.6}
\definecolor{purple}{rgb}{0.4,.2,0.7}

\newcommand{\be}{\begin{equation}}
\newcommand{\ee}{\end{equation}}

\usepackage[colorlinks=true,citecolor=darkgreen,linkcolor=black,urlcolor=purple]{hyperref}

\usepackage{pdfsync}

\makeatletter
\newcommand*{\defeq}{\mathrel{\rlap{%
                     \raisebox{0.3ex}{$\m@th\cdot$}}%
                     \raisebox{-0.3ex}{$\m@th\cdot$}}%
                     =} 
\makeatother

\def\be{\begin{eqnarray}}
\def\ee{\end{eqnarray}}

\newcommand{\bea}{\begin{eqnarray}}
\newcommand{\eea}{\end{eqnarray}}
\def\ben{\begin{equation}}
\def\een{\end{equation}}

     \let\r=v

\let\C=\Chi 

\def\be{\begin{equation}}
\def\ee{\end{equation}}
\def\ba{\begin{array}}
\def\ea{\end{array}}

\def\ba#1\ea{\begin{align}#1\end{align}}
\def\bs#1\es{\begin{split}#1\end{split}}

\definecolor{vert}{rgb}{0.1367 0.543 0.1367}

\newcommand{\id}{\mathbb{1}}

\allowdisplaybreaks  

\numberwithin{equation}{section}

\def \be {\begin{equation}}
\def \ee {\end{equation}}
	
\def \JM#1 {{\color{blue}  JM: #1 }}
\def \AAl#1 {{\color{red}  AA: #1 }}

\begin{document}
\onehalfspacing

\begin{center}

{\LARGE  {
Euclidean wormholes for individual 2d CFTs
}}

\vskip1cm

Jeevan Chandra

\vskip5mm
Department of Physics, Cornell University, Ithaca, New York, USA

\vskip5mm

{\tt jn539@cornell.edu }

\end{center}

\vspace{4mm}

\begin{abstract}

We interpret appropriate families of Euclidean wormhole solutions of AdS$_3$ gravity in individual 2d CFTs as replica wormholes described by branching around the time-symmetric apparent horizons of black holes sourced by the backreaction of heavy point particles. These wormholes help describe a rich formalism to coarse grain pure states in 2d CFTs dual to the black hole geometries because the wormhole amplitudes match with the Renyi entropies of CFT states obtained by decohering the pure states in a specific way. This formalism can be generalised to coarse grain pure states in several copies of the dual CFT dual to multi-boundary black holes using wormhole solutions with higher genus boundaries using which we illustrate that coarse graining away the interior of multi-boundary black holes sets the mutual information between any two copies of the dual CFT to zero. Furthermore, this formalism of coarse graining pure states can be extended to decohere transition matrices between pure states which helps interpret more general families of wormhole solutions including those with non replica-symmetric boundary conditions in individual CFTs. The pseudo entropy of the decohered transition matrices has interesting holographic interpretation in terms of the area of minimal surfaces on appropriate black hole or wormhole geometries. The wormhole solutions which show up in the coarse graining formalism also compute the Renyi entropies of Hawking radiation after the Page time in a setup which generalizes the West Coast model to 3d gravity. Using this setup, we discuss the evaporation of one-sided black holes sourced by massive point particles and multi-boundary black holes in 3d gravity.
\noindent

 \end{abstract}

\pagebreak
\pagestyle{plain} 

\setcounter{tocdepth}{2}
{}
\vfill

\ \vspace{-2cm}
\renewcommand{\baselinestretch}{1}\small
\tableofcontents
\renewcommand{\baselinestretch}{1.15}\normalsize

\newcommand{\brho}{\bar{\rho}}
\newcommand{\Svn}{S_{\rm vN}}
\newcommand{\N}{{\cal N}}
\newcommand{\C}{{\cal C}}
\renewcommand{\H}{{\cal H}}

\section{Introduction}

It was shown by Saad, Shenker and Stanford \cite{Saad:2019lba} that JT gravity is dual to an ensemble of 1d quantum systems described by a random matrix Hamiltonian. This is an example of an averaged holographic duality. More recently, it was proposed by the authors of \cite{Chandra:2022bqq} that semiclassical 3d gravity is an average of large-$c$ CFTs. Such dualities are to be contrasted with the class of dualities predicted by the AdS/CFT correspondence where a theory of quantum gravity in asymptotically AdS$_{d+1}$ spacetime is conjectured to be dual to a particular CFT in $d$ spacetime dimensions. Spacetime wormholes have played a key role in these recent developments. They are connected contributions to the gravitational path integral joining otherwise disjoint boundary components. When viewed in the light of AdS/CFT correspondence, such connected contributions present a factorisation paradox \cite{Maldacena:2004rf} because the gravitational path integral on spacetime wormhole geometries gives a connected contribution to a product of CFT observables which obstructs factorization. However, there is no factorization paradox in the context of JT/RMT duality and in the context of semiclassical 3d gravity/ Large-$c$ CFT correspondence because spacetime wormholes are interpreted in terms of ensemble averages of boundary observables.

This however raises an important question: \emph{How do we interpret spacetime wormholes in individual boundary theories?} To answer this question, it is necessary to understand how the interpretation of spacetime wormholes in individual boundary theories does not obstruct factorisation. It is possible that factorisation can be restored by the addition of UV-sensitive contributions to the gravitational path integral. This idea has recently been explored in the context of the SYK model where factorisation in an individual theory described by a fixed choice of coupling constants is restored by non self-averaging contributions called half wormholes \cite{Saad:2021rcu}. See also \cite{Mukhametzhanov:2021nea,Mukhametzhanov:2021hdi} for related work on half wormholes in the SYK model.

A different interpretation for spacetime wormholes in individual CFTs was proposed in \cite{Chandra:2022fwi}. They proposed that spacetime wormholes obtained by branching around the time-symmetric apparent horizon of a black hole geometry dual to a pure state in the CFT compute overlaps between GHZ-like entangled states constructed from multiple copies of the pure state in the dual CFT. This quite naturally results in the re-interpretation of spacetime wormholes as replica wormholes which compute the Renyi entropies of a coarse grained state defined by a diagonal projection of the pure state dual to the black hole geometry in the energy basis. (See also \cite{Liu:2020jsv} for a discussion of diagonal ensembles in the context of equilibrated pure states and their relation to replica wormholes.) This proposal was tested by constructing wormhole solutions in AdS$_{d+1}$ sourced by EOW branes or collapsing thin shells and using them to coarse grain the microstates dual to black hole geometries in those models. Those models are particularly simple to study because the black hole solutions are spherically symmetric and are locally described by the metric on an eternal black hole. The restriction of spherical symmetry can be lifted in the context of 3d gravity using the black hole and wormhole solutions sourced by massive point particles constructed in \cite{Chandra:2022bqq}. In the present work, we show that these wormhole solutions provide a richer formalism to coarse grain pure states generalising the formalism described in \cite{Chandra:2022fwi} in 2d CFTs. This in turn shall provide an interpretation of these wormhole solutions in individual 2d CFTs which is also consistent with the ensemble interpretation provided in \cite{Chandra:2022bqq}. 

The coarse graining formalism applies to wormhole solutions with replica symmetric boundary conditions in individual CFTs. However, there are interesting families of wormhole solutions in 3d gravity with non-replica symmetric boundary conditions. To interpret these solutions in individual CFTs, we generalise the idea of decohering CFT states used in our coarse graining formalism to decohere transition matrices between pure states which don't necessarily have a non-vanishing overlap. For the case where the states have a vanishing overlap, one may also interpret the resulting wormhole amplitudes to be quantifying the extent of global charge violation in quantum gravity following the idea presented in \cite{Bah:2022uyz}. Furthermore, we extend the idea of decohering CFT states (resp. transition matrices) to states (resp. transition matrices) in several copies of the CFT Hilbert space to interpret the wormhole solutions with higher genus boundaries in individual CFTs. We illustrate that our coarse graining procedure amounts to getting rid of both classical correlations and quantum entanglement between the copies of the CFT thereby setting the mutual information between any two copies of the CFT to zero.

Interestingly, it turns out that the wormhole solutions which appear in the coarse graining formalism are also responsible for computing the entropy of Hawking radiation \cite{Penington:2019npb,Almheiri:2019psf,Almheiri:2019qdq,Penington:2019kki} after the Page time in a setup which generalizes the West Coast model of JT gravity \cite{Penington:2019kki} to 3d gravity. This was also observed in \cite{Chandra:2022fwi} where the authors generalized the West Coast model to higher dimensions to discuss evaporation of spherically symmetric black holes sourced by the backreaction of EOW branes or thin shells. However, in this paper, we discuss evaporation of more general black hole solutions which are not necessarily spherically symmetric in 3d gravity \cite{Chandra:2023dgq} by constructing flavoured states dual to the black holes obtained by exciting the CFT vacuum by charged CFT operators. We explore the possibility of selectively charging the CFT operators creating the pure state dual to the black hole in our setup, and illustrate that the entanglement island which develops on the time-symmetric slice after the Page time must necessarily contain the charged particle. By discussing ``evaporation" of black holes whose time-symmetric spatial slice has several minimal surfaces resembling a `python' with several lunch regions \cite{Brown:2019rox} using this setup where we selectively charge the CFT operators, we illustrate that the Page curve describing the entropy of Hawking radiation would saturate to the black hole entropy at late times only if the outermost minimal surface i.e the apparent horizon has smaller area than the other non-trivial minimal surfaces. For this reason, we interpret the setup to be describing an old black hole made young again by infalling matter. We also discuss evaporation of multi-boundary black holes in 3d gravity and illustrate that at late times, the entropy of the radiation matches with the coarse grained entropy of the interior of the black hole defined in section \ref{CGmulti}.

The plan of the paper is as follows: In section \ref{secCG}, we describe the formalism to coarse grain pure states in 2d CFT and illustrate various features of the formalism using examples. In section \ref{secTM}, we generalize the idea of decohering pure states described in section 2 to decohere transition matrices between pure states and illustrate the idea using examples. In section \ref{secevap}, we discuss evaporation of the various types of black hole solutions described in section 2 and compute the Page curve in each case. Finally, in section \ref{secDisc}, we summarise the paper and comment on some future directions.

\section{Coarse graining pure states in 2d CFT} \label{secCG}

In this section, we describe a replica formalism to coarse grain pure states in 2d CFTs using the Euclidean wormhole solutions of 3d gravity. The idea of using wormholes to coarse grain pure states was introduced in \cite{Chandra:2022fwi} where pure states dual to Euclidean black hole geometries in AdS$_{\text{d+1}}$ formed by the backreaction of spherically symmetric matter sources were coarse grained using multi-boundary wormhole solutions. Restricting to AdS$_3$/CFT$_2$, the formalism that we describe here is applicable for more general CFT states which could be dual to non-spherically symmetric black hole geometries. From the bulk point of view, this formalism helps interpret a class of $k$-boundary wormhole solutions with $\mathbb{Z}_k$-symmetric boundary conditions in individual CFTs. We illustrate using perhaps the simplest one-sided black hole geometry in sections \ref{CGrepWh} and \ref{CGprobe} that the coarse grained state defined by a diagonal projection in the Virasoro representation space of the pure state expanded using an OPE channel has the following two features: Its entanglement spectrum matches with the holographic expectation from replica wormholes constructed by branching around a time-symmetric apparent horizon; It preserves the correlations of boundary gravitons and geodesic probes outside the horizon. In section \ref{CGinner}, we generalise our coarse graining formalism to coarse grain different portions of the interior when there are multiple minimal surfaces on the time-symmetric slice of the black hole geometry. Finally, in section \ref{CGmulti}, we discuss how to coarse grain pure states in several copies of the CFT using wormhole solutions with higher genus boundaries mainly by coarse graining the Partially Entangled Thermal State (PETS) obtained by exciting the thermofield double by a local operator insertion.

\subsubsection*{Virasoro primaries vs energy eigenstates}

Note that coarse graining the pure states discussed subsequently by decohering them in the energy basis would not give Renyi entropies that match with the holographic expectation from wormholes. This is because the pure states in discussion are asymmetric so the contribution from Virasoro descendents is comparable to that of the Virasoro primaries at the saddlepoint. This is also related to the fact that energy eigenstates cannot be treated to be pseudorandom while applying ETH for these asymmetric states in 2d CFT because Virasoro symmetry completely determines the statistics of Virasoro descendents given the statistics of Virasoro primaries. This was observed in \cite{Chandra:2023dgq} where it played a crucial role in turning these asymmetric semiclassical CFT states truncated at the semiclassical saddlepoint into random tensor networks. In the present work, we use this reasoning to motivate our prescription to coarse grain these states by decohering the Virasoro primary contribution alone retaining the correlations of Virasoro descendents.

\subsection{Coarse graining using replica wormholes} \label{CGrepWh}

\subsubsection*{The black hole and wormhole geometries}

We consider non-spherically symmetric one-sided black holes in 3d gravity with a time-symmetric apparent horizon formed by the backreaction of heavy point particles. We restrict to the class of such black hole geometries which are prepared by a Euclidean path integral. We refer the reader to \cite{Chandra:2023dgq} for more details on the construction of these geometries from quotients of $\mathbb{H}_3$. In the dual CFT, these black holes correspond to pure states with the Euclidean geometry being dual to a saddle in the norm of the pure state. For example, the black hole geometry formed from two sufficiently heavy conical defects is dual to the pure state\footnote{A sufficient condition for this state to be dual to a black hole is that the two operators are above the multi-twist theshold of $\Delta>\frac{c}{16}$ so that the $\mathcal{O}_1\mathcal{O}_2$ OPE is dominated by black hole states at the saddle \cite{Collier:2018exn}.}, $\ket{\Psi}=\mathcal{O}_1\mathcal{O}_2\ket{0}$ where $\mathcal{O}_{1,2}$ are scalar primary operators dual to heavy point particles in the bulk,
\begin{equation} \label{2defBH}
    \langle \Psi\ket{\Psi}=\vcenter{\hbox{
\begin{overpic}[width=1in]{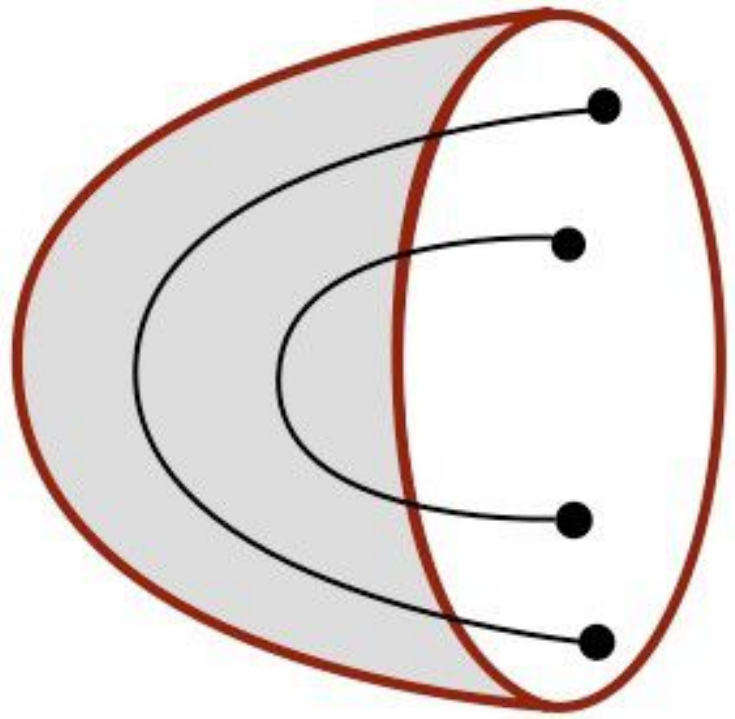}
\put(0,45){\tikz \draw[dashed,red] (1,5)--(5,5);}
\end{overpic}
}}
\end{equation}
with the boundary being $S^2$ and the CFT state is obtained by cutting the Euclidean path integral along the dashed line. The dashed line corresponds to the time-symmetric spatial slice of the black hole geometry. The geometry of this spatial slice is identical to that of a hyperbolic twice-punctured disk and is sketched below,
\begin{equation}
    \vcenter{\hbox{
\begin{overpic}[width=1.5in,grid=false]{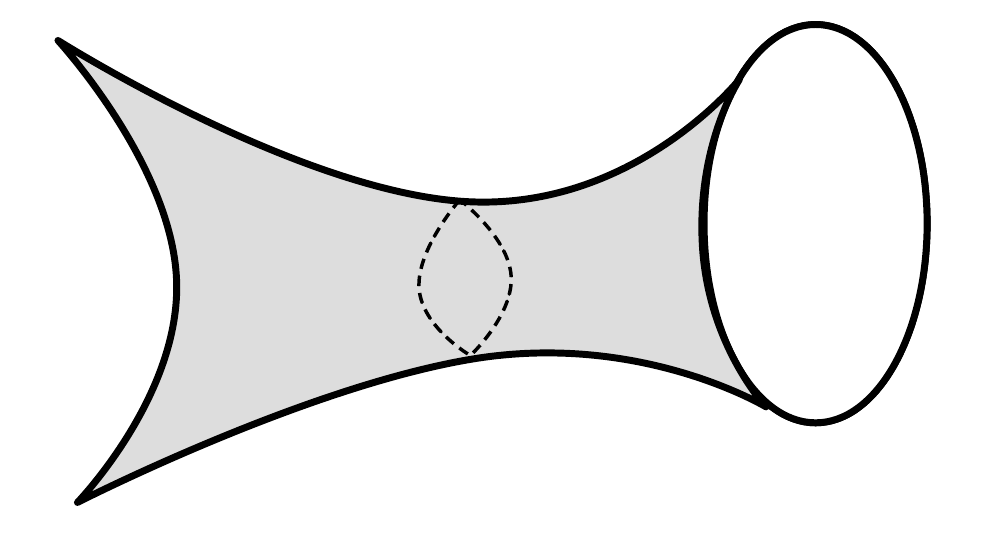}
\put (47,10) {$\Gamma$}
\end{overpic}
}}
\end{equation}
There is an extremal surface denoted $\Gamma$ on the figure which is locally minimal under space-like deformations which we identify to be the apparent horizon of the corresponding Lorentzian black hole geometry. 

Since we are working with a one-sided black hole geometry, the dominant extremal surface (in the RT sense) homologous to the boundary is the trivial surface which has vanishing fine-grained entropy. This observation can be easily reconciled with the dual CFT because the black hole geometry is dual to a pure state, $\rho=\ket{\Psi}\bra{\Psi}$ which has vanishing von-Neumann entropy, $S(\rho)=-\text{Tr}(\rho \log \rho)=0$. However, there is a non-zero coarse grained entropy attributed to the area of the apparent horizon. We seek a CFT interpretation for this coarse grained entropy which in turn would provide a microscopic description for the area of the time-symmetric apparent horizon. Using the idea of \cite{Chandra:2022fwi}, we use wormholes to answer this question i.e, to define a coarse graining map for holographic 2d CFTs. We use the gravitational replica method of Lewkowycz and Maldacena \cite{Lewkowycz:2013nqa} to construct these wormholes. Specifically, we introduce a conical defect of opening angle $\frac{2\pi}{k}$ at the apparent horizon of the black hole geometry described above and allow it to backreact. We interpret the resulting singular geometry as a $\mathbb{Z}_k$-orbifold and cyclically glue $k$ copies of this geometry across the interior to get a smooth $k$-boundary wormhole geometry. Since all solutions to Einstein's equations in 3d are locally $\mathbb{H}_3$ (away from defect trajectories in the case of orbifolds), these wormhole geometries can be constructed by taking appropriate quotients of $\mathbb{H}_3$. 

For example, the two-boundary wormhole topologies obtained by branching around the apparent horizon of the pure state black hole in (\ref{2defBH}) correspond to the $\mathbb{Z}_2$-symmetric `Fuchsian' wormhole geometries with four defects going across the wormhole solving the boundary conditions corresponding to $(\langle \Psi \ket{\Psi})^2$. This geometry was constructed in \cite{Chandra:2022bqq} with the metric,
\begin{equation}
   ds^2=d\rho^2+\cosh^2\rho e^{\Phi}|dz|^2
\end{equation}
where $\Phi$ solves the Liouville equation on the 4-punctured sphere. The gravitational partition function denoted $Z_2$ on this wormhole was also calculated in \cite{Chandra:2022bqq} and shown to be the square of a Liouville correlator between the same vertex operators, $Z_2 \approx |G_L|^2$. The partition functions $Z_k$ on $k>2$ wormhole topologies can be calculated using the recently proposed Virasoro TQFT prescription \cite{Collier:2023fwi} and they match with the prediction from the large-$c$ ensemble of \cite{Chandra:2022bqq} in the semiclassical limit \cite{CollierInProgress}. The expression for $Z_k$ is written in (\ref{WHGHZ}).

\subsubsection*{The microstate statistics from the bulk}

Now, we shall use the above wormhole solutions to coarse grain the pure state $\ket{\Psi}$ with the idea being that given the pure state $\rho=\ket{\Psi}\bra{\Psi}$, we can find a mixed state $\overline{\rho}$ whose Renyi entropies can be computed holographically using the wormhole solutions. For simplicity, we shall illustrate the idea using $\ket{\Psi}=\mathcal{O}_1\mathcal{O}_2\ket{0}$. First, expand the product using the Virasoro OPE blocks \cite{Fitzpatrick:2016mtp, Chandra:2023dgq},
\begin{equation} \label{Psi1}
    \ket{\Psi}=\sum_p c_{12p}\left| {\cal B}\left[  \vcenter{\hbox{\vspace{0.12in}
	\begin{tikzpicture}[scale=.75]
	\draw[thick,<-] (-1/2,0) -- (-3/2,0);
	\node[below,scale=0.75] at (-1,0) {$h_p$};
	\draw[thick] (-3/2,0) -- (-3/2,0.8);
	\node[above,scale=0.75] at (-3/2,0.8) {$h_2$};
        \draw[thick] (-3/2,0) -- (-5/2,0);
        \node[left,scale=0.75] at (-5/2,0) {$h_1$};
	\end{tikzpicture}
	}} \right] \right|^2 \ket{p}
\end{equation}
where $c_{12p}$ is the OPE coefficient between the external primaries $\mathcal{O}_1,\mathcal{O}_2$ with the primary $\mathcal{O}_p$ running in the intermediate channel. ${\cal B}\left[  \vcenter{\hbox{\vspace{0.12in}
	\begin{tikzpicture}[scale=.75]
	\draw[thick,<-] (-1/2,0) -- (-3/2,0);
	\node[below,scale=0.75] at (-1,0) {$h_p$};
	\draw[thick] (-3/2,0) -- (-3/2,0.8);
	\node[above,scale=0.75] at (-3/2,0.8) {$h_2$};
        \draw[thick] (-3/2,0) -- (-5/2,0);
        \node[left,scale=0.75] at (-5/2,0) {$h_1$};
	\end{tikzpicture}
	}} \right] $ is a shorthand for the chiral Virasoro OPE block which encodes the contribution from the Virasoro descendents of $\mathcal{O}_p$ in the $\mathcal{O}_1\mathcal{O}_2$ OPE. The sum runs over all primaries labelled by $p$. The norm of this state can be computed from the 4-point function, $\bra{0}\mathcal{O}_2(z_2,\overline{z}_2)^{\dagger}\mathcal{O}_1(z_1,\overline{z}_1)^{\dagger}\mathcal{O}_1(z_1,\overline{z}_1)\mathcal{O}_2(z_2,\overline{z}_2)\ket{0}$ with the operators inserted in a reflection symmetric fashion on the $S^2$, expanded for instance in the $(12) \to (12)$ channel 4-point conformal blocks\footnote{Recall that in the Euclidean signature, the adjoint of an operator inserted on the $S^2$ is given by $ \mathcal{O}(z,\overline{z})^{\dagger}=z^{-2\overline{h}}\overline{z}^{-2h}\mathcal{O}(\frac{1}{\overline{z}},\frac{1}{z})$ where $(h,\overline{h})$ are the conformal weights of the primary operator $\mathcal{O}$. The cross ratio between the insertion points is $ x=\frac{z_{12}z_{34}}{z_{13}z_{24}}=\frac{|z_1-z_2|^2}{|1-\overline{z}_1z_2|^2}$ is real and positive for the kinematics used to compute the norm.}. From the bulk point of view, we interpret the OPE coefficents $c_{12p}$ to be describing a black hole microstate which is dressed by boundary gravitons whose contribution is captured by the Virasoro OPE block. Statistical properties of the microstate coefficients $c_{12p}$ can be read off by matching the norm of the pure state averaged over a suitable energy window with the gravitational partition function $Z_1$ of the dual geometry (\ref{2defBH}) which agrees with the Virasoro identity block in the semiclassical limit \cite{Hartman:2013mia, Faulkner:2013yia},
\begin{equation}
    Z_1 \approx \left | \vcenter{\hbox{\vspace{0.12in}
	\begin{tikzpicture}[scale=.75]
        \draw[thick] (-1/2,0) -- (-1/2,0.8);
        \node[above, scale=0.75] at (-1/2,0.8) {$2$};
        \draw[thick] (-1/2,0) -- (1/2,0);
        \node[right, scale=0.75] at (1/2,0) {$2$};
	\draw[thick] (-1/2,0) -- (-3/2,0);
	\node[below,scale=0.75] at (-1,0) {$\mathbb{1}$};
	\draw[thick] (-3/2,0) -- (-3/2,0.8);
	\node[above,scale=0.75] at (-3/2,0.8) {$1$};
        \draw[thick] (-3/2,0) -- (-5/2,0);
        \node[left,scale=0.75] at (-5/2,0) {$1$};
	\end{tikzpicture}
	}} \right |^2 = \left | \int dh_p\rho_0(h_p)C_0(h_1,h_2,h_p)\vcenter{\hbox{\vspace{0.12in}
	\begin{tikzpicture}[scale=.75]
        \draw[thick] (-1/2,0) -- (-1/2,0.8);
        \node[above, scale=0.75] at (-1/2,0.8) {$2$};
        \draw[thick] (-1/2,0) -- (1/2,0);
        \node[right, scale=0.75] at (1/2,0) {$1$};
	\draw[thick] (-1/2,0) -- (-3/2,0);
	\node[below,scale=0.75] at (-1,0) {$p$};
	\draw[thick] (-3/2,0) -- (-3/2,0.8);
	\node[above,scale=0.75] at (-3/2,0.8) {$2$};
        \draw[thick] (-3/2,0) -- (-5/2,0);
        \node[left,scale=0.75] at (-5/2,0) {$1$};
	\end{tikzpicture}
	}} \right |^2
\end{equation}
The second equality corresponds to expanding the identity block in the dual channel with the fusion kernel expressed in terms of a smooth function $C_0$ of conformal weights.\footnote{This function is written down in closed form for instance in \cite{Collier:2019weq}. However, the explicit expression for $C_0$ is not going to be used in this paper.} Thus, requiring that\footnote{We emphasize that this is a match between a family of saddles labelled by the cross ratio between the insertion points which is real and positive for the chosen kinematics.} $\langle \Psi_1\ket{\Psi_1}\approx Z_1$ with an ansatz for the microstate coefficients, 
\begin{equation} \label{OPEansatz}
   c_{12p}=e^{g(h_p,\overline{h}_p)}c_p
\end{equation}
where $g(h,\overline{h})$ is a smooth function of the primary weights (The dependence on external weights is implicit in this ansatz) and $c_p$ is a factor which captures fine-grained details of the microstate with $|c_p|^2\approx 1$ when averaged over a suitable energy window. By reading off the smooth function from the gravitational saddle, we see that when averaged over an energy window, the microstate coefficients satisfy,
\begin{equation}
    \overline{|c_{12p}|^2}=C_0(h_1,h_2,h_p)C_0(\overline{h}_1,\overline{h}_2,\overline{h}_p)
\end{equation}

\subsubsection*{Wormholes and GHZ-like states}

Now, we use the expansion of the microstate in (\ref{Psi1}) to define an entangled state in $k$ copies of the CFT,
\begin{equation}
    \ket{\Psi_k}=\sum_p c_{12p}^k \left(\left | {\cal B}\left[  \vcenter{\hbox{\vspace{0.12in}
	\begin{tikzpicture}[scale=.75]
	\draw[thick,<-] (-1/2,0) -- (-3/2,0);
	\node[below,scale=0.75] at (-1,0) {$h_p$};
	\draw[thick] (-3/2,0) -- (-3/2,0.8);
	\node[above,scale=0.75] at (-3/2,0.8) {$h_2$};
        \draw[thick] (-3/2,0) -- (-5/2,0);
        \node[left,scale=0.75] at (-5/2,0) {$h_1$};
	\end{tikzpicture}
	}} \right] \right |^2\ket{p} \right)^{\otimes k}
\end{equation}
This state is analogous to a $k$-copy GHZ state studied in the context of quantum information for qubit systems \cite{greenberger1989going}, in the sense that the microstate coefficients have a diagonal pattern of entanglement. However, due to the dressing by boundary gravitons, the state exhibits a richer pattern of entanglement between the low energy degrees of freedom. We claim that in the semiclassical limit, the $k$-boundary wormhole amplitude $Z_k$ agrees with the saddle in the norm of $\ket{\Psi_k}$ when averaged over a window,
\begin{equation} \label{WHGHZ}
    Z_k\approx  \langle \Psi_k\ket{\Psi_k} \approx  \left | \int dh_p\rho_0(h_p)C_0(h_1,h_2,h_p)^k\left (\vcenter{\hbox{\vspace{0.12in}
	\begin{tikzpicture}[scale=.75]
        \draw[thick] (-1/2,0) -- (-1/2,0.8);
        \node[above, scale=0.75] at (-1/2,0.8) {$2$};
        \draw[thick] (-1/2,0) -- (1/2,0);
        \node[right, scale=0.75] at (1/2,0) {$1$};
	\draw[thick] (-1/2,0) -- (-3/2,0);
	\node[below,scale=0.75] at (-1,0) {$p$};
	\draw[thick] (-3/2,0) -- (-3/2,0.8);
	\node[above,scale=0.75] at (-3/2,0.8) {$2$};
        \draw[thick] (-3/2,0) -- (-5/2,0);
        \node[left,scale=0.75] at (-5/2,0) {$1$};
	\end{tikzpicture}
	}} \right )^k \right  |^2
\end{equation}
The match at $k=2$ can be easily verified using the action for the 2-boundary wormhole computed in \cite{Chandra:2022bqq}. For $k>2$, the wormhole amplitudes can be computed using the Virasoro TQFT prescription \cite{Collier:2023fwi} whenever the wormhole exists as a saddle,\footnote{We expect that for a given cross ratio $x$, there is a $k_{\text{max}}(x)$ such that only wormholes with $k<k_{\text{max}}(x)$ exists as saddles. This expectation is analogous to a similar observation in \cite{Chandra:2022fwi} that wormholes constructed by branching around the horizon of B-state and thin shell black holes exists as saddles only for $k<k_{\text{max}}$ which is determined by the Euclidean time at the brane/ shell insertion on the boundary which plays the role of cross-ratio in that setup.} in which case the semiclassical partition function of the wormhole matches with the norm of the GHZ-like state at the saddlepoint.

\subsubsection*{The coarse grained state and its spectrum}

The density matrix associated with the pure state (\ref{Psi1}) is given by,
\begin{equation} \label{rhoBH}
    \rho=\ket{\Psi}\bra{\Psi}=\sum_{p,q}c_{12p}c^*_{12q}\left | {\cal B}\left[  \vcenter{\hbox{\vspace{0.12in}
	\begin{tikzpicture}[scale=.75]
	\draw[thick,<-] (-1/2,0) -- (-3/2,0);
	\node[below,scale=0.75] at (-1,0) {$h_p$};
	\draw[thick] (-3/2,0) -- (-3/2,0.8);
	\node[above,scale=0.75] at (-3/2,0.8) {$h_2$};
        \draw[thick] (-3/2,0) -- (-5/2,0);
        \node[left,scale=0.75] at (-5/2,0) {$h_1$};
	\end{tikzpicture}
	}} \right] \right |^2\ket{p}\bra{q}
\left | {\cal B}\left[  \vcenter{\hbox{\vspace{0.12in}
	\begin{tikzpicture}[scale=.75]
	\draw[thick,<-] (-1/2,0) -- (-3/2,0);
	\node[below,scale=0.75] at (-1,0) {$h_q$};
	\draw[thick] (-3/2,0) -- (-3/2,0.8);
	\node[above,scale=0.75] at (-3/2,0.8) {$h_2$};
        \draw[thick] (-3/2,0) -- (-5/2,0);
        \node[left,scale=0.75] at (-5/2,0) {$h_1$};
	\end{tikzpicture}
	}} \right] ^{\dagger}\right|^2
\end{equation}
Motivated by the interesting and non-trivial match between wormhole amplitudes and the norms of GHZ-like entangled states (\ref{WHGHZ}), we define a coarse grained version $\overline{\rho}$ of the pure state (\ref{rhoBH}),
\begin{equation} \label{CGstate}
    \overline{\rho}=\sum_p |c_{12p}|^2\left | {\cal B}\left[  \vcenter{\hbox{\vspace{0.12in}
	\begin{tikzpicture}[scale=.75]
	\draw[thick,<-] (-1/2,0) -- (-3/2,0);
	\node[below,scale=0.75] at (-1,0) {$h_p$};
	\draw[thick] (-3/2,0) -- (-3/2,0.8);
	\node[above,scale=0.75] at (-3/2,0.8) {$h_2$};
        \draw[thick] (-3/2,0) -- (-5/2,0);
        \node[left,scale=0.75] at (-5/2,0) {$h_1$};
	\end{tikzpicture}
	}} \right] \right |^2\ket{p}\bra{p}
\left | {\cal B}\left[  \vcenter{\hbox{\vspace{0.12in}
	\begin{tikzpicture}[scale=.75]
	\draw[thick,<-] (-1/2,0) -- (-3/2,0);
	\node[below,scale=0.75] at (-1,0) {$h_p$};
	\draw[thick] (-3/2,0) -- (-3/2,0.8);
	\node[above,scale=0.75] at (-3/2,0.8) {$h_2$};
        \draw[thick] (-3/2,0) -- (-5/2,0);
        \node[left,scale=0.75] at (-5/2,0) {$h_1$};
	\end{tikzpicture}
	}} \right] ^{\dagger}\right|^2
\end{equation}
such that the Renyi entropy of the coarse grained state matches with the norm of the GHZ-like state. So, we have an identity,
\begin{equation}
   Z_k \approx \text{Tr}(\overline{\rho}^k)=\langle \Psi_k \ket{\Psi_k}
\end{equation}
which indicates that the wormholes compute the Renyi entropy of the coarse grained state holographically. For example, the four-boundary wormhole computing the fourth Renyi of the coarse grained state is sketched below,
\begin{align}
      \text{Tr}(\overline{\rho}^k)\approx \vcenter{\hbox{
\begin{overpic}[width=1.8in,grid=false]{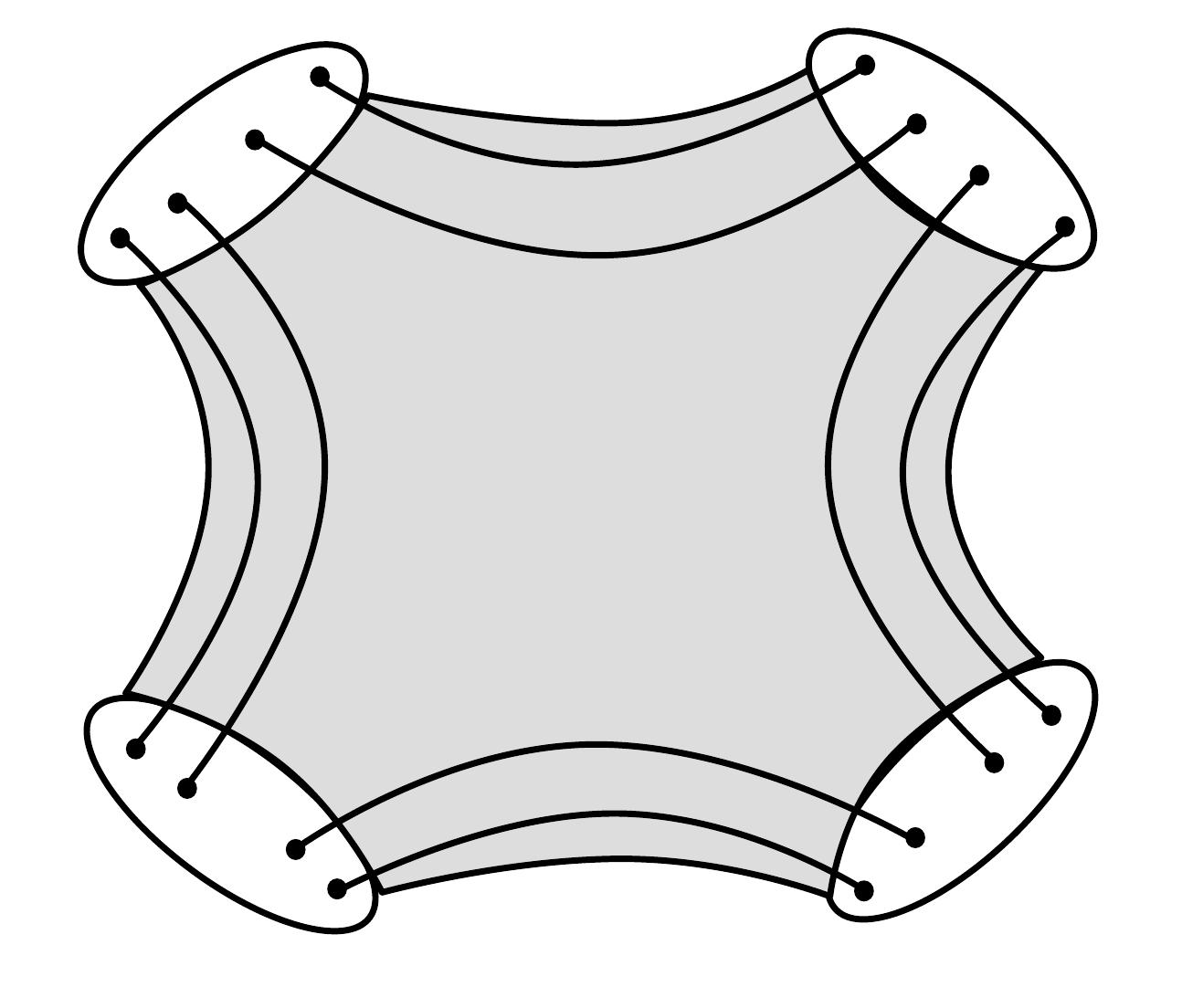}
\end{overpic}
}}
\end{align}
By construction, the wormholes are gravitational replicas which compute the coarse grained entropy of the apparent horizon. Thus, we have a CFT formula for the coarse grained entropy of apparent horizons,
\begin{equation}
    S(\overline{\rho})=-\partial_k\log \left(\frac{Z_k}{Z_1^k} \right)\bigg|_{k=1}
\end{equation}
where $S(\overline{\rho})$ is the von-Neumann entropy of the coarse grained state,
\begin{equation}
    S(\overline{\rho})=-\text{Tr}(\hat{\overline{\rho}} \log\hat{\overline{\rho}}), \quad \quad \hat{\overline{\rho}}=\frac{\overline{\rho}}{\text{Tr}\overline{\rho}}
\end{equation}
We evaluate the entropy of the coarse grained state using the saddle point equation for the partition function. To this end, we introduce the following large-$c$ conventions,
\begin{equation}
 \begin{split}
    & \rho_0(h)\approx e^{\frac{\pi c\gamma}{6}}\\
    & \sqrt{\rho_0(h)} C_0(h_1,h_2,h)\approx e^{-\frac{cw(\gamma)}{6}}\\
    & \mathcal{F}_{1221}(h;x)\approx e^{-\frac{c f(\gamma;x)}{6}}
  \end{split}    
\end{equation}
where $\gamma$ is related to the weight $h$ by the relation $h=\frac{c}{24}(1+\gamma^2)$ and $\rho_0(h)$ is the Cardy density.
With these conventions, the saddle point momentum denoted $\gamma_k$ (which is a function of the cross ratio $x$) in the evaluation of $Z_k$ can be determined implicitly by solving the saddle-point equation,
\begin{equation} \label{saddlepoint}
    (1-\frac{k}{2})\pi=kw'(\gamma_k)+kf'(\gamma_k;x)
\end{equation}
There is a similar equation for the other chiral half of the partition function. The saddle in the partition function for the $k$-boundary wormhole is expected to land on a scalar primary for the chosen kinematics since the cross ratio $x$ is real. It is straightforward to check that the entropy of the coarse grained state satisfies,
\begin{equation} \label{Cgent}
   S(\overline{\rho})=\frac{c}{6}(2\pi\gamma_1(x))
\end{equation}
where $\Delta=\frac{c}{12}(1+\gamma^2)$ is the scaling dimension of the scalar primary operator running in the $11 \to 22$ channel obtained by dualising the 4-point identity block. It was shown in \cite{Chandra:2023dgq} by uniformising the time-symmetric spatial slice that the area of the apparent horizon for the black hole geometry in (\ref{2defBH}) is given by $A_{\text{hor}}=2\pi \gamma_1$. Thus, we see that,
\begin{equation}
    S(\overline{\rho})=\frac{A_{\text{hor}}}{4G}
\end{equation}
In fact, we see that the modular Renyi entropy which is a natural generalisation of the von Neumann entropy with an interesting holographic interpretation introduced in \cite{Dong:2016fnf} takes a form similar to (\ref{Cgent}) using the saddlepoint equation (\ref{saddlepoint}) for the partition function of the $k$-boundary wormhole\footnote{The usual Renyi entropy $S_k(\overline{\rho})=\frac{1}{1-k}\log\left(\text{Tr}(\hat{\overline{\rho}}^k)\right)$ has a more complicated form because of the explicit dependence on the structure constants and conformal blocks. Note that even the modular Renyi entropy depends on the structure constants and conformal blocks but only implicitly through the saddlepoint equation (\ref{saddlepoint}).},
\begin{equation}
   \Tilde{S}_k(\overline{\rho})=-k^2\partial_n\left(\frac{\text{Tr}(\overline{\rho}^n)}{n (\text{Tr}\overline{\rho})^n} \right)\bigg |_{n=k}=\frac{c}{6}(2\pi \gamma_k(x))
\end{equation}
Based on the general arguments presented in \cite{Dong:2016fnf}, it is expected that the modular Renyi entropy computed above matches with the area of a minimal surface on the replica wormhole. In \cite{Chandra:2022bqq}, it was shown using the results of \cite{Hadasz:2005gk} that $\frac{A_2}{4G_N}=\frac{c}{6}(2\pi\gamma_2(x))$ where $A_2$ is the length of a minimal geodesic around the waist separating the two pairs of identical operators of the 4-punctured sphere on the reflection symmetric slice of the Maldacena-Maoz wormhole (the $k=2$ wormhole in the present setup). $\gamma_2(x)$ is the saddle point momentum in the Liouville 4-point function of the two pairs of identical operators when expanded in the $(12)\to (12)$ channel. Thus, we have explicitly verified that
\begin{equation}
   \Tilde{S}_k(\overline{\rho})=\frac{A_k}{4G_N}
\end{equation}
for $k=1$ and $k=2$. Here $A_k$ is the length of a minimal geodesic on the $k$-boundary wormhole. It would be interesting to understand if this match can be explicitly verified for $k\geq 3$.

Now, we make some general comments about the coarse graining formalism illustrated above. Note that the holographic coarse graining map $\overline{\rho}=\mathcal{C}(\rho)$ illustrated by (\ref{CGstate}) is more generally implemented as follows: Decohere the correlations between Virasoro primaries in the black hole microstate $\rho$ retaining the correlations described by Virasoro descendents. Intuitively, this corresponds to coarse graining over stuff behind the apparent horizon which is to be treated as UV physics inaccessible to an IR observer while retaining correlations between stuff outside the apparent horizon. It would be interesting to understand if the map $\mathcal{C}$ can be implemented using a quantum channel. Following \cite{Chandra:2023dgq}, it can shown that when the CFT state is truncated to a window around a semiclassical saddlepoint, it can be coarse grained using a block dephasing quantum channel with the Kraus operators being projectors onto Virasoro representations in the window. Within this window, the Virasoro OPE block acts isometrically (upto normalisation) on the truncated microstate. However, in the present case where we don't truncate the OPE, it is not clear if the coarse graining map can be implemented using a quantum channel.

\subsection{Coarse graining in the presence of a probe particle} \label{CGprobe}

In this section, we shall illustrate how the coarse graining map changes in the presence of a probe particle propagating on the black hole background. We add a probe particle to the setup described in the previous section, for which we may neglect the gravitational backreaction to leading order in the mass of the particle so that it propagates along a geodesic on the geometry dual to the pure state $\ket{\Psi}$. We denote the CFT state in the presence of the probe particle by $\ket{\Tilde{\Psi}}=\mathcal{O}\mathcal{O}_1\mathcal{O}_2\ket{0}$ where $\mathcal{O}$ is a scalar primary operator with scaling dimension $1 \ll \Delta_{\mathcal{O}} \ll c$ dual to the probe particle. The density matrix of the state when expanded using Virasoro OPE blocks is given by $\Tilde{\rho}=\ket{\Tilde{\Psi}}\bra{\Tilde{\Psi}}$,
\begin{equation}
   \Tilde{\rho}=\sum_{p,p',q,q'}c_{12q} c^*_{12q}c_{\mathcal{O}pq} c^*_{\mathcal{O}p'q} \left|\mathcal{B}\left[\vcenter{\hbox{\vspace{0.12in}
	\begin{tikzpicture}[scale=.75]
	\draw[thick,->] (-1/2,0) -- (1/2,0);
	\node[below,scale=0.75] at (0,0) {$h_p$};
	\draw[thick] (-1/2,0) -- (-3/2,0);
	\node[below,scale=0.75] at (-1,0) {$h_q$};
	\draw[thick] (-1/2,0) -- (-1/2,1*0.8);
	\node[above,scale=0.75] at (-1/2,0.8) {$h_{\mathcal{O}}$};
	\draw[thick] (-3/2,0) -- (-3/2,0.8);
	\node[above,scale=0.75] at (-3/2,0.8) {$h_2$};
	\draw[thick] (-3/2,0) -- (-3/2-0.8,0);
	\node[left,scale=0.75] at (-3/2-0.8,0) {$h_1$};
	\end{tikzpicture}
	}} \right]\right|^2 \ket{p}\bra{p'} \left|\mathcal{B}\left[\vcenter{\hbox{\vspace{0.12in}
	\begin{tikzpicture}[scale=.75]
	\draw[thick,->] (-1/2,0) -- (1/2,0);
	\node[below,scale=0.75] at (0,0) {$h_{p'}$};
	\draw[thick] (-1/2,0) -- (-3/2,0);
	\node[below,scale=0.75] at (-1,0) {$h_{q'}$};
	\draw[thick] (-1/2,0) -- (-1/2,1*0.8);
	\node[above,scale=0.75] at (-1/2,0.8) {$h_{\mathcal{O}}$};
	\draw[thick] (-3/2,0) -- (-3/2,0.8);
	\node[above,scale=0.75] at (-3/2,0.8) {$h_2$};
	\draw[thick] (-3/2,0) -- (-3/2-0.8,0);
	\node[left,scale=0.75] at (-3/2-0.8,0) {$h_1$};
	\end{tikzpicture}
	}} \right]^\dagger\right|^2
\end{equation}
On the time-symmetric slice in the bulk, the particle could either be outside or inside the apparent horizon depending on the location of the insertion point of $\mathcal{O}$ on the boundary $S^2$ as sketched respectively in the pair of figures below,
\begin{align}
     \vcenter{\hbox{
\begin{overpic}[width=1.8in,grid=false]{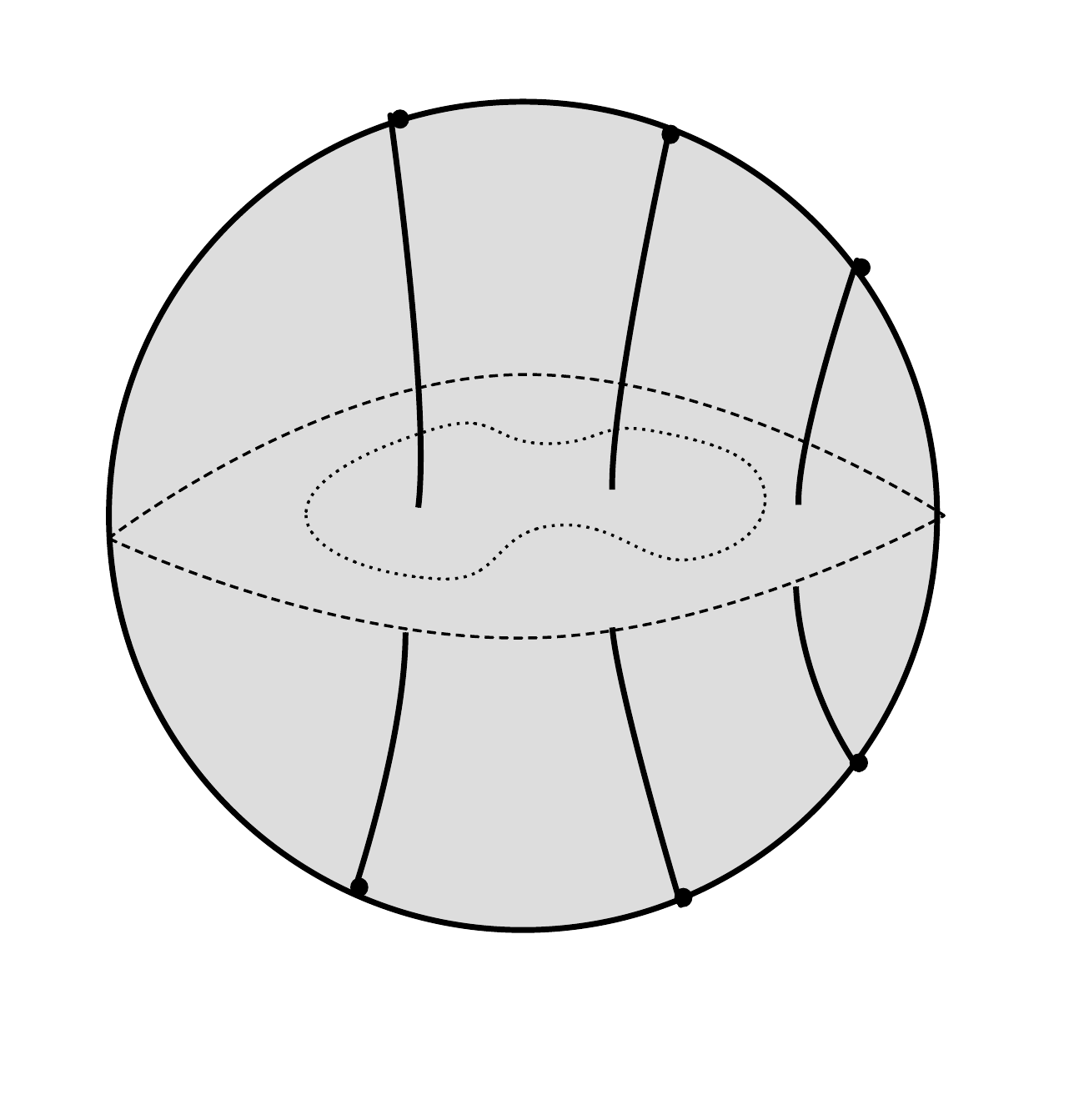}
\put (46,44) {$\Gamma_{\text{out}}$}
\put (33,93) {$1$}
\put (30,12) {$1$}
\put (60,92) {$2$}
\put (60,12) {$2$}
\put (76,78) {$\mathcal{O}$}
\put (76,26) {$\mathcal{O}$}
\end{overpic}
}} \qquad \qquad 
  \vcenter{\hbox{
\begin{overpic}[width=1.7in,grid=false]{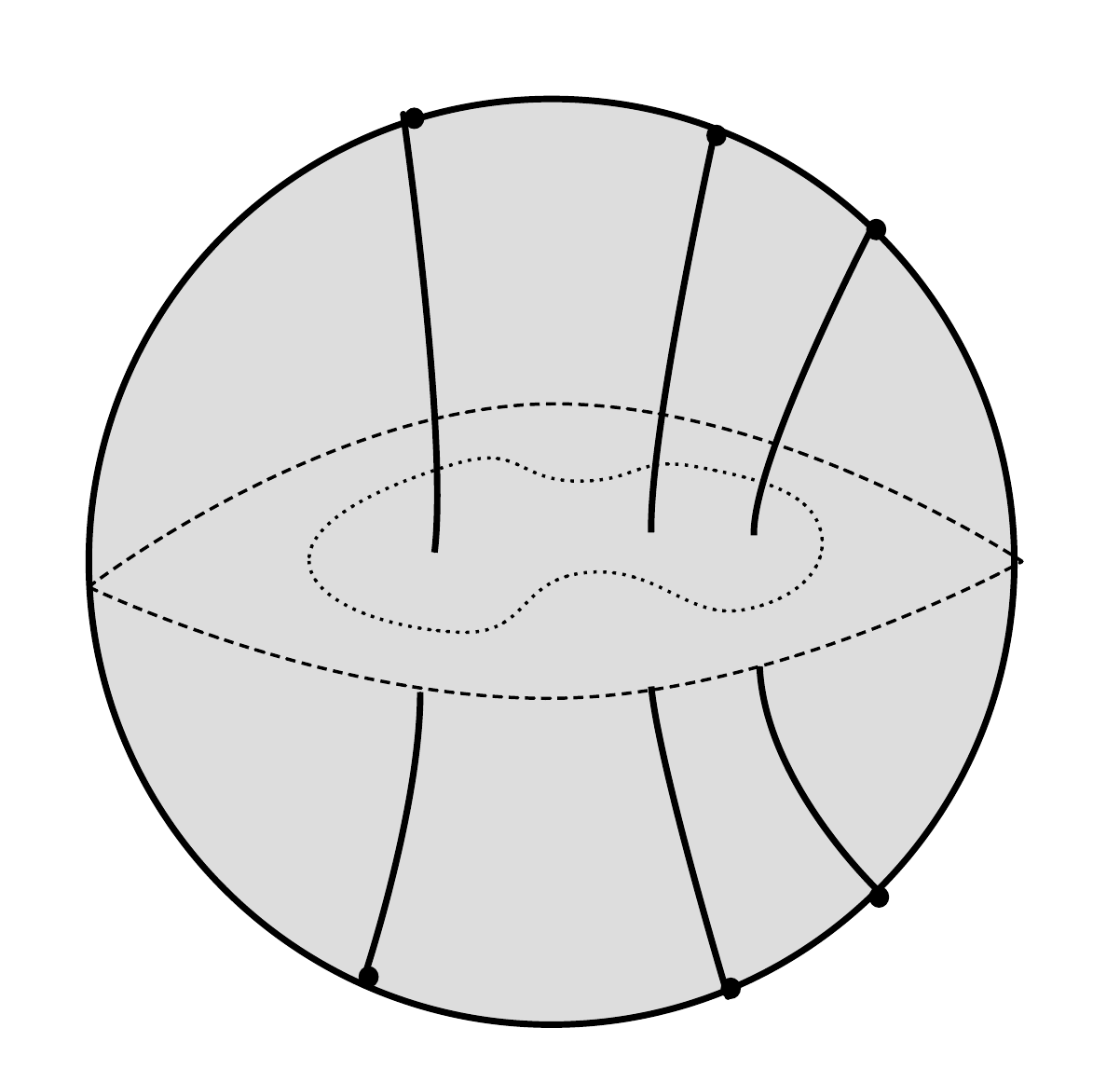}
\put (48,36) {$\Gamma_{\text{in}}$}
\put (33,93) {$1$}
\put (30,0) {$1$}
\put (60,92) {$2$}
\put (64,0) {$2$}
\put (76,78) {$\mathcal{O}$}
\put (76,10) {$\mathcal{O}$}
\end{overpic}
}} 
\end{align}
where $\Gamma_{\text{out}}$ and $\Gamma_{\text{in}}$ are the apparent horizons respectively for the cases where the probe particle is outside or inside the horizon. They are in general different from the apparent horizon for the black hole without the probe particle because of backreaction by the probe particle.
When the particle is outside the horizon, we interpret it to be dressing the microstate by IR correlations in which case we define the dressed coarse-grained state,
\begin{equation}
 \begin{split}
   \overline{\rho}_{\text{out}} &=\sum_p |c_{12p}|^2 \mathcal{O}\left | {\cal B}\left[  \vcenter{\hbox{\vspace{0.12in}
	\begin{tikzpicture}[scale=.75]
	\draw[thick,<-] (-1/2,0) -- (-3/2,0);
	\node[below,scale=0.75] at (-1,0) {$h_p$};
	\draw[thick] (-3/2,0) -- (-3/2,0.8);
	\node[above,scale=0.75] at (-3/2,0.8) {$h_2$};
        \draw[thick] (-3/2,0) -- (-5/2,0);
        \node[left,scale=0.75] at (-5/2,0) {$h_1$};
	\end{tikzpicture}
	}} \right] \right |^2\ket{p}\bra{p}
\left | {\cal B}\left[  \vcenter{\hbox{\vspace{0.12in}
	\begin{tikzpicture}[scale=.75]
	\draw[thick,<-] (-1/2,0) -- (-3/2,0);
	\node[below,scale=0.75] at (-1,0) {$h_p$};
	\draw[thick] (-3/2,0) -- (-3/2,0.8);
	\node[above,scale=0.75] at (-3/2,0.8) {$h_2$};
        \draw[thick] (-3/2,0) -- (-5/2,0);
        \node[left,scale=0.75] at (-5/2,0) {$h_1$};
	\end{tikzpicture}
	}} \right] ^{\dagger}\right|^2\mathcal{O}^{\dagger} \\
    &= \sum_{p,p',q}|c_{12q}|^2 c_{\mathcal{O}pq} c^*_{\mathcal{O}p'q} \left|\mathcal{B}\left[\vcenter{\hbox{\vspace{0.12in}
	\begin{tikzpicture}[scale=.75]
	\draw[thick,->] (-1/2,0) -- (1/2,0);
	\node[below,scale=0.75] at (0,0) {$h_p$};
	\draw[thick] (-1/2,0) -- (-3/2,0);
	\node[below,scale=0.75] at (-1,0) {$h_q$};
	\draw[thick] (-1/2,0) -- (-1/2,1*0.8);
	\node[above,scale=0.75] at (-1/2,0.8) {$h_{\mathcal{O}}$};
	\draw[thick] (-3/2,0) -- (-3/2,0.8);
	\node[above,scale=0.75] at (-3/2,0.8) {$h_2$};
	\draw[thick] (-3/2,0) -- (-3/2-0.8,0);
	\node[left,scale=0.75] at (-3/2-0.8,0) {$h_1$};
	\end{tikzpicture}
	}} \right]\right|^2 \ket{p}\bra{p'} \left|\mathcal{B}\left[\vcenter{\hbox{\vspace{0.12in}
	\begin{tikzpicture}[scale=.75]
	\draw[thick,->] (-1/2,0) -- (1/2,0);
	\node[below,scale=0.75] at (0,0) {$h_{p'}$};
	\draw[thick] (-1/2,0) -- (-3/2,0);
	\node[below,scale=0.75] at (-1,0) {$h_q$};
	\draw[thick] (-1/2,0) -- (-1/2,1*0.8);
	\node[above,scale=0.75] at (-1/2,0.8) {$h_{\mathcal{O}}$};
	\draw[thick] (-3/2,0) -- (-3/2,0.8);
	\node[above,scale=0.75] at (-3/2,0.8) {$h_2$};
	\draw[thick] (-3/2,0) -- (-3/2-0.8,0);
	\node[left,scale=0.75] at (-3/2-0.8,0) {$h_1$};
	\end{tikzpicture}
	}} \right]^\dagger\right|^2
  \end{split}
\end{equation}
Observe that $\overline{\rho}_{\text{out}}$ is obtained from the original state $\Tilde{\rho}$ by introducing a diagonal projection $\delta_{qq'}$ in the sum over primaries.
On the other hand, when the particle is inside the horizon, it should be treated as being part of a new black hole microstate and hence should be coarse grained over. Hence, for this case, we define the coarse grained state,
\begin{equation}
    \overline{\rho}_{\text{in}}=\sum_{p,q}|c_{12q}|^2 |c_{\mathcal{O}qp}|^2 \left|\mathcal{B}\left[\vcenter{\hbox{\vspace{0.12in}
	\begin{tikzpicture}[scale=.75]
	\draw[thick,->] (-1/2,0) -- (1/2,0);
	\node[below,scale=0.75] at (0,0) {$h_p$};
	\draw[thick] (-1/2,0) -- (-3/2,0);
	\node[below,scale=0.75] at (-1,0) {$h_q$};
	\draw[thick] (-1/2,0) -- (-1/2,1*0.8);
	\node[above,scale=0.75] at (-1/2,0.8) {$h_{\mathcal{O}}$};
	\draw[thick] (-3/2,0) -- (-3/2,0.8);
	\node[above,scale=0.75] at (-3/2,0.8) {$h_2$};
	\draw[thick] (-3/2,0) -- (-3/2-0.8,0);
	\node[left,scale=0.75] at (-3/2-0.8,0) {$h_1$};
	\end{tikzpicture}
	}} \right]\right|^2 \ket{p}\bra{p} \left|\mathcal{B}\left[\vcenter{\hbox{\vspace{0.12in}
	\begin{tikzpicture}[scale=.75]
	\draw[thick,->] (-1/2,0) -- (1/2,0);
	\node[below,scale=0.75] at (0,0) {$h_p$};
	\draw[thick] (-1/2,0) -- (-3/2,0);
	\node[below,scale=0.75] at (-1,0) {$h_q$};
	\draw[thick] (-1/2,0) -- (-1/2,1*0.8);
	\node[above,scale=0.75] at (-1/2,0.8) {$h_{\mathcal{O}}$};
	\draw[thick] (-3/2,0) -- (-3/2,0.8);
	\node[above,scale=0.75] at (-3/2,0.8) {$h_2$};
	\draw[thick] (-3/2,0) -- (-3/2-0.8,0);
	\node[left,scale=0.75] at (-3/2-0.8,0) {$h_1$};
	\end{tikzpicture}
	}} \right]^\dagger\right|^2
\end{equation}
We see that $\overline{\rho}_{\text{out}}$ is obtained from the original state $\Tilde{\rho}$ by introducing two diagonal projections $\delta_{qq'}\delta_{pp'}$ in the sum over primaries with the idea that all the correlations in the interior have to be decohered while coarse graining.
Observe that with either coarse graining prescription, the trace of the coarse grained state evaluates to the square of the 6-point identity block in the star channel with the operators fusing with their respective adjoints into the identity representation,
\begin{equation}
   \text{Tr}(\overline{\rho}_{\text{in}}) \approx \left | \vcenter{\hbox{
	\begin{tikzpicture}[scale=0.75]
	\draw[thick] (0,0) -- (0,1);
	\draw[thick](0,1) -- (-0.866*0.8,1+0.8*0.5);
	\draw[thick] (0,1) -- (0.866*0.8,1+0.8*0.5);
	\draw[thick] (0,0) -- (0.866,-1/2);
	\draw[thick] (0,0) -- (-0.866,-1/2);
	\draw[thick] (0.866,-1/2) -- (0.866+0.8*0.866,-1/2+0.8*1/2);
	\draw[thick] (0.866,-1/2) -- (0.866,-1/2-0.8);
	\draw[thick] (-0.866,-1/2) -- (-0.866-0.8*0.866,-1/2+0.8*1/2);
	\draw[thick] (-0.866,-1/2) -- (-0.866,-1/2-0.8);
	\node[left,scale=0.75] at (-0.866*0.8,1+0.8*0.5) {$2$};
	\node[right,scale=0.75] at (0.866*0.8,1+0.8*0.5) {$2$};
	\node[right,scale=0.75] at (0.866+0.8*0.866,-1/2+0.8*1/2) {$\mathcal{O}$};
	\node[below,scale=0.75] at (0.866,-1/2-0.8) {$\mathcal{O}$};
	\node[left, scale=0.75] at (-0.866-0.8*0.866,-1/2+0.8*1/2) {$1$};
	\node[below, scale=0.75] at (-0.866,-1/2-0.8) {$1$};
	\node[left, scale=0.75] at (0,1/2) {$\id$};
	\node[above, scale=0.75] at (1/2*0.866,-1/2*1/2) {$\id$};
	\node[above, scale=0.75] at (-1/2*0.866,-1/2*1/2) {$\id$};
	\end{tikzpicture}
	}} \right |^2 \approx \text{Tr}(\overline{\rho}_{\text{out}})
\end{equation}
 This is verified using the following identity derived using appropriate fusion moves \cite{Chandra:2022bqq},
\begin{equation}
  \vcenter{\hbox{
	\begin{tikzpicture}[scale=0.75]
	\draw[thick] (0,0) -- (0,1);
	\draw[thick](0,1) -- (-0.866*0.8,1+0.8*0.5);
	\draw[thick] (0,1) -- (0.866*0.8,1+0.8*0.5);
	\draw[thick] (0,0) -- (0.866,-1/2);
	\draw[thick] (0,0) -- (-0.866,-1/2);
	\draw[thick] (0.866,-1/2) -- (0.866+0.8*0.866,-1/2+0.8*1/2);
	\draw[thick] (0.866,-1/2) -- (0.866,-1/2-0.8);
	\draw[thick] (-0.866,-1/2) -- (-0.866-0.8*0.866,-1/2+0.8*1/2);
	\draw[thick] (-0.866,-1/2) -- (-0.866,-1/2-0.8);
	\node[left,scale=0.75] at (-0.866*0.8,1+0.8*0.5) {$2$};
	\node[right,scale=0.75] at (0.866*0.8,1+0.8*0.5) {$2$};
	\node[right,scale=0.75] at (0.866+0.8*0.866,-1/2+0.8*1/2) {$\mathcal{O}$};
	\node[below,scale=0.75] at (0.866,-1/2-0.8) {$\mathcal{O}$};
	\node[left, scale=0.75] at (-0.866-0.8*0.866,-1/2+0.8*1/2) {$1$};
	\node[below, scale=0.75] at (-0.866,-1/2-0.8) {$1$};
	\node[left, scale=0.75] at (0,1/2) {$\id$};
	\node[above, scale=0.75] at (1/2*0.866,-1/2*1/2) {$\id$};
	\node[above, scale=0.75] at (-1/2*0.866,-1/2*1/2) {$\id$};
	\end{tikzpicture}
	}}  = \int dh_pdh_q \rho_0(h_p)\rho(h_q) C_0(h_1,h_2,h_q)C_0(h_{\mathcal{O}},h_p,h_q)
\vcenter{\hbox{
	\begin{tikzpicture}[scale=.75]
	\draw[thick] (-1/2,0) -- (1/2,0);
	\node[below,scale=0.75] at (0,0) {$p$};
	\draw[thick] (-1/2,0) -- (-3/2,0);
	\node[below,scale=0.75] at (-1,0) {$q$};
	\draw[thick] (-1/2,0) -- (-1/2,1*0.8);
	\node[above,scale=0.75] at (-1/2,0.8) {$\mathcal{O}$};
	\draw[thick] (-3/2,0) -- (-3/2,0.8);
	\node[above,scale=0.75] at (-3/2,0.8) {$2$};
	\draw[thick] (-3/2,0) -- (-3/2-0.8,0);
	\node[left,scale=0.75] at (-3/2-0.8,0) {$1$};
	\draw[thick] (1/2,0) -- (1/2,0.8);
	\node[above,scale=0.75] at (1/2,0.8) {$\mathcal{O}$};
	\draw[thick] (1/2,0) -- (3/2,0);
	\node[below,scale=0.75] at (1,0) {$q$};
	\draw[thick] (3/2,0) -- (3/2,0.8);
	\node[above,scale=0.75] at (3/2,0.8) {$2$};
	\draw[thick] (3/2,0) -- (3/2+0.8,0);
	\node[right,scale=0.75] at (3/2+0.8,0) {$1$};
	\end{tikzpicture}
	}}
\end{equation}
So, we cannot distinguish between the two coarse graining prescriptions at this level. However, the Renyis of the two states can distinguish between the two prescriptions,
\begin{equation}
      \text{Tr}(\overline{\rho}_{\text{in}}^k) \approx \left |\int dh_p \rho_0(h_p) \int \prod_{i=1}^kdh_{q_i}\rho_0(h_{q_i})  C_0(h_1,h_2,h_{q_i})  C_0(h_{\mathcal{O}},h_p,h_{q_i}) \vcenter{\hbox{
	\begin{tikzpicture}[scale=.75]
	\draw[thick] (-1/2,0) -- (1/2,0);
	\node[below,scale=0.75] at (0,0) {$p$};
	\draw[thick] (-1/2,0) -- (-3/2,0);
	\node[below,scale=0.75] at (-1,0) {$q_i$};
	\draw[thick] (-1/2,0) -- (-1/2,1*0.8);
	\node[above,scale=0.75] at (-1/2,0.8) {$\mathcal{O}$};
	\draw[thick] (-3/2,0) -- (-3/2,0.8);
	\node[above,scale=0.75] at (-3/2,0.8) {$2$};
	\draw[thick] (-3/2,0) -- (-3/2-0.8,0);
	\node[left,scale=0.75] at (-3/2-0.8,0) {$1$};
	\draw[thick] (1/2,0) -- (1/2,0.8);
	\node[above,scale=0.75] at (1/2,0.8) {$\mathcal{O}$};
	\draw[thick] (1/2,0) -- (3/2,0);
	\node[below,scale=0.75] at (1,0) {$q_{i+1}$};
	\draw[thick] (3/2,0) -- (3/2,0.8);
	\node[above,scale=0.75] at (3/2,0.8) {$2$};
	\draw[thick] (3/2,0) -- (3/2+0.8,0);
	\node[right,scale=0.75] at (3/2+0.8,0) {$1$};
	\end{tikzpicture}
	}} \right |^2
\end{equation}
and 
\begin{multline}
      \text{Tr}(\overline{\rho}_{\text{out}}^k) \approx \left |\int dh_p \rho_0(h_p) \int \prod_{i=1}^kdh_{q_i}\rho_0(h_{q_i})  C_0(h_1,h_2,h_{q_i})  C_0(h_{\mathcal{O}},h_p,h_{q_i}) \vcenter{\hbox{
	\begin{tikzpicture}[scale=.75]
	\draw[thick] (-1/2,0) -- (1/2,0);
	\node[below,scale=0.75] at (0,0) {$p$};
	\draw[thick] (-1/2,0) -- (-3/2,0);
	\node[below,scale=0.75] at (-1,0) {$q_i$};
	\draw[thick] (-1/2,0) -- (-1/2,1*0.8);
	\node[above,scale=0.75] at (-1/2,0.8) {$\mathcal{O}$};
	\draw[thick] (-3/2,0) -- (-3/2,0.8);
	\node[above,scale=0.75] at (-3/2,0.8) {$2$};
	\draw[thick] (-3/2,0) -- (-3/2-0.8,0);
	\node[left,scale=0.75] at (-3/2-0.8,0) {$1$};
	\draw[thick] (1/2,0) -- (1/2,0.8);
	\node[above,scale=0.75] at (1/2,0.8) {$\mathcal{O}$};
	\draw[thick] (1/2,0) -- (3/2,0);
	\node[below,scale=0.75] at (1,0) {$q_{i+1}$};
	\draw[thick] (3/2,0) -- (3/2,0.8);
	\node[above,scale=0.75] at (3/2,0.8) {$2$};
	\draw[thick] (3/2,0) -- (3/2+0.8,0);
	\node[right,scale=0.75] at (3/2+0.8,0) {$1$};
	\end{tikzpicture}
	}} \right |^2 \\
    +  \left |\int dh_q \rho_0(h_q) C_0^k(h_1,h_2,h_q)  \int \prod_{i=1}^kdh_{p_i}\rho_0(h_{p_i})  C_0(h_{\mathcal{O}},h_{p_i},h_q) \vcenter{\hbox{
	\begin{tikzpicture}[scale=.75]
	\draw[thick] (-1/2,0) -- (1/2,0);
	\node[below,scale=0.75] at (0,0) {$p_i$};
	\draw[thick] (-1/2,0) -- (-3/2,0);
	\node[below,scale=0.75] at (-1,0) {$q$};
	\draw[thick] (-1/2,0) -- (-1/2,1*0.8);
	\node[above,scale=0.75] at (-1/2,0.8) {$\mathcal{O}$};
	\draw[thick] (-3/2,0) -- (-3/2,0.8);
	\node[above,scale=0.75] at (-3/2,0.8) {$2$};
	\draw[thick] (-3/2,0) -- (-3/2-0.8,0);
	\node[left,scale=0.75] at (-3/2-0.8,0) {$1$};
	\draw[thick] (1/2,0) -- (1/2,0.8);
	\node[above,scale=0.75] at (1/2,0.8) {$\mathcal{O}$};
	\draw[thick] (1/2,0) -- (3/2,0);
	\node[below,scale=0.75] at (1,0) {$q$};
	\draw[thick] (3/2,0) -- (3/2,0.8);
	\node[above,scale=0.75] at (3/2,0.8) {$2$};
	\draw[thick] (3/2,0) -- (3/2+0.8,0);
	\node[right,scale=0.75] at (3/2+0.8,0) {$1$};
	\end{tikzpicture}
	}} \right |^2
\end{multline}
where the two terms correspond to different sets of Wick contractions between the OPE coefficients.\footnote{We have implicitly assumed that products of OPE coefficients are self-averaging in this derivation so the coarse graining prescription agrees with the Gaussian ensemble average at large-$c$ \cite{Chandra:2022bqq}. It would be interesting to verify this assumption by computing non-Gaussianities and checking if they are suppressed relative to the Gaussian contribution.} Note that there are other Wick contractions which correspond to non-replica symmetric contributions to $ \text{Tr}(\overline{\rho}_{\text{out}}^k)$ which we have not written in the above expression. These contributions have fewer $\rho_0$ factors in them so we expect that they are sub-dominant in the semiclassical limit in some region of the moduli space. In the subsequent discussions, we shall however assume that these contributions are not important in the $k\to 1$ limit.

Now, we wish to match the Renyi entropies of either coarse grained states with the partition functions of appropriate multi-boundary wormholes. For the case where the particle is outside the horizon, we observe that by branching around the horizon and replicating, we get multi-boundary wormholes with the heavy defects going across the wormhole cyclically between different asymptotic boundaries while the probe particle's trajectories start and end on the same boundary.
On the other hand, for the case where the particle is inside the horizon, the wormholes obtained by branching around the backreacted horizon have both the heavy defects and the probe particle going across the wormhole.
The wormhole amplitudes in either case can be calculated using the Virasoro TQFT prescription \cite{Collier:2023fwi} or using the large-$c$ ensemble \cite{Chandra:2022bqq} and read,
\begin{multline} \label{Zksame}
     Z_k^{\text{same}} \approx \left |\int dh_q \rho_0(h_q) C_0^k(h_1,h_2,h_q)  \int \prod_{i=1}^kdh_{p_i}\rho_0(h_{p_i})  C_0(h_{\mathcal{O}},h_{p_i},h_q) \vcenter{\hbox{
	\begin{tikzpicture}[scale=.75]
	\draw[thick] (-1/2,0) -- (1/2,0);
	\node[below,scale=0.75] at (0,0) {$p_i$};
	\draw[thick] (-1/2,0) -- (-3/2,0);
	\node[below,scale=0.75] at (-1,0) {$q$};
	\draw[thick] (-1/2,0) -- (-1/2,1*0.8);
	\node[above,scale=0.75] at (-1/2,0.8) {$\mathcal{O}$};
	\draw[thick] (-3/2,0) -- (-3/2,0.8);
	\node[above,scale=0.75] at (-3/2,0.8) {$2$};
	\draw[thick] (-3/2,0) -- (-3/2-0.8,0);
	\node[left,scale=0.75] at (-3/2-0.8,0) {$1$};
	\draw[thick] (1/2,0) -- (1/2,0.8);
	\node[above,scale=0.75] at (1/2,0.8) {$\mathcal{O}$};
	\draw[thick] (1/2,0) -- (3/2,0);
	\node[below,scale=0.75] at (1,0) {$q$};
	\draw[thick] (3/2,0) -- (3/2,0.8);
	\node[above,scale=0.75] at (3/2,0.8) {$2$};
	\draw[thick] (3/2,0) -- (3/2+0.8,0);
	\node[right,scale=0.75] at (3/2+0.8,0) {$1$};
	\end{tikzpicture}
	}} \right |^2
\end{multline}
\begin{multline} \label{Zkacross}
     Z_k^{\text{across}} \approx \left |\int dh_p \rho_0(h_p) \int \prod_{i=1}^kdh_{q_i}\rho_0(h_{q_i})  C_0(h_1,h_2,h_{q_i})  C_0(h_{\mathcal{O}},h_p,h_{q_i}) \vcenter{\hbox{
	\begin{tikzpicture}[scale=.75]
	\draw[thick] (-1/2,0) -- (1/2,0);
	\node[below,scale=0.75] at (0,0) {$p$};
	\draw[thick] (-1/2,0) -- (-3/2,0);
	\node[below,scale=0.75] at (-1,0) {$q_i$};
	\draw[thick] (-1/2,0) -- (-1/2,1*0.8);
	\node[above,scale=0.75] at (-1/2,0.8) {$\mathcal{O}$};
	\draw[thick] (-3/2,0) -- (-3/2,0.8);
	\node[above,scale=0.75] at (-3/2,0.8) {$2$};
	\draw[thick] (-3/2,0) -- (-3/2-0.8,0);
	\node[left,scale=0.75] at (-3/2-0.8,0) {$1$};
	\draw[thick] (1/2,0) -- (1/2,0.8);
	\node[above,scale=0.75] at (1/2,0.8) {$\mathcal{O}$};
	\draw[thick] (1/2,0) -- (3/2,0);
	\node[below,scale=0.75] at (1,0) {$q_{i+1}$};
	\draw[thick] (3/2,0) -- (3/2,0.8);
	\node[above,scale=0.75] at (3/2,0.8) {$2$};
	\draw[thick] (3/2,0) -- (3/2+0.8,0);
	\node[right,scale=0.75] at (3/2+0.8,0) {$1$};
	\end{tikzpicture}
	}} \right |^2
\end{multline}
where the superscripts denote whether the probe particle goes `across' the wormhole or starts and ends on the `same' boundary.
Importantly, we observe that the Renyi entropies of the two coarse grained states match with the above multi-boundary wormhole amplitudes in the semiclassical limit,
\begin{equation}
     \text{Tr}(\overline{\rho}_{\text{in}}^k) \approx Z_k^{\text{same}}, \qquad \qquad  \text{Tr}(\overline{\rho}_{\text{out}}^k) \approx Z_k^{\text{same}}+Z_k^{\text{across}}
\end{equation}
 For example, the four-boundary wormholes contributing to the fourth Renyi of the coarse grained state in either case are sketched below,
\begin{align}
 \begin{split}
 \text{Tr}(\overline{\rho}_{\text{out}}^4) \approx    & \vcenter{\hbox{
\begin{overpic}[width=1.8in,grid=false]{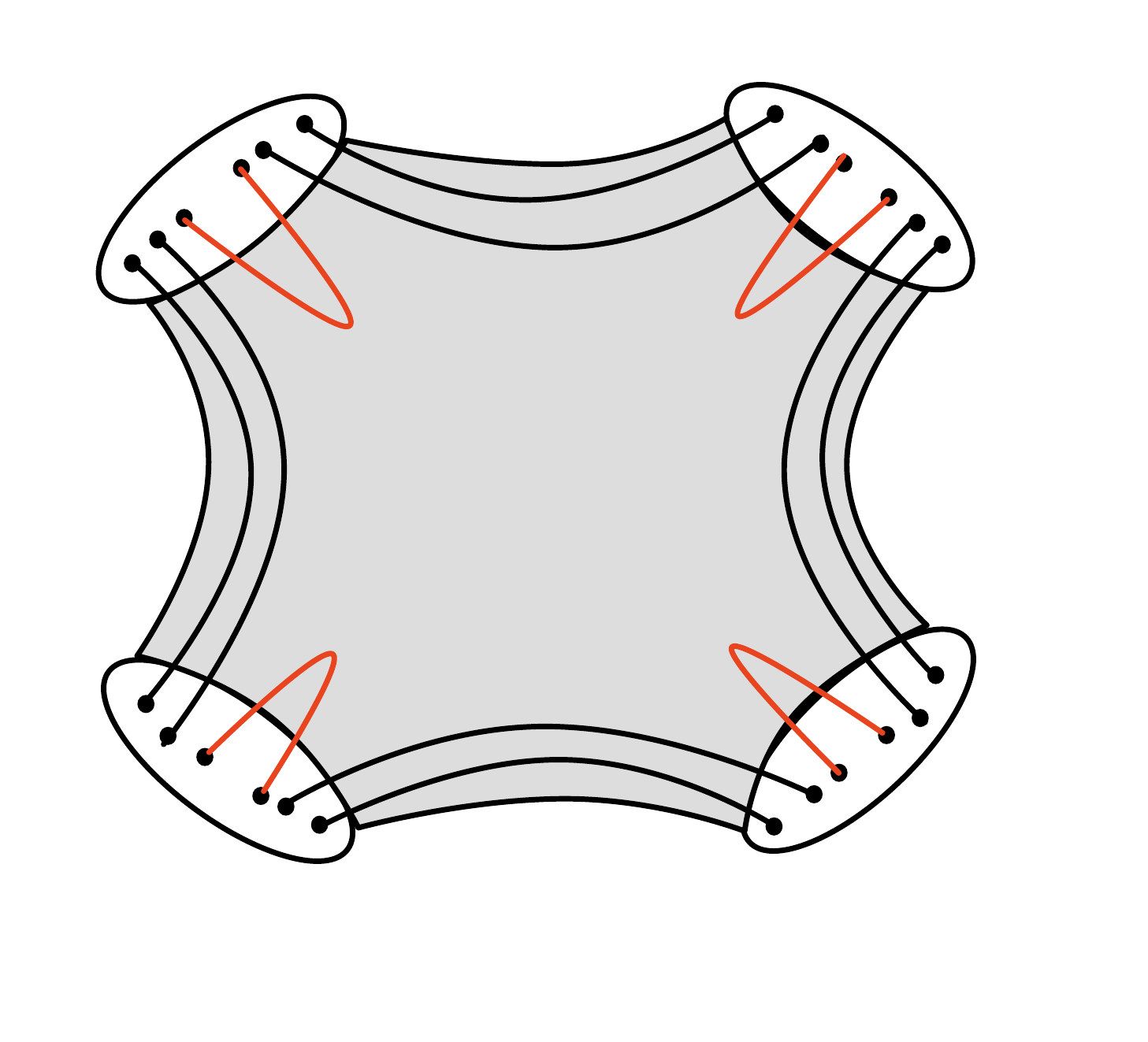}
\end{overpic}
}} + \vcenter{\hbox{
\begin{overpic}[width=1.7in,grid=false]{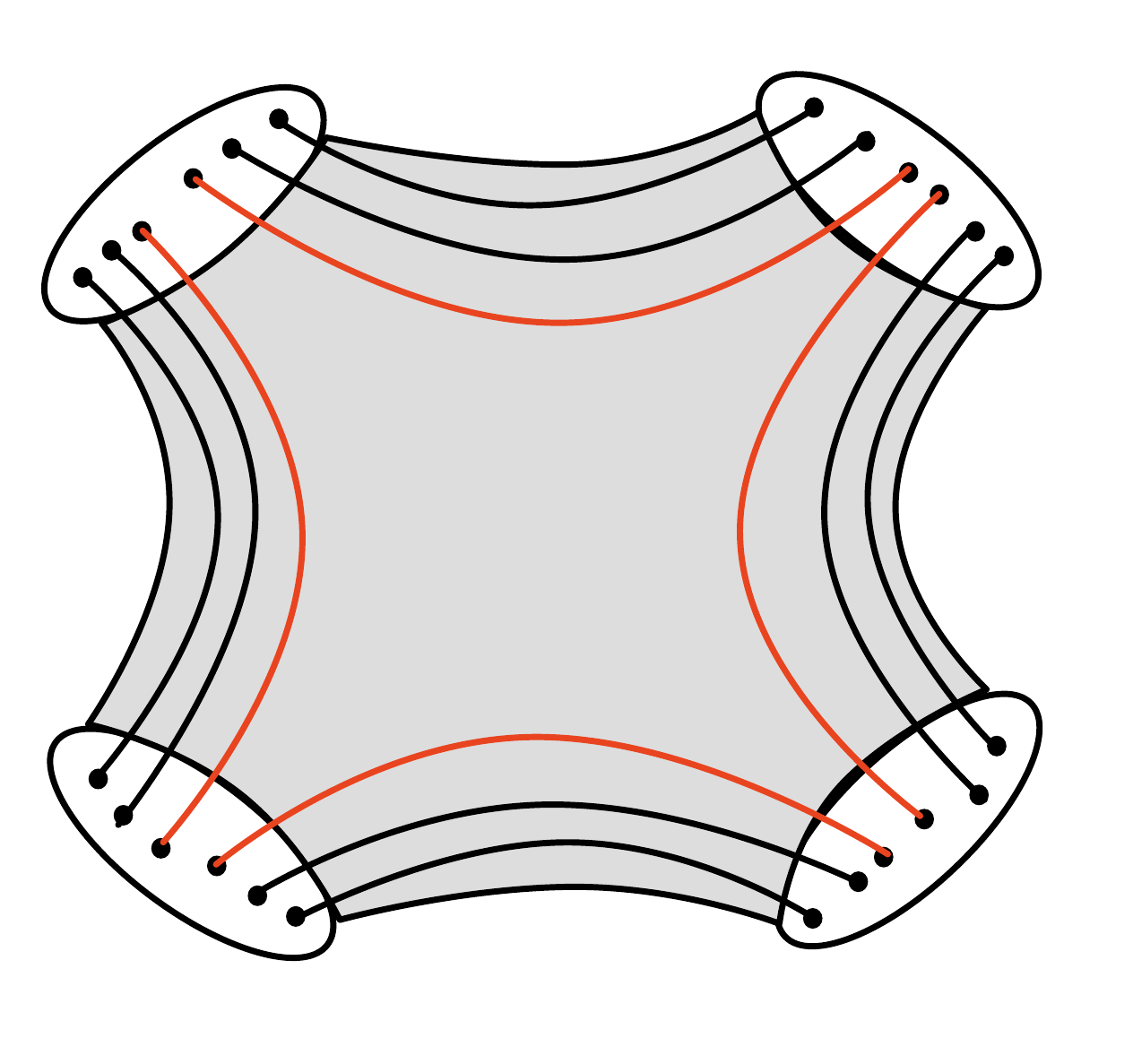}
\end{overpic}
}} \\
 \text{Tr}(\overline{\rho}_{\text{in}}^4) \approx  &\vcenter{\hbox{
\begin{overpic}[width=1.7in,grid=false]{more-figures/fourpointfourbdryprobein.pdf}
\end{overpic}
}} 
 \end{split}
\end{align}
with the trajectory of the probe sketched in red. We can exploit the fact that we are imposing replica-symmetric boundary conditions corresponding to norms of CFT states to simplify the expressions for the wormhole amplitudes at the saddlepoint. For the case where the probe is outside the horizon, the bulk sets the internal primaries $q$ to be the same on all the boundaries also identifying them on the same boundary while the primaries $p_i$ are in general different. However, for the present case where we choose the cross ratios to be real and same on each boundary due to replica symmetry, the saddlepoint equations for each of the internal primaries $p_i$ are the same, so the saddle point weights of $p_i$ are expected to match on the $k$ copies. For the case where the probe is inside the horizon, the bulk sets the internal primaries $p$ to match on all the boundaries while $q_i$ are cyclically identified between successive boundaries. Again, for the chosen kinematics, the saddle point equations for $q_i$ are symmetrical so that at the saddle, the weights of $q_i$ match. Thus, for the chosen replica-symmetric kinematics with real cross ratios, the wormhole amplitudes at the saddlepoint can be expressed as
\begin{equation}
  \begin{split}
      & Z_k^{\text{across}} \approx \left | \rho_0(h_p)\rho_0(h_q)^k C_0^k(h_1,h_2,h_q) C_0^k(h_{\mathcal{O}},h_p,h_q) \left ( \vcenter{\hbox{
	\begin{tikzpicture}[scale=.75]
	\draw[thick] (-1/2,0) -- (1/2,0);
	\node[below,scale=0.75] at (0,0) {$p$};
	\draw[thick] (-1/2,0) -- (-3/2,0);
	\node[below,scale=0.75] at (-1,0) {$q$};
	\draw[thick] (-1/2,0) -- (-1/2,1*0.8);
	\node[above,scale=0.75] at (-1/2,0.8) {$\mathcal{O}$};
	\draw[thick] (-3/2,0) -- (-3/2,0.8);
	\node[above,scale=0.75] at (-3/2,0.8) {$2$};
	\draw[thick] (-3/2,0) -- (-3/2-0.8,0);
	\node[left,scale=0.75] at (-3/2-0.8,0) {$1$};
	\draw[thick] (1/2,0) -- (1/2,0.8);
	\node[above,scale=0.75] at (1/2,0.8) {$\mathcal{O}$};
	\draw[thick] (1/2,0) -- (3/2,0);
	\node[below,scale=0.75] at (1,0) {$q$};
	\draw[thick] (3/2,0) -- (3/2,0.8);
	\node[above,scale=0.75] at (3/2,0.8) {$2$};
	\draw[thick] (3/2,0) -- (3/2+0.8,0);
	\node[right,scale=0.75] at (3/2+0.8,0) {$1$};
	\end{tikzpicture}
	}} \right )^k \right |^2 \\
    & Z_k^{\text{same}} \approx \left | \rho_0(h_p)^k\rho_0(h_q) C_0^k(h_1,h_2,h_q) C_0^k(h_{\mathcal{O}},h_p,h_q) \left ( \vcenter{\hbox{
	\begin{tikzpicture}[scale=.75]
	\draw[thick] (-1/2,0) -- (1/2,0);
	\node[below,scale=0.75] at (0,0) {$p$};
	\draw[thick] (-1/2,0) -- (-3/2,0);
	\node[below,scale=0.75] at (-1,0) {$q$};
	\draw[thick] (-1/2,0) -- (-1/2,1*0.8);
	\node[above,scale=0.75] at (-1/2,0.8) {$\mathcal{O}$};
	\draw[thick] (-3/2,0) -- (-3/2,0.8);
	\node[above,scale=0.75] at (-3/2,0.8) {$2$};
	\draw[thick] (-3/2,0) -- (-3/2-0.8,0);
	\node[left,scale=0.75] at (-3/2-0.8,0) {$1$};
	\draw[thick] (1/2,0) -- (1/2,0.8);
	\node[above,scale=0.75] at (1/2,0.8) {$\mathcal{O}$};
	\draw[thick] (1/2,0) -- (3/2,0);
	\node[below,scale=0.75] at (1,0) {$q$};
	\draw[thick] (3/2,0) -- (3/2,0.8);
	\node[above,scale=0.75] at (3/2,0.8) {$2$};
	\draw[thick] (3/2,0) -- (3/2+0.8,0);
	\node[right,scale=0.75] at (3/2+0.8,0) {$1$};
	\end{tikzpicture}
	}} \right )^k \right |^2
  \end{split}
\end{equation}
where all the conformal weights in the above expression are saddlepoint weights and  $\rho_0, C_0$ and the block should be replaced by their large-$c$ exponentiated expressions.
Now, we can determine the von-Neumann entropies of the two coarse grained states from the Renyi entropies calcuated from wormhole amplitudes. To this end, we observe that\footnote{In the calculation described in (\ref{entrprobe}), we have written the terms involving the Liouville structure constants and the blocks collectively after factoring out the explicit $k$ dependence in terms of functions $\alpha$ which depend on $k$ only through the dependence of the saddle point weights. These terms cancel against the terms coming from the derivatives of the Cardy entropy when the saddlepoint equation is imposed. },
\begin{equation} \label{entrprobe}
   \begin{split}
       & \partial_k \log\left (\frac{Z^{\text{across}}_k}{Z_1^k}\right)\bigg |_{k=1} =- \partial_k \left((\frac{k}{2}-1)S_{p_k}+k\frac{S_{p_1}}{2}+k\alpha_{p_k,q_k}-k\alpha_{p_1,q_1}\right )\bigg |_{k=1} = - S_{p_1} \\
       & \partial_k \log\left (\frac{Z^{\text{same}}_k}{Z_1^k}\right)\bigg |_{k=1} = -\partial_k \left((k-1)S_{q_k}-\frac{k}{2}S_{p_k}+k\frac{S_{p_1}}{2}+k\alpha_{p_k,q_k}-k\alpha_{p_1,q_1}\right )\bigg |_{k=1} = - S_{q_1}
   \end{split}
\end{equation}
where $S_p$ and $S_q$ denote the Cardy entropies\footnote{Recall the Cardy entropy at large-$c$ takes the functional form, $S_0(h,\overline{h})=2\pi\sqrt{\frac{c}{6}(h-\frac{c}{24})}+2\pi\sqrt{\frac{c}{6}(\overline{h}-\frac{c}{24})}$} at the saddle-point weights of the internal primaries $p$ and $q$, which are complicated functions of the two cross ratios between the 6 insertion points on the boundary $S^2$ for either case. However, their explicit form is not important for our purpose. Note that for the chosen kinematics, the saddle lands on scalar primaries in each of the internal legs. Also note that the calculation in (\ref{entrprobe}) makes sense only if the saddlepoint equations for $Z_k$ have a solution as $k\to 1$. When the probe is outside the horizon, we expect that there is no solution to the saddlepoint equation for $Z_k^{\text{across}}$ simply because there is no extremal surface on the time-symmetric slice surrounding the probe particle. Thus, the entropies of the two coarse grained states read,
\begin{equation}
  \begin{split}
      & S(\overline{\rho}_{\text{in}}) = S_{p_1}=\frac{\text{Area}(\Gamma_{\text{in}})}{4G_N} \\
      &  S(\overline{\rho}_{\text{out}}) = S_{q_1}=\frac{\text{Area}(\Gamma_{\text{out}})}{4G_N}
   \end{split}
\end{equation}
Importantly, we observe that the entropies of the coarse grained states for either case match with the entropies of the corresponding backreacted horizons justifying our coarse graining prescriptions. Note that the relation between $S_{p_1}$ and $S_{q_1}$ for either case can be calculated using Zamolodchikov's monodromy method for the conformal block \cite{ZamoRecursion} as was done in \cite{Chandra:2023dgq} in the context of deriving an isometric transition for holographic codes across the horizon. In particular, it was shown in \cite{Chandra:2023dgq} that the probe outside the horizon acts isometrically which in the present notation would correspond to $S_{p_1}>S_{q_1}$ and the probe inside the horizon acts co-isometrically meaning $S_{p_1}<S_{q_1}$.

Thus, combining the results of sections (\ref{CGrepWh}) and (\ref{CGprobe}), we have shown that the coarse grained state (\ref{CGstate}) has the following features:
\begin{enumerate}
  \item Its Renyi entropies match with the partition functions of replica wormhole geometries described by branching around the time-symmetric apparent horizon of the black hole. 
\item It preserves the correlations of geodesic probes outside the horizon and of boundary gravitons.
\end{enumerate}
So far, we have been cavalier about the use of the term `coarse grained state'. We use the above two features to more generally define a coarse grained state outside the apparent horizon on the time-symmetric spatial slice of a black hole geometry. It would be interesting to understand how this holographic coarse graining prescription motivated by wormholes relates to the complexity coarse graining preocedure of Engelhardt-Wall \cite{Engelhardt:2018kcs,Engelhardt:2017aux}.

\subsection{Coarse graining and inner minimal surfaces} \label{CGinner}

So far, we illustrated our coarse graining formalism using pure state black hole geometries with a single time-symmetric apparent horizon. However, we could consider more general black hole geometries formed by the backreaction of several heavy conical defects. These geometries are expected to have multiple minimal geodesics on the time-symmetric spatial slice which we shall refer to as inner horizons. Precisely, when all the operators dual to the conical defects are above the multi-twist theshold ($h=\overline{h}>\frac{c}{24}$), then at the saddle, all the internal states in any OPE channel are above the black hole threshold \cite{Collier:2018exn}. The scaling dimensions of these states is related to the length of a corresponding minimal geodesic by the relation $\Delta_i=\frac{c}{12}(1+(\frac{\ell_i}{2\pi})^2)$ which is equivalent to saying that the area matches with the Cardy entropy at the saddle-point weights i.e, $\frac{\ell_i}{4G_N}=S_0(h_i^*,\overline{h}_i^*)$ \cite{Chandra:2023dgq}. So, there is a family of nested locally minimal geodesics which we shall refer to as inner minimal surfaces on the time-symmetric slice of the dual black hole geometry in any OPE channel. We can provide a coarse grained intepretation for the area of each of these inner minimal surfaces using the replica formalism illustrated in the previous section with the wormholes constructed by branching around the inner horizon of interest. We illustrate this with the example of a pure state obtained by exciting the vacuum by three scalar primary operators above the multi-twist threshold,
\begin{equation}
   \ket{\Psi}=\mathcal{O}_3\mathcal{O}_2\mathcal{O}_1\ket{0}= \sum_{p,q}c_{12q}c_{q3p} \left | \mathcal{B}\left[\vcenter{\hbox{\vspace{0.12in}
	\begin{tikzpicture}[scale=.75]
	\draw[thick,->] (-1/2,0) -- (1/2,0);
	\node[below,scale=0.75] at (0,0) {$h_p$};
	\draw[thick] (-1/2,0) -- (-3/2,0);
	\node[below,scale=0.75] at (-1,0) {$h_q$};
	\draw[thick] (-1/2,0) -- (-1/2,1*0.8);
	\node[above,scale=0.75] at (-1/2,0.8) {$h_3$};
	\draw[thick] (-3/2,0) -- (-3/2,0.8);
	\node[above,scale=0.75] at (-3/2,0.8) {$h_2$};
	\draw[thick] (-3/2,0) -- (-3/2-0.8,0);
	\node[left,scale=0.75] at (-3/2-0.8,0) {$h_1$};
	\end{tikzpicture}
	}} \right]\right|^2 \ket{p}
\end{equation}
where we have expanded the OPE in a particular channel just for illustration. The spatial slice of the dual black hole geometry is sketched below:
\begin{align} 
 \vcenter{\hbox{
\begin{overpic}[width=1.5in,grid=false]{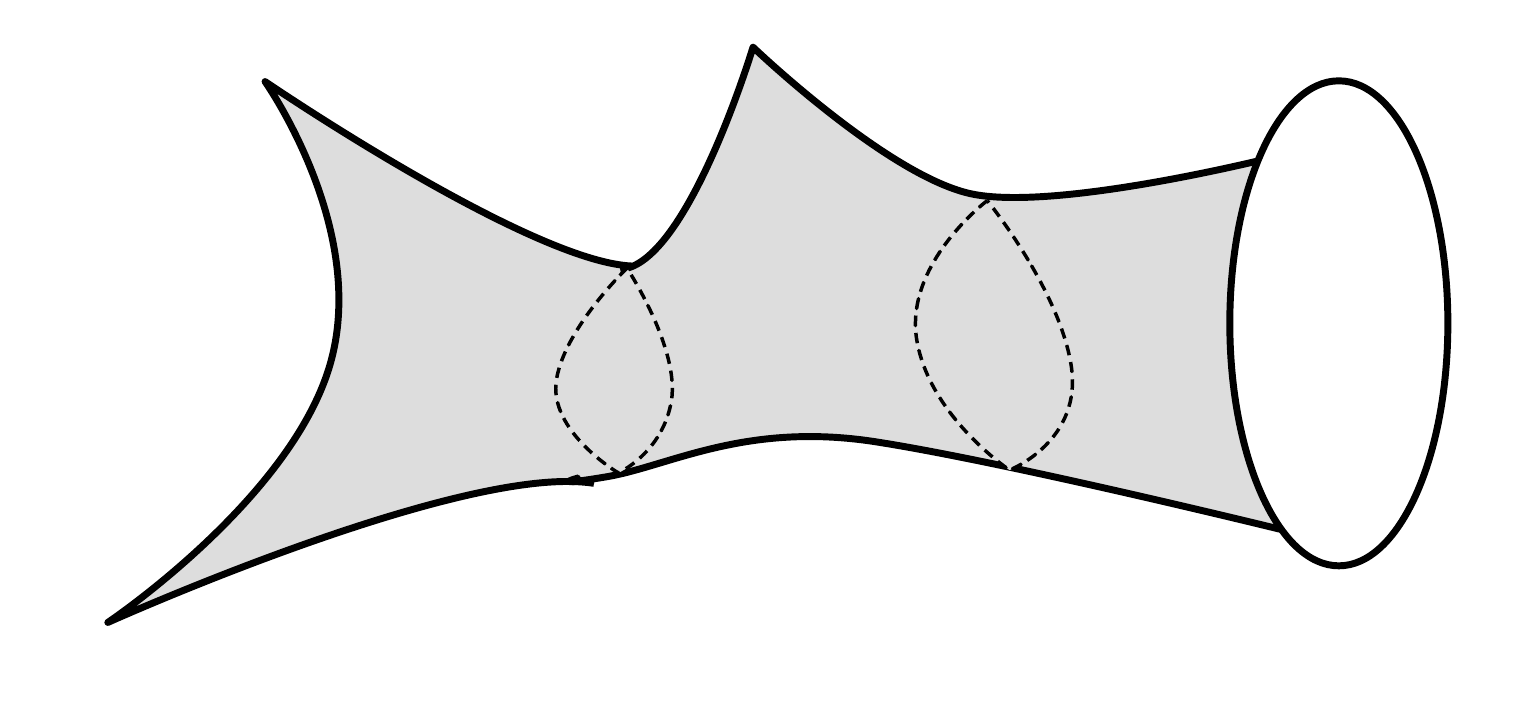}
\put (38,5) {{$\Gamma_{\text{inner}}$}}
\put (62,5) {$\Gamma_{\text{outer}}$}
\put (3,3) {$1$}
\put (15,42) {$2$}
\put (47,44) {$3$}
\end{overpic}
}}
\end{align}
The two minimal geodesics marked are the inner and outer minimal surfaces corresponding to the OPE channel described above. The two coarse grained states obtained by decohering the microstate in the interior of the outer and inner minimal surfaces are respectively given by,
\begin{equation} \label{CGstates}
 \begin{split}
   & \overline{\rho}_{\text{outer}} =\sum_{p,q}|c_{12q}|^2 |c_{3qp}|^2 \left|\mathcal{B}\left[\vcenter{\hbox{\vspace{0.12in}
	\begin{tikzpicture}[scale=.75]
	\draw[thick,->] (-1/2,0) -- (1/2,0);
	\node[below,scale=0.75] at (0,0) {$h_p$};
	\draw[thick] (-1/2,0) -- (-3/2,0);
	\node[below,scale=0.75] at (-1,0) {$h_q$};
	\draw[thick] (-1/2,0) -- (-1/2,1*0.8);
	\node[above,scale=0.75] at (-1/2,0.8) {$h_3$};
	\draw[thick] (-3/2,0) -- (-3/2,0.8);
	\node[above,scale=0.75] at (-3/2,0.8) {$h_2$};
	\draw[thick] (-3/2,0) -- (-3/2-0.8,0);
	\node[left,scale=0.75] at (-3/2-0.8,0) {$h_1$};
	\end{tikzpicture}
	}} \right]\right|^2 \ket{p}\bra{p} \left|\mathcal{B}\left[\vcenter{\hbox{\vspace{0.12in}
	\begin{tikzpicture}[scale=.75]
	\draw[thick,->] (-1/2,0) -- (1/2,0);
	\node[below,scale=0.75] at (0,0) {$h_p$};
	\draw[thick] (-1/2,0) -- (-3/2,0);
	\node[below,scale=0.75] at (-1,0) {$h_q$};
	\draw[thick] (-1/2,0) -- (-1/2,1*0.8);
	\node[above,scale=0.75] at (-1/2,0.8) {$h_3$};
	\draw[thick] (-3/2,0) -- (-3/2,0.8);
	\node[above,scale=0.75] at (-3/2,0.8) {$h_2$};
	\draw[thick] (-3/2,0) -- (-3/2-0.8,0);
	\node[left,scale=0.75] at (-3/2-0.8,0) {$h_1$};
	\end{tikzpicture}
	}} \right]^\dagger\right|^2 \\
    &  \overline{\rho}_{\text{inner}}= \sum_{p,p',q}|c_{12q}|^2 c_{3pq} c^*_{3p'q} \left|\mathcal{B}\left[\vcenter{\hbox{\vspace{0.12in}
	\begin{tikzpicture}[scale=.75]
	\draw[thick,->] (-1/2,0) -- (1/2,0);
	\node[below,scale=0.75] at (0,0) {$h_p$};
	\draw[thick] (-1/2,0) -- (-3/2,0);
	\node[below,scale=0.75] at (-1,0) {$h_q$};
	\draw[thick] (-1/2,0) -- (-1/2,1*0.8);
	\node[above,scale=0.75] at (-1/2,0.8) {$h_3$};
	\draw[thick] (-3/2,0) -- (-3/2,0.8);
	\node[above,scale=0.75] at (-3/2,0.8) {$h_2$};
	\draw[thick] (-3/2,0) -- (-3/2-0.8,0);
	\node[left,scale=0.75] at (-3/2-0.8,0) {$h_1$};
	\end{tikzpicture}
	}} \right]\right|^2 \ket{p}\bra{p'} \left|\mathcal{B}\left[\vcenter{\hbox{\vspace{0.12in}
	\begin{tikzpicture}[scale=.75]
	\draw[thick,->] (-1/2,0) -- (1/2,0);
	\node[below,scale=0.75] at (0,0) {$h_{p'}$};
	\draw[thick] (-1/2,0) -- (-3/2,0);
	\node[below,scale=0.75] at (-1,0) {$h_q$};
	\draw[thick] (-1/2,0) -- (-1/2,1*0.8);
	\node[above,scale=0.75] at (-1/2,0.8) {$h_3$};
	\draw[thick] (-3/2,0) -- (-3/2,0.8);
	\node[above,scale=0.75] at (-3/2,0.8) {$h_2$};
	\draw[thick] (-3/2,0) -- (-3/2-0.8,0);
	\node[left,scale=0.75] at (-3/2-0.8,0) {$h_1$};
	\end{tikzpicture}
	}} \right]^\dagger\right|^2
  \end{split}
\end{equation}
so the Renyi entropies of these states in the semiclassical limit read,
\begin{equation}
      \text{Tr}(\overline{\rho}_{\text{outer}}^k) \approx \left |\int dh_p \rho_0(h_p) \int \prod_{i=1}^kdh_{q_i}\rho_0(h_{q_i})  C_0(h_1,h_2,h_{q_i})  C_0(h_{\mathcal{O}},h_p,h_{q_i}) \vcenter{\hbox{
	\begin{tikzpicture}[scale=.75]
	\draw[thick] (-1/2,0) -- (1/2,0);
	\node[below,scale=0.75] at (0,0) {$p$};
	\draw[thick] (-1/2,0) -- (-3/2,0);
	\node[below,scale=0.75] at (-1,0) {$q_i$};
	\draw[thick] (-1/2,0) -- (-1/2,1*0.8);
	\node[above,scale=0.75] at (-1/2,0.8) {$\mathcal{O}$};
	\draw[thick] (-3/2,0) -- (-3/2,0.8);
	\node[above,scale=0.75] at (-3/2,0.8) {$2$};
	\draw[thick] (-3/2,0) -- (-3/2-0.8,0);
	\node[left,scale=0.75] at (-3/2-0.8,0) {$1$};
	\draw[thick] (1/2,0) -- (1/2,0.8);
	\node[above,scale=0.75] at (1/2,0.8) {$\mathcal{O}$};
	\draw[thick] (1/2,0) -- (3/2,0);
	\node[below,scale=0.75] at (1,0) {$q_{i+1}$};
	\draw[thick] (3/2,0) -- (3/2,0.8);
	\node[above,scale=0.75] at (3/2,0.8) {$2$};
	\draw[thick] (3/2,0) -- (3/2+0.8,0);
	\node[right,scale=0.75] at (3/2+0.8,0) {$1$};
	\end{tikzpicture}
	}} \right |^2
\end{equation}
and 
\begin{multline}
      \text{Tr}(\overline{\rho}_{\text{inner}}^k) \approx \left |\int dh_p \rho_0(h_p) \int \prod_{i=1}^kdh_{q_i}\rho_0(h_{q_i})  C_0(h_1,h_2,h_{q_i})  C_0(h_{\mathcal{O}},h_p,h_{q_i}) \vcenter{\hbox{
	\begin{tikzpicture}[scale=.75]
	\draw[thick] (-1/2,0) -- (1/2,0);
	\node[below,scale=0.75] at (0,0) {$p$};
	\draw[thick] (-1/2,0) -- (-3/2,0);
	\node[below,scale=0.75] at (-1,0) {$q_i$};
	\draw[thick] (-1/2,0) -- (-1/2,1*0.8);
	\node[above,scale=0.75] at (-1/2,0.8) {$\mathcal{O}$};
	\draw[thick] (-3/2,0) -- (-3/2,0.8);
	\node[above,scale=0.75] at (-3/2,0.8) {$2$};
	\draw[thick] (-3/2,0) -- (-3/2-0.8,0);
	\node[left,scale=0.75] at (-3/2-0.8,0) {$1$};
	\draw[thick] (1/2,0) -- (1/2,0.8);
	\node[above,scale=0.75] at (1/2,0.8) {$\mathcal{O}$};
	\draw[thick] (1/2,0) -- (3/2,0);
	\node[below,scale=0.75] at (1,0) {$q_{i+1}$};
	\draw[thick] (3/2,0) -- (3/2,0.8);
	\node[above,scale=0.75] at (3/2,0.8) {$2$};
	\draw[thick] (3/2,0) -- (3/2+0.8,0);
	\node[right,scale=0.75] at (3/2+0.8,0) {$1$};
	\end{tikzpicture}
	}} \right |^2 \\
    +  \left |\int dh_q \rho_0(h_q) C_0^k(h_1,h_2,h_q)  \int \prod_{i=1}^kdh_{p_i}\rho_0(h_{p_i})  C_0(h_{\mathcal{O}},h_{p_i},h_q) \vcenter{\hbox{
	\begin{tikzpicture}[scale=.75]
	\draw[thick] (-1/2,0) -- (1/2,0);
	\node[below,scale=0.75] at (0,0) {$p_i$};
	\draw[thick] (-1/2,0) -- (-3/2,0);
	\node[below,scale=0.75] at (-1,0) {$q$};
	\draw[thick] (-1/2,0) -- (-1/2,1*0.8);
	\node[above,scale=0.75] at (-1/2,0.8) {$\mathcal{O}$};
	\draw[thick] (-3/2,0) -- (-3/2,0.8);
	\node[above,scale=0.75] at (-3/2,0.8) {$2$};
	\draw[thick] (-3/2,0) -- (-3/2-0.8,0);
	\node[left,scale=0.75] at (-3/2-0.8,0) {$1$};
	\draw[thick] (1/2,0) -- (1/2,0.8);
	\node[above,scale=0.75] at (1/2,0.8) {$\mathcal{O}$};
	\draw[thick] (1/2,0) -- (3/2,0);
	\node[below,scale=0.75] at (1,0) {$q$};
	\draw[thick] (3/2,0) -- (3/2,0.8);
	\node[above,scale=0.75] at (3/2,0.8) {$2$};
	\draw[thick] (3/2,0) -- (3/2+0.8,0);
	\node[right,scale=0.75] at (3/2+0.8,0) {$1$};
	\end{tikzpicture}
	}} \right |^2
\end{multline}
At a conceptual level, the important distinction between the analysis done in the previous section and in the current section is that we are coarse graining over different portions of the same geometry in this section while we coarse grained over similar portions of different geometries in the previous section. Using the results for the wormhole amplitudes from the previous section which have the same functional form even if the probe operator is replaced by a heavy defect operator, we can easily see that the Renyi entropies of $\overline{\rho}_{\text{outer}}$ match with the partition function of wormholes where all the three defects are going across the wormhole. On the other hand, the Renyi entropies of $\overline{\rho}_{\text{inner}}$ match with the sum of the partition function of wormholes where defects 1 and 2 go across the wormhole while defect 3 starts and ends on the same boundary, and the partition function of wormholes where all the three defects are going across the wormhole,
\begin{equation}
     \text{Tr}(\overline{\rho}_{\text{outer}}^k) \approx Z_k^{\text{same}}, \qquad \qquad  \text{Tr}(\overline{\rho}_{\text{inner}}^k) \approx Z_k^{\text{same}}+Z_k^{\text{across}}
\end{equation}
For example, the 4-boundary wormholes computing the fourth Renyis for the two cases respectively are sketched below,
\begin{align}
 \begin{split}
 & \text{Tr}(\overline{\rho}_{\text{outer}}^4) \approx     \vcenter{\hbox{
\begin{overpic}[width=1.8in,grid=false]{more-figures/fourpointfourbdryprobein.pdf}
\end{overpic}
}} \\
&  \text{Tr}(\overline{\rho}_{\text{inner}}^4) \approx  \vcenter{\hbox{
\begin{overpic}[width=2in,grid=false]{more-figures/fourpointfourbdryprobeout.pdf}
\end{overpic}
}} + \vcenter{\hbox{
\begin{overpic}[width=1.8in,grid=false]{more-figures/fourpointfourbdryprobein.pdf}
\end{overpic}
}}
  \end{split}
\end{align}
where the trajectory of defect $3$ is marked in red. Note that in the present setup, there is a saddlepoint in the partition function of both type of wormhole topologies because there is also a minimal surface surrounding all the three defects on the time-symmetric slice which was unlike the situation in the previous section where the probe particle outside the horizon was not heavy enough to create a new minimal surface.  Using the results for the wormhole amplitudes in (\ref{Zksame}) and (\ref{Zkacross}), we can compute the von Neumann entropies of the two coarse grained states (\ref{CGstates}),
\begin{equation}
  \begin{split}
   & S(\overline{\rho}_{\text{inner}})=\text{min}\left\{\frac{\text{Area}(\Gamma_{\text{inner}})}{4G_N}, \frac{\text{Area}(\Gamma_{\text{outer}})}{4G_N} \right \} \\
& S(\overline{\rho}_{\text{outer}})=\frac{\text{Area}(\Gamma_{\text{outer}})}{4G_N}
   \end{split}
\end{equation}
This result is intuitively expected because the entropy associated with coarse graining away a region in the interior of an extremal surface should match with the area of the globally minimal extremal surface outside the region. In general, the area of the inner minimal surface could be greater or smaller than the area of the outer minimal surface because a heavy defect could possibly act isometrically even if hidden behind a horizon as shown in \cite{Chandra:2023dgq} unlike a probe particle which always acts co-ismetrically (lowers the saddlepoint primary energy) when it is behind the horizon.

\subsection{Coarse graining multi-copy CFT states}  \label{CGmulti}

We can readily generalise the coarse graining formalism described in the previous parts of this section to coarse grain pure states in several copies of the CFT. The replica wormholes that inform us about the coarse graining map would have higher genus boundaries in this case. So, the generalisation of the coarse graining formalism to multi-copy CFT states would help interpret wormholes with higher genus boundaries in individual CFTs. For concreteness, we illustrate the formalism with the example of a particular two-copy CFT state called the partially entangled thermal state (PETS) introduced in the context of the SYK model in \cite{Goel:2018ubv} and generalised to 2d CFTs in \cite{Chandra:2023dgq} where it was discussed mainly from the perspective of tensor networks. The PETS state is obtained by exciting the thermofield double by a local operation insertion,
\begin{equation} \label{twocopystate}
  \ket{\text{PETS}}= \mathcal{O}\ket{\text{TFD}}= \sum_{p,q} c_{\mathcal{O}pq} \left | {\cal B}\left[  \vcenter{\hbox{\vspace{0.12in}
	\begin{tikzpicture}[scale=.75]
	\draw[thick,<-] (-1/2,0) -- (-3/2,0);
	\node[below,scale=0.75] at (-1,0) {$h_p$};
	\draw[thick] (-3/2,0) -- (-3/2,0.8);
	\node[above,scale=0.75] at (-3/2,0.8) {$h_{\mathcal{O}}$};
        \draw[thick,->] (-3/2,0) -- (-5/2,0);
        \node[below,scale=0.75] at (-2,0) {$h_q$};
	\end{tikzpicture}
	}} \right] \right|^2 \ket{p}_R\ket{q}_L
\end{equation}
where the subscripts beneath the primaries label the two copies of the CFT. The OPE block is defined by organising the sum over states in $\mathcal{O}\ket{\text{TFD}}$ into a sum over Virasoro representations. 
For the subsequent discussion, we consider $\mathcal{O}$ to be a scalar primary operator below the black hole threshold with scaling dimension $\Delta=O(c)$. The norm of the PETS state can be computed using the thermal 2-point function of $\mathcal{O}$ expanded for instance using the two-point conformal blocks on the torus in the `necklace channel',
\begin{equation}
   \langle \text{PETS} \ket{\text{PETS}}=\langle \mathcal{O}^{\dagger}\mathcal{O} \rangle _{\beta}= \sum_{p,q} |c_{\mathcal{O}pq}|^2  \left|\vcenter{\hbox{
	\begin{tikzpicture}[scale=0.75]
	\draw[thick] (0,0) circle (1);
	\draw[thick] (-1,0) -- (-2,0);
	\node[left] at (-2,0) {$\mathcal{O}$};
	\node[above] at (0,1) {$p$};
	\node[below] at (0,-1) {$q$};
	\draw[thick] (1,0) -- (2,0);
	\node[right] at (2,0) {$\mathcal{O}$};
        \node[scale=0.75] at (0,0) {$\tau=\frac{i\beta}{2\pi}$};
	\end{tikzpicture}
	}} \right|^2
\end{equation}
Therefore, for consistency with the expansion of the PETS state in (\ref{twocopystate}), the OPE block must satisfy
\begin{equation}
   \bra{q'}_2\bra{p'}_1 \left | {\cal B}\left[  \vcenter{\hbox{\vspace{0.12in}
	\begin{tikzpicture}[scale=.75]
	\draw[thick,<-] (-1/2,0) -- (-3/2,0);
	\node[below,scale=0.75] at (-1,0) {$h_p$};
	\draw[thick] (-3/2,0) -- (-3/2,0.8);
	\node[above,scale=0.75] at (-3/2,0.8) {$h_{\mathcal{O}}$};
        \draw[thick,->] (-3/2,0) -- (-5/2,0);
        \node[below,scale=0.75] at (-2,0) {$h_q$};
	\end{tikzpicture}
	}} \right]^\dagger \right |^2 \left | {\cal B}\left[  \vcenter{\hbox{\vspace{0.12in}
	\begin{tikzpicture}[scale=.75]
	\draw[thick,<-] (-1/2,0) -- (-3/2,0);
	\node[below,scale=0.75] at (-1,0) {$h_p$};
	\draw[thick] (-3/2,0) -- (-3/2,0.8);
	\node[above,scale=0.75] at (-3/2,0.8) {$h_{\mathcal{O}}$};
        \draw[thick,->] (-3/2,0) -- (-5/2,0);
        \node[below,scale=0.75] at (-2,0) {$h_q$};
	\end{tikzpicture}
	}} \right] \right |^2 \ket{p}_1 \ket{q}_2 = \left | \vcenter{\hbox{
	\begin{tikzpicture}[scale=0.75]
	\draw[thick] (0,0) circle (1);
	\draw[thick] (-1,0) -- (-2,0);
	\node[left] at (-2,0) {$\mathcal{O}$};
	\node[above] at (0,1) {$p$};
	\node[below] at (0,-1) {$q$};
	\draw[thick] (1,0) -- (2,0);
	\node[right] at (2,0) {$\mathcal{O}$};
        \node[scale=0.75] at (0,0) {$\tau$};
	\end{tikzpicture}
	}} \right |^2 \delta{pp'}\delta{qq'}
\end{equation}
In the large-$c$ limit, the above norm can be computed holographically using the partition function on the static Euclidean BTZ black hole geometry backreacted by the propagation of a conical defect dual to $\mathcal{O}$ which is dominated by the identity block in the channel where the external operator $\mathcal{O}$ fuses with its adjoint into the identity representation,
\begin{equation}
   Z_1 \approx \left | \vcenter{\hbox{
	\begin{tikzpicture}[scale=0.75]
	\draw[thick] (0,0) circle (1);
	\draw[thick] (0,1) -- (0,2);
	\draw[thick] (0,2) -- (0.866,2+1/2);
	\draw[thick] (0,2) -- (-0.866,2+1/2);
	\node[left] at (0,3/2) {$\id$};
	\node[left] at (-1,0) {$\id$};
	\node[left] at (-0.866,2+1/2) {$\mathcal{O}$};
	\node[right] at (0.866,2+1/2) {$\mathcal{O}$};
	\node[scale=0.75] at (0,0) {$-1/\tau$};
	\end{tikzpicture}
	}}  \right |^2 = \left | \int dh_p dh_q \rho_0(h_p) \rho_0(h_q) C_0(h_{\mathcal{O}},h_p,h_q) \vcenter{\hbox{
	\begin{tikzpicture}[scale=0.75]
	\draw[thick] (0,0) circle (1);
	\draw[thick] (-1,0) -- (-2,0);
	\node[left] at (-2,0) {$\mathcal{O}$};
	\node[above] at (0,1) {$p$};
	\node[below] at (0,-1) {$q$};
	\draw[thick] (1,0) -- (2,0);
	\node[right] at (2,0) {$\mathcal{O}$};
        \node[scale=0.75] at (0,0) {$\tau=\frac{i\beta}{2\pi}$};
	\end{tikzpicture}
	}} \right |^2
\end{equation}
The second equality is derived by composing a fusion move with two modular-S moves as shown for instance in \cite{Collier:2019weq}. As observed in \cite{Chandra:2023dgq}, in the hyperbolic slicing coordinates, the time symmetric spatial slice of the dual geometry is a punctured hyperbolic cylinder. When $\mathcal{O}$ is sufficiently heavy which we shall be assuming for the rest of this discussion, there are two minimal geodesics on this spatial slice corresponding to the two apparent horizons as shown in the figure below,
\begin{align}\label{btzdf}
    \vcenter{\hbox{
\begin{overpic}[width=1.8in,grid=false]{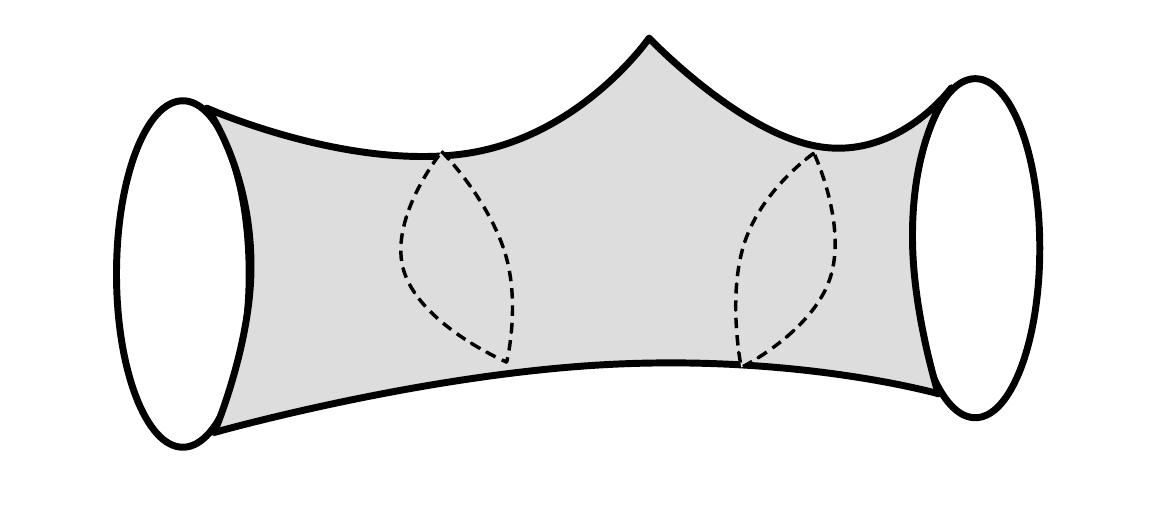}
\put (63,4) {$\Gamma_R$}
\put (42,4)  {$\Gamma_L$}
\end{overpic}
}}
\end{align}
Note that the areas of the two minimal surfaces match with the Cardy entropies at the two saddlepoint weights in the calculation of the norm of the PETS state as observed in \cite{Chandra:2023dgq}.
\begin{equation}
   S_0(h_p,\overline{h}_p^*)=\frac{\text{Area}(\Gamma_R)}{4G_N} \quad \quad \quad S_0(h_q,\overline{h}_q^*)=\frac{\text{Area}(\Gamma_L)}{4G_N}
\end{equation}
Writing an ansatz for the OPE coefficients $c_{\mathcal{O}pq}$ interpreted as black hole microstate coefficients analogous to (\ref{OPEansatz}), we use the fact that $ \langle \text{PETS} \ket{\text{PETS}}\approx Z_1$ to read off the smooth part of the microstate coefficients which is again given by the function $C_0$ in a different parameter regime where one of the weights in the argument is below the black hole threshold while the other two are above the black hole threshold. 

We coarse grain multi-copy CFT states by doing a diagonal projection in Virasoro representation space in each copy of the CFT. In other words, we completely decohere the correlations between Virasoro primaries in each copy retaining the correlations between descendents. This choice is informed to us by the wormhole solutions that we consider. For example, the replica-symmetric $k$-boundary wormholes with boundaries being twice-punctured tori inform us about how to coarse grain the PETS state. We define the coarse grained PETS state by,
\begin{equation} \label{CGPETS}
    \overline{\rho}=\sum_{p,q} |c_{\mathcal{O}pq}|^2 \left | {\cal B}\left[  \vcenter{\hbox{\vspace{0.12in}
	\begin{tikzpicture}[scale=.75]
	\draw[thick,<-] (-1/2,0) -- (-3/2,0);
	\node[below,scale=0.75] at (-1,0) {$h_p$};
	\draw[thick] (-3/2,0) -- (-3/2,0.8);
	\node[above,scale=0.75] at (-3/2,0.8) {$h_{\mathcal{O}}$};
        \draw[thick,->] (-3/2,0) -- (-5/2,0);
        \node[below,scale=0.75] at (-2,0) {$h_q$};
	\end{tikzpicture}
	}} \right] \right|^2 \ket{p}_R\ket{q}_L \bra{q}_L \bra{p}_R
	\left | {\cal B}\left[  \vcenter{\hbox{\vspace{0.12in}
	\begin{tikzpicture}[scale=.75]
	\draw[thick,<-] (-1/2,0) -- (-3/2,0);
	\node[below,scale=0.75] at (-1,0) {$h_p$};
	\draw[thick] (-3/2,0) -- (-3/2,0.8);
	\node[above,scale=0.75] at (-3/2,0.8) {$h_{\mathcal{O}}$};
        \draw[thick,->] (-3/2,0) -- (-5/2,0);
        \node[below,scale=0.75] at (-2,0) {$h_q$};
	\end{tikzpicture}
	}} \right]^\dagger \right|^2
\end{equation}
Intuitively, this prescription amounts to coarse graining away the `python's lunch' \cite{Brown:2019rox} region in between the two horizons. We shall quantify this statement soon using mutual information. The Renyi entropies of the coarse grained PETS state match with the partition functions of the wormholes described above solving the replica symmetric boundary conditions corresponding to $(\langle \text{PETS} \ket{\text{PETS}})^k$,
\begin{equation}
   \text{Tr}(\overline{\rho}^k) \approx \left | \int dh_p dh_q \rho_0(h_p) \rho_0(h_q) C_0(h_{\mathcal{O}},h_p,h_q)^k \left ( \vcenter{\hbox{
	\begin{tikzpicture}[scale=0.75]
	\draw[thick] (0,0) circle (1);
	\draw[thick] (-1,0) -- (-2,0);
	\node[left] at (-2,0) {$\mathcal{O}$};
	\node[above] at (0,1) {$p$};
	\node[below] at (0,-1) {$q$};
	\draw[thick] (1,0) -- (2,0);
	\node[right] at (2,0) {$\mathcal{O}$};
        \node[scale=0.75] at (0,0) {$\tau=\frac{i\beta}{2\pi}$};
	\end{tikzpicture}
	}} \right )^k\right |^2 \approx Z_k
\end{equation}
By construction, these wormholes solve the boundary conditions corresponding to $(\langle \text{PETS} \ket{\text{PETS}})^k$.
For example, the 4-boundary wormhole computing the fourth Renyi of the coarse grained PETS state is sketched below,
\begin{align}
    \text{Tr}(\overline{\rho}^4) \approx \vcenter{\hbox{
\begin{overpic}[width=1.8in,grid=false]{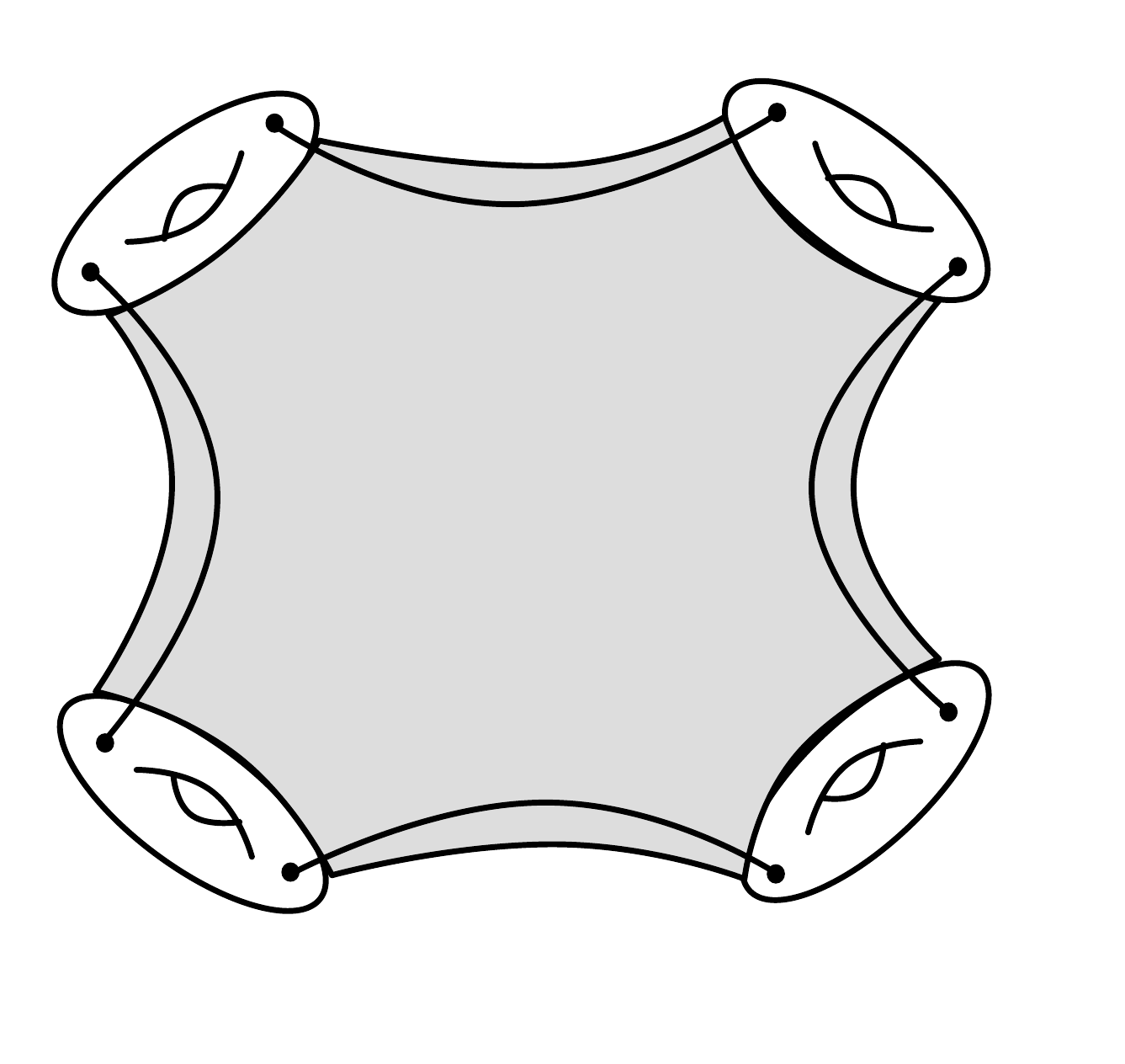}
\end{overpic}
}}
\end{align}
For the calculation of the von Neumann entropy, it is convenient to write the above expression schematically as $\text{Tr}(\overline{\rho}^k)\approx \left | \int dh_p dh_q \rho_0(h_p) \rho_0(h_q) ( \dots )^k\right |^2$ where the dots denote terms which only depend on $k$ implicitly throught the dependence of the saddlepoint weights and hence drop out as $k\to 1$ when the equations of motion are imposed. Therefore, the von Neumann entropy of the coarse grained PETS state computed using the replica trick is given by the sum of Cardy entropies at the two saddle point weights which corresponds geometrically to the sum of the areas of the two apparent horizons i.e,
\begin{equation}
    S(\overline{\rho})=S_0(h^*_p,\overline{h}^*_p)+S_0(h^*_q,\overline{h}^*_q)=\frac{(\text{Area}(\Gamma_R)+\text{Area}(\Gamma_L))}{4G_N}
\end{equation}
We shall now show that the effect of coarse graining away the `lunch' region between $\Gamma_R$ and $\Gamma_L$ in (\ref{btzdf}) is to set the mutual information between the two copies of the CFT to zero. To this end, we take a partial trace of the coarse grained PETS state over one of the copies, $ \overline{\rho}_{R,L}=\text{Tr}_{L,R}(\overline{\rho})$,
\begin{equation}
   \begin{split}
      & \overline{\rho}_R=\sum_{p,q} |c_{\mathcal{O}pq}|^2 \ket{p}_R \left | \mathcal{B}\left [ \vcenter{\hbox{\vspace{0.12in}
	\begin{tikzpicture}[scale=.75]
        \draw[thick] (-1/2,0) -- (-1/2,0.8);
        \node[above, scale=0.75] at (-1/2,0.8) {$h_{\mathcal{O}}$};
        \draw[thick,->] (-1/2,0) -- (1/2,0);
        \node[below, scale=0.75] at (0,0) {$h_p$};
	\draw[thick] (-1/2,0) -- (-3/2,0);
	\node[below,scale=0.75] at (-1,0) {$h_q$};
	\draw[thick] (-3/2,0) -- (-3/2,0.8);
	\node[above,scale=0.75] at (-3/2,0.8) {$h_{\mathcal{O}}$};
        \draw[thick,->] (-3/2,0) -- (-5/2,0);
        \node[below,scale=0.75] at (-2,0) {$h_p$};
	\end{tikzpicture}
	}} \right ] \right |^2 \bra{p}_R \\
 & \overline{\rho}_L=\sum_{p,q} |c_{\mathcal{O}pq}|^2 \ket{q}_L \left |\mathcal{B}\left [ \vcenter{\hbox{\vspace{0.12in}
	\begin{tikzpicture}[scale=.75]
        \draw[thick] (-1/2,0) -- (-1/2,0.8);
        \node[above, scale=0.75] at (-1/2,0.8) {$h_{\mathcal{O}}$};
        \draw[thick,->] (-1/2,0) -- (1/2,0);
        \node[below, scale=0.75] at (0,0) {$h_q$};
	\draw[thick] (-1/2,0) -- (-3/2,0);
	\node[below,scale=0.75] at (-1,0) {$h_p$};
	\draw[thick] (-3/2,0) -- (-3/2,0.8);
	\node[above,scale=0.75] at (-3/2,0.8) {$h_{\mathcal{O}}$};
        \draw[thick,->] (-3/2,0) -- (-5/2,0);
        \node[below,scale=0.75] at (-2,0) {$h_q$};
	\end{tikzpicture}
	}} \right ] \right |^2 \bra{q}_L
   \end{split}
\end{equation}
The OPE block is implicitly labelled by two moduli which we can take to be the lengths of the two minimal surfaces. The difference between the two expressions above is that for the other copy, the dependence of the OPE block on the two moduli is reversed.
Notice that for the PETS state, taking a partial trace commutes with the coarse graining operation i.e, we would arrive at the same state by first tracing out one of the copies and then doing a diagonal projection on primaries for the resulting state. The Renyi entropies of the reduced coarse grained state satisfy
\begin{equation} \label{trblock}
  \begin{split}
   \text{Tr}(\overline{\rho}_R^k) & \approx \left |\int dh_p \rho_0(h_p) \prod_{i=1}^k \left (\int dh_{q_i}\rho_0(h_{q_i}) C_0(h_{\mathcal{O}},h_p,h_{q_i})\right)  \vcenter{\hbox{
	\begin{tikzpicture}[scale=0.75]
	\draw[thick] (0,0) circle (1);
	\draw[thick] (-1,0) -- (-2,0);
        \draw[thick] (-0.5,0.866) -- (-1,1.732);
        \draw[thick] (0.5,0.866) -- (1,1.732);
        \draw[thick] (-0.5,-0.866) -- (-1,-1.732);
	\node[left] at (-2,0) {$\mathcal{O}$};
	\node[above] at (0,1) {$p$};
	\node[below, scale=2] at (0,-1) {$.$};
        \node[below, scale=2] at  (0.8,-0.666) {$.$};
        \node[below, scale=2] at  (0.4,-0.82) {$.$};
	\draw[thick] (1,0) -- (2,0);
	\node[right] at (2,0) {$\mathcal{O}$};
        \node[scale=1] at (-1.2,1.932) {$\mathcal{O}$};
        \node[scale=1] at (1.2,1.932) {$\mathcal{O}$};
        \node[scale=1] at (-1.2,-1.932) {$\mathcal{O}$};
        \node[scale=0.75] at (0,0) {$\tau_{Rk}$};
        \node[left] at (-0.7,0.666) {$q_1$};
        \node[left] at (-0.8,-0.666) {$p$};
        \node[right] at (0.7,0.666) {$q_k$};
        \node[right] at (0.9,-0.466) {$p$};
	\end{tikzpicture}
	}}
\right |^2 \\
 & =  \left |\int dh_p \rho_0(h_p) \vcenter{\hbox{
	\begin{tikzpicture}[scale=0.75]
	\draw[thick] (0,0) circle (1);
	\draw[thick] (-1,0) -- (-2,0);
        \draw[thick] (-2,0) -- (-2-0.866,0.5);
        \draw[thick] (-2,0) -- (-2-0.866,-0.5);
        \draw[thick] (2,0) -- (2+0.866,0.5);
        \draw[thick] (2,0) -- (2+0.866,-0.5);
        \draw[thick] (0,1) -- (0,2);
        \draw[thick] (0,2) -- (0.5,2+0.866);
        \draw[thick] (0,2) -- (-0.5,2+0.866);
	\node[scale=1] at  (0.7,2+0.866+0.2) {$\mathcal{O}$};
        \node[scale=1] at  (-0.7,2+0.866+0.2) {$\mathcal{O}$};
        \node[scale=1] at  (2+0.866+0.2,0.5+0.2) {$\mathcal{O}$};
        \node[scale=1] at  (2+0.866+0.2,-0.5-0.2) {$\mathcal{O}$};
        \node[scale=1] at  (-2-0.866-0.2,0.5+0.2) {$\mathcal{O}$};
        \node[scale=1] at  (-2-0.866-0.2,-0.5-0.2) {$\mathcal{O}$};
	\node[right] at (0,1.5) {$\id$};
        \node[above] at (-1.5,0) {$\id$};
        \node[above] at (1.5,0) {$\id$};
        \node[left] at (-0.7,0.666) {$p$};
        \node[right] at (0.7,0.666) {$p$};
        \node[left] at (-0.7,-0.666) {$p$};
        \node[right] at (0.7,-0.666) {$p$};
	\node[scale=2] at (0,-1.3) {$.$};
        \node[scale=2] at (0.5,-1.1) {$.$};
        \node[scale=2] at (-0.5,-1.1) {$.$};
	\draw[thick] (1,0) -- (2,0);
        \node[scale=0.75] at (0,0) {$\tau_{Rk}$};
	\end{tikzpicture}
	}}
\right |^2 \\
& = \left | \vcenter{\hbox{
	\begin{tikzpicture}[scale=0.75]
	\draw[thick] (0,0) circle (1);
	\draw[thick] (-1,0) -- (-2,0);
        \draw[thick] (-2,0) -- (-2-0.866,0.5);
        \draw[thick] (-2,0) -- (-2-0.866,-0.5);
        \draw[thick] (2,0) -- (2+0.866,0.5);
        \draw[thick] (2,0) -- (2+0.866,-0.5);
        \draw[thick] (0,1) -- (0,2);
        \draw[thick] (0,2) -- (0.5,2+0.866);
        \draw[thick] (0,2) -- (-0.5,2+0.866);
	\node[scale=1] at  (0.7,2+0.866+0.2) {$\mathcal{O}$};
        \node[scale=1] at  (-0.7,2+0.866+0.2) {$\mathcal{O}$};
        \node[scale=1] at  (2+0.866+0.2,0.5+0.2) {$\mathcal{O}$};
        \node[scale=1] at  (2+0.866+0.2,-0.5-0.2) {$\mathcal{O}$};
        \node[scale=1] at  (-2-0.866-0.2,0.5+0.2) {$\mathcal{O}$};
        \node[scale=1] at  (-2-0.866-0.2,-0.5-0.2) {$\mathcal{O}$};
	\node[right] at (0,1.5) {$\id$};
        \node[above] at (-1.5,0) {$\id$};
        \node[above] at (1.5,0) {$\id$};
        \node[left] at (-0.7,0.666) {$\id$};
        \node[right] at (0.7,0.666) {$\id$};
        \node[left] at (-0.7,-0.666) {$\id$};
        \node[right] at (0.7,-0.666) {$\id$};
	\node[scale=2] at (0,-1.3) {$.$};
        \node[scale=2] at (0.5,-1.1) {$.$};
        \node[scale=2] at (-0.5,-1.1) {$.$};
	\draw[thick] (1,0) -- (2,0);
        \node[scale=0.75] at (0,0) {$-1/\tau_{Rk}$};
	\end{tikzpicture}
	}}
\right |^2
\end{split}
\end{equation}
In going from the first to the second line, we did a fusion on each of the internal legs with primaries $q_i$ following which we did an S-transform on the torus component of the block to go to the third line. Thus, we have expressed the Renyi entropy in terms of a $2k$-point identity block on the torus.  Similarly, for the other copy, the Renyi entropy of the reduced coarse grained state can be computed from a similar identity block evaluated at a different modulus,
\begin{equation}
      \text{Tr}(\overline{\rho}_L^k) \approx \left | \vcenter{\hbox{
	\begin{tikzpicture}[scale=0.75]
	\draw[thick] (0,0) circle (1);
	\draw[thick] (-1,0) -- (-2,0);
        \draw[thick] (-2,0) -- (-2-0.866,0.5);
        \draw[thick] (-2,0) -- (-2-0.866,-0.5);
        \draw[thick] (2,0) -- (2+0.866,0.5);
        \draw[thick] (2,0) -- (2+0.866,-0.5);
        \draw[thick] (0,1) -- (0,2);
        \draw[thick] (0,2) -- (0.5,2+0.866);
        \draw[thick] (0,2) -- (-0.5,2+0.866);
	\node[scale=1] at  (0.7,2+0.866+0.2) {$\mathcal{O}$};
        \node[scale=1] at  (-0.7,2+0.866+0.2) {$\mathcal{O}$};
        \node[scale=1] at  (2+0.866+0.2,0.5+0.2) {$\mathcal{O}$};
        \node[scale=1] at  (2+0.866+0.2,-0.5-0.2) {$\mathcal{O}$};
        \node[scale=1] at  (-2-0.866-0.2,0.5+0.2) {$\mathcal{O}$};
        \node[scale=1] at  (-2-0.866-0.2,-0.5-0.2) {$\mathcal{O}$};
	\node[right] at (0,1.5) {$\id$};
        \node[above] at (-1.5,0) {$\id$};
        \node[above] at (1.5,0) {$\id$};
        \node[left] at (-0.7,0.666) {$\id$};
        \node[right] at (0.7,0.666) {$\id$};
        \node[left] at (-0.7,-0.666) {$\id$};
        \node[right] at (0.7,-0.666) {$\id$};
	\node[scale=2] at (0,-1.3) {$.$};
        \node[scale=2] at (0.5,-1.1) {$.$};
        \node[scale=2] at (-0.5,-1.1) {$.$};
	\draw[thick] (1,0) -- (2,0);
        \node[scale=0.75] at (0,0) {$-1/\tau_{Lk}$};
	\end{tikzpicture}
	}}
\right |^2
\end{equation}
Note that the moduli (locations of the operator insertions and the complex structure modulus of the torus) in general depend on which CFT copy we consider and on the Renyi index $k$.
It is expected that these Renyi entropies can be computed holographically from the partition functions of replica wormholes obtained by branching around the corresponding apparent horizons. However, in the present work, we shall not be explicitly constructing these geometries which would involve determining how the moduli change as a function of the Renyi index $k$.
The von Neumann entropies of the reduced coarse grained density matrices match with the areas of the corresponding apparent horizons.\footnote{To see this, it is convenient to rewrite the expression for the Renyi entropy of the coarse grained state as 
\begin{equation}
 \text{Tr}(\overline{\rho}_1^k)\approx \int \left |dh_p \rho_0(h_p) \right |^2\left (\int \left |dh_q \rho_0(h_q) C_0(h_{\mathcal{O}},h_p,h_q) \right |^2 \bra{p} \left | \mathcal{B}\left [ \vcenter{\hbox{\vspace{0.12in}
	\begin{tikzpicture}[scale=.75]
        \draw[thick] (-1/2,0) -- (-1/2,0.8);
        \node[above, scale=0.75] at (-1/2,0.8) {$h_{\mathcal{O}}$};
        \draw[thick,->] (-1/2,0) -- (1/2,0);
        \node[below, scale=0.75] at (0,0) {$h_p$};
	\draw[thick] (-1/2,0) -- (-3/2,0);
	\node[below,scale=0.75] at (-1,0) {$h_q$};
	\draw[thick] (-3/2,0) -- (-3/2,0.8);
	\node[above,scale=0.75] at (-3/2,0.8) {$h_{\mathcal{O}}$};
        \draw[thick,->] (-3/2,0) -- (-5/2,0);
        \node[below,scale=0.75] at (-2,0) {$h_p$};
	\end{tikzpicture}
	}} \right ] \right |^2 \ket{p}\right)^k 
\end{equation}
In the calculation of the von Neumann entropy, the terms scaling with the Renyi index drop out in the $k\to1$ limit when the saddlepoint equations are satisfied. It is interesting to note that requiring that the two ways of expressing the Renyi of the coarse grained state match seems to suggest that the $2k$-point block on the torus in (\ref{trblock}) can be expressed in a factorised form for the choice of moduli relevant to the problem,
\begin{equation}
   \left | \vcenter{\hbox{
	\begin{tikzpicture}[scale=0.75]
	\draw[thick] (0,0) circle (1);
	\draw[thick] (-1,0) -- (-2,0);
        \draw[thick] (-0.5,0.866) -- (-1,1.732);
        \draw[thick] (0.5,0.866) -- (1,1.732);
        \draw[thick] (-0.5,-0.866) -- (-1,-1.732);
	\node[left] at (-2,0) {$\mathcal{O}$};
	\node[above] at (0,1) {$p$};
	\node[below, scale=2] at (0,-1) {$.$};
        \node[below, scale=2] at  (0.8,-0.666) {$.$};
        \node[below, scale=2] at  (0.4,-0.82) {$.$};
	\draw[thick] (1,0) -- (2,0);
	\node[right] at (2,0) {$\mathcal{O}$};
        \node[scale=1] at (-1.2,1.932) {$\mathcal{O}$};
        \node[scale=1] at (1.2,1.932) {$\mathcal{O}$};
        \node[scale=1] at (-1.2,-1.932) {$\mathcal{O}$};
        \node[scale=0.75] at (0,0) {$\tau_{Rk}$};
        \node[left] at (-0.7,0.666) {$q_1$};
        \node[left] at (-0.8,-0.666) {$p$};
        \node[right] at (0.7,0.666) {$q_k$};
        \node[right] at (0.9,-0.466) {$p$};
	\end{tikzpicture}
	}} \right |^2 = \prod_{i=1}^k\bra{p} \left | \mathcal{B}\left [ \vcenter{\hbox{\vspace{0.12in}
	\begin{tikzpicture}[scale=.75]
        \draw[thick] (-1/2,0) -- (-1/2,0.8);
        \node[above, scale=0.75] at (-1/2,0.8) {$h_{\mathcal{O}}$};
        \draw[thick,->] (-1/2,0) -- (1/2,0);
        \node[below, scale=0.75] at (0,0) {$h_p$};
	\draw[thick] (-1/2,0) -- (-3/2,0);
	\node[below,scale=0.75] at (-1,0) {$h_{q_i}$};
	\draw[thick] (-3/2,0) -- (-3/2,0.8);
	\node[above,scale=0.75] at (-3/2,0.8) {$h_{\mathcal{O}}$};
        \draw[thick,->] (-3/2,0) -- (-5/2,0);
        \node[below,scale=0.75] at (-2,0) {$h_p$};
	\end{tikzpicture}
	}} \right ] \right |^2 \ket{p}
\end{equation}
} 
Thus, we have 
\begin{equation} \label{CGPETS}
   S(\overline{\rho}_{L,R})=\frac{\text{Area}(\Gamma_{L,R})}{4G_N} 
\end{equation}
Using this result, we observe that the mutual information defined as $I(A:B)=S(A)+S(B)-S(AB)$ between the two copies of the CFT vanishes in the coarse grained PETS state,
\begin{equation}
   I(\text{CFT}_L:\text{CFT}_R)=S(\overline{\rho}_L)+S(\overline{\rho}_R)-S(\overline{\rho})=0
\end{equation}
so the two copies become uncorrelated (no classical correlations nor quantum entanglement) upon coarse graining. From (\ref{CGPETS}), we see that our coarse graining prescription helps provide a CFT interpretation for the subdominant extremal surface in the sense of the RT formula. In the absence of coarse graining, $\rho=\ket{\text{PETS}}\bra{\text{PETS}}$ is an entangled pure state in two copies of the CFT so the entanglement entropy calculated for instance using the RT prescription would only capture the globally minimal extremal surface,
\begin{equation}
  S(\rho_R)=S(\rho_L)=\frac{\text{min}(\text{Area}(\Gamma_R),\text{Area}(\Gamma_L))}{4G_N}
\end{equation} 
We also see that without coarse graining, the mutual information between the 2 copies in the PETS state is twice the entanglement entropy.

\subsubsection*{Relation to the Thermo-mixed double (TMD) state}

In the limit $\Delta_{\mathcal{O}} \to 0$, the PETS state reduces to the TFD state which has a trivial lunch region, so our prescription for coarse graining would not apply. However, it is interesting to note that in this limit, the Renyis of the reduced coarse grained state give the vacuum character in the dual channel i.e, $\text{Tr}(\overline{\rho}_R^k)=|\mathcal{\chi}_{\id}(-1/\tau_{Rk})|^2$. In this limit, by symmetry we expect that the Renyis of the two reduced coarse grained states match.\footnote{Note that when the two insertions of $\mathcal{O}$ on the thermal circle are diametrically opposite, then the particle dual to $\mathcal{O}$ propagates symmetrically in between the two boundaries of the backreacted eternal black hole. For this case, there are two horizons of equal area created on either side of the trajectory for any non-zero value of $\Delta_{\cal O}$. In this fine-tuned case, by symmetry, we expect $\text{Tr}(\overline{\rho}_1^k)=\text{Tr}(\overline{\rho}_2^k)$ although $\overline{\rho}$ is not pure.} Setting $\tau_{Rk}=\tau_{Lk}=\frac{ik\beta}{2\pi}$, these Renyi entropies match with the Renyis of the thermo-mixed double state (TMD) which was intoduced in \cite{Verlinde:2020upt} by decohering the TFD state in the energy basis,
\begin{equation} \label{TMD}
  \rho_{\text{TMD}}=\sum_n e^{-\beta E_n} \ket{n}_R \ket{n}_L \bra{n}_L\bra{n}_R
\end{equation}
Holographically, these Renyi entropies are calculated from the Gibbons-Hawking type replica geometries obtained by branching around the horizon of the eternal BTZ geometry. It is important to contrast the TMD state from the state obtained by taking the $\Delta_{\mathcal{O}}\to 0$ limit of the coarse grained PETS state (\ref{CGPETS}) which we shall call the $\widetilde{\text{TMD}}$ state,
\begin{equation} \label{TMDtilde}
   \rho_{\widetilde{\text{TMD}}}=\sum_p  \left | {\cal B}\left[  \vcenter{\hbox{\vspace{0.12in}
	\begin{tikzpicture}[scale=.75]
        \draw[thick] (-3/2,0.2) -- (-3/2,0.2);
	\draw[thick,<-] (-1/2,-1/2) -- (-3/2,-1/2);
	\node[below,scale=0.75] at (-3/2,-1/2) {$h_p$};
        \draw[thick,->] (-3/2,-1/2) -- (-5/2,-1/2);
	\end{tikzpicture}
	}} \right] \right|^2 \ket{p}_R\ket{p}_L \bra{p}_L \bra{p}_R
	\left | {\cal B}\left[  \vcenter{\hbox{\vspace{0.12in}
	\begin{tikzpicture}[scale=.75]
        \draw[thick] (-3/2,0.2) -- (-3/2,0.2);
	\draw[thick,<-] (-1/2,-1/2) -- (-3/2,-1/2);
	\node[below,scale=0.75] at (-3/2,-1/2) {$h_p$};
        \draw[thick,->] (-3/2,-1/2) -- (-5/2,-1/2);
	\end{tikzpicture}
	}} \right]^\dagger \right|^2
\end{equation}
where the OPE block related to the torus character by the relation,
\begin{equation}
   \bra{q}_L \bra{q}_R \left | {\cal B}\left[  \vcenter{\hbox{\vspace{0.12in}
	\begin{tikzpicture}[scale=.75]
        \draw[thick] (-3/2,0.2) -- (-3/2,0.2);
	\draw[thick,<-] (-1/2,-1/2) -- (-3/2,-1/2);
	\node[below,scale=0.75] at (-3/2,-1/2) {$h_q$};
        \draw[thick,->] (-3/2,-1/2) -- (-5/2,-1/2);
	\end{tikzpicture} 
	}} \right]^\dagger \right|^2 \left | {\cal B}\left[  \vcenter{\hbox{\vspace{0.12in}
	\begin{tikzpicture}[scale=.75]
        \draw[thick] (-3/2,0.2) -- (-3/2,0.2);
	\draw[thick,<-] (-1/2,-1/2) -- (-3/2,-1/2);
	\node[below,scale=0.75] at (-3/2,-1/2) {$h_p$};
        \draw[thick,->] (-3/2,-1/2) -- (-5/2,-1/2);
	\end{tikzpicture}
	}} \right] \right |^2 \ket{p}_R \ket{p}_L = |\chi_{h_p}(\tau)|^2 \delta_{pq}
\end{equation}
The $\widetilde{\text{TMD}}$ state is obtained by decohering the primary contribution in the TFD state retaining the correlations between Virasoro descendents. However, as we have noted, the Renyis of the TMD and $\widetilde{\text{TMD}}$ match at the saddle. This is possibly because we can ignore the contribution from Virasoro descendents for spherically symmetric states at the saddle. 

\subsubsection*{The probe limit}

Now, we discuss the limit where the particle dual to $\mathcal{O}$ is a geodesic probe. In this case, $\mathcal{O}$ is not heavy enough to create a new minimal surface on the time-symmetric spatial slice, so the resulting PETS state should be viewed as a dressed TFD state instead of a two-sided black hole microstate. So, the coarse graining procedure that we described for the PETS black hole would not directly apply for this case since there is no interior region to coarse grain away. A sketch of the spatial slice is shown below,
\begin{align}\label{btzdf}
    \vcenter{\hbox{
\begin{overpic}[width=1.8in,grid=false]{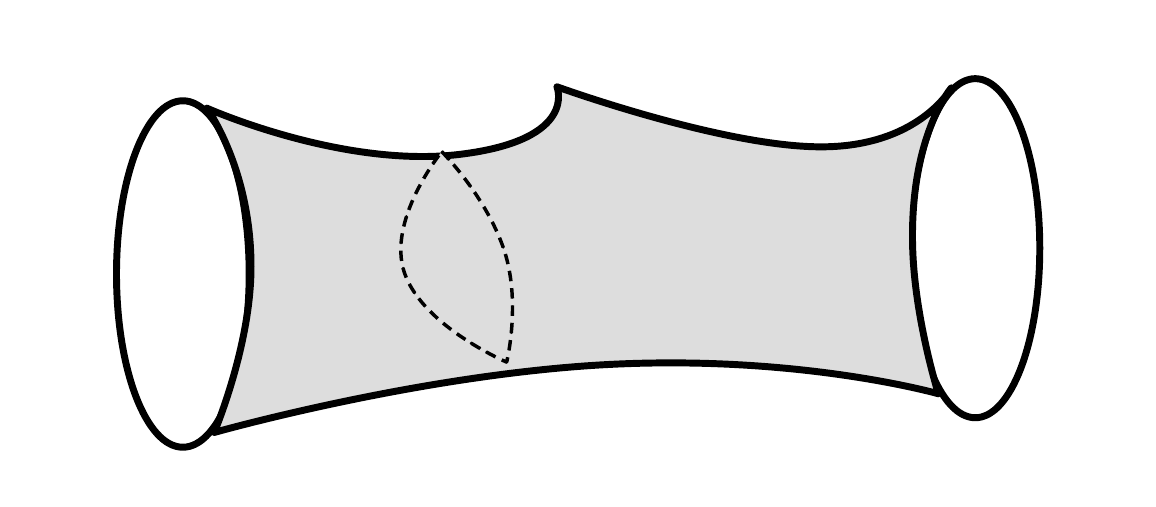}
\put (42,4)  {$\Gamma$}
\end{overpic}
}}
\end{align}
We wish to understand what is the coarse grained state on either side of the horizon $\Gamma$ given the two-copy state,
\begin{equation}
      \rho=\sum_{p,q,p',q'} c_{\mathcal{O}pq}c^*_{\mathcal{O}p'q'} \left | {\cal B}\left[  \vcenter{\hbox{\vspace{0.12in}
	\begin{tikzpicture}[scale=.75]
	\draw[thick,<-] (-1/2,0) -- (-3/2,0);
	\node[below,scale=0.75] at (-1,0) {$h_p$};
	\draw[thick] (-3/2,0) -- (-3/2,0.8);
	\node[above,scale=0.75] at (-3/2,0.8) {$h_{\mathcal{O}}$};
        \draw[thick,->] (-3/2,0) -- (-5/2,0);
        \node[below,scale=0.75] at (-2,0) {$h_q$};
	\end{tikzpicture}
	}} \right] \right|^2 \ket{p}_R\ket{q}_L\bra{q'}_L \bra{p'}_R
	\left | {\cal B}\left[  \vcenter{\hbox{\vspace{0.12in}
	\begin{tikzpicture}[scale=.75]
	\draw[thick,<-] (-1/2,0) -- (-3/2,0);
	\node[below,scale=0.75] at (-1,0) {$h_{p'}$};
	\draw[thick] (-3/2,0) -- (-3/2,0.8);
	\node[above,scale=0.75] at (-3/2,0.8) {$h_{\mathcal{O}}$};
        \draw[thick,->] (-3/2,0) -- (-5/2,0);
        \node[below,scale=0.75] at (-2,0) {$h_{q'}$};
	\end{tikzpicture}
	}} \right]^\dagger \right|^2
\end{equation}
We claim that the coarse grained state to the right of the horizon is same as the one obtained by taking a partial trace of the above state wrt the left copy,
\begin{equation}
   \overline{\rho}_R=\rho_R=\sum_{p,p',q} c_{\mathcal{O}pq}c^*_{\mathcal{O}p'q} \ket{p}_R \left | \mathcal{B}\left [ \vcenter{\hbox{\vspace{0.12in}
	\begin{tikzpicture}[scale=.75]
        \draw[thick] (-1/2,0) -- (-1/2,0.8);
        \node[above, scale=0.75] at (-1/2,0.8) {$h_{\mathcal{O}}$};
        \draw[thick,->] (-1/2,0) -- (1/2,0);
        \node[below, scale=0.75] at (0,0) {$h_{p'}$};
	\draw[thick] (-1/2,0) -- (-3/2,0);
	\node[below,scale=0.75] at (-1,0) {$h_q$};
	\draw[thick] (-3/2,0) -- (-3/2,0.8);
	\node[above,scale=0.75] at (-3/2,0.8) {$h_{\mathcal{O}}$};
        \draw[thick,->] (-3/2,0) -- (-5/2,0);
        \node[below,scale=0.75] at (-2,0) {$h_p$};
	\end{tikzpicture}
	}} \right ] \right |^2 \bra{p'}_R
\end{equation}
whereas the coarse grained state to the left of the horizon is obtained by decohering the state obtained by taking a partial trace wrt the right copy,
\begin{equation}
 \overline{\rho}_L=\sum_{p,q} |c_{\mathcal{O}pq}|^2 \ket{q}_L\left |\mathcal{B}\left [ \vcenter{\hbox{\vspace{0.12in}
	\begin{tikzpicture}[scale=.75]
        \draw[thick] (-1/2,0) -- (-1/2,0.8);
        \node[above, scale=0.75] at (-1/2,0.8) {$h_{\mathcal{O}}$};
        \draw[thick,->] (-1/2,0) -- (1/2,0);
        \node[below, scale=0.75] at (0,0) {$h_q$};
	\draw[thick] (-1/2,0) -- (-3/2,0);
	\node[below,scale=0.75] at (-1,0) {$h_p$};
	\draw[thick] (-3/2,0) -- (-3/2,0.8);
	\node[above,scale=0.75] at (-3/2,0.8) {$h_{\mathcal{O}}$};
        \draw[thick,->] (-3/2,0) -- (-5/2,0);
        \node[below,scale=0.75] at (-2,0) {$h_q$};
	\end{tikzpicture}
	}} \right ] \right |^2 \bra{q}_L
\end{equation}
A non-trivial check of this coarse graining prescription is that the coarse grained entropies of both the states match with the area of the horizon,
\begin{equation}
  S(\overline{\rho}_L)=S(\overline{\rho}_R)=\frac{\text{Area}(\Gamma)}{4G_N}
\end{equation}
The derivation of the above result follows similar logic to that used in the derivation of the coarse grained entropy for the pure state black hole dressed by a probe particle as described in section \ref{CGprobe}.

\subsubsection*{Note on coarse graining a three-copy CFT state}

The coarse graining formalism that we have illustrated using the PETS state is robust and can be applied to coarse grain other multi-copy CFT states which in turn would help interpret other families of multi-boundary wormholes with higher genus boundaries including wormhole solutions in pure 3d gravity in individual CFTs. Consider for example, the three-copy state discussed for instance in \cite{Chandra:2023dgq}, 
\begin{equation} \label{threecopystate}
  \ket{\Psi}= \sum_{p,q,r} c_{pqr} \left | {\cal B}\left[  \vcenter{\hbox{\vspace{0.12in}
	\begin{tikzpicture}[scale=.75]
	\draw[thick,<-] (-1/2,0) -- (-3/2,0);
	\node[below,scale=0.75] at (-1,0) {$h_p$};
	\draw[thick,->] (-3/2,0) -- (-3/2,1);
	\node[right,scale=0.75] at (-3/2,0.5) {$h_{r}$};
        \draw[thick,->] (-3/2,0) -- (-5/2,0);
        \node[below,scale=0.75] at (-2,0) {$h_q$};
	\end{tikzpicture}
	}} \right] \right|^2 \ket{p}_1\ket{q}_2\ket{r}_3
\end{equation}
whose norm can be computed by expanding the genus-two partition function using the sunset channel conformal blocks on a genus-two Riemann surface. So, this state is holographically dual to the three-boundary black hole \cite{Balasubramanian:2014hda} which is a genus-two handlebody obtained by making the cycles in the dumbbell channel contractible in the bulk, sliced in a way that the time-symmetric spatial slice is topologically a pair of pants with the Liouville metric induced on it with ZZ \cite{Zamolodchikov:2001ah} boundary conditions for the Liouville field on the three boundaries. In this metric, there are three minimal geodesics labelled $\Gamma_i$ around each cuff of the pair of pants as sketched below,
\begin{align} \label{prpants}
    \vcenter{\hbox{
\begin{overpic}[width=1.8in,grid=false]{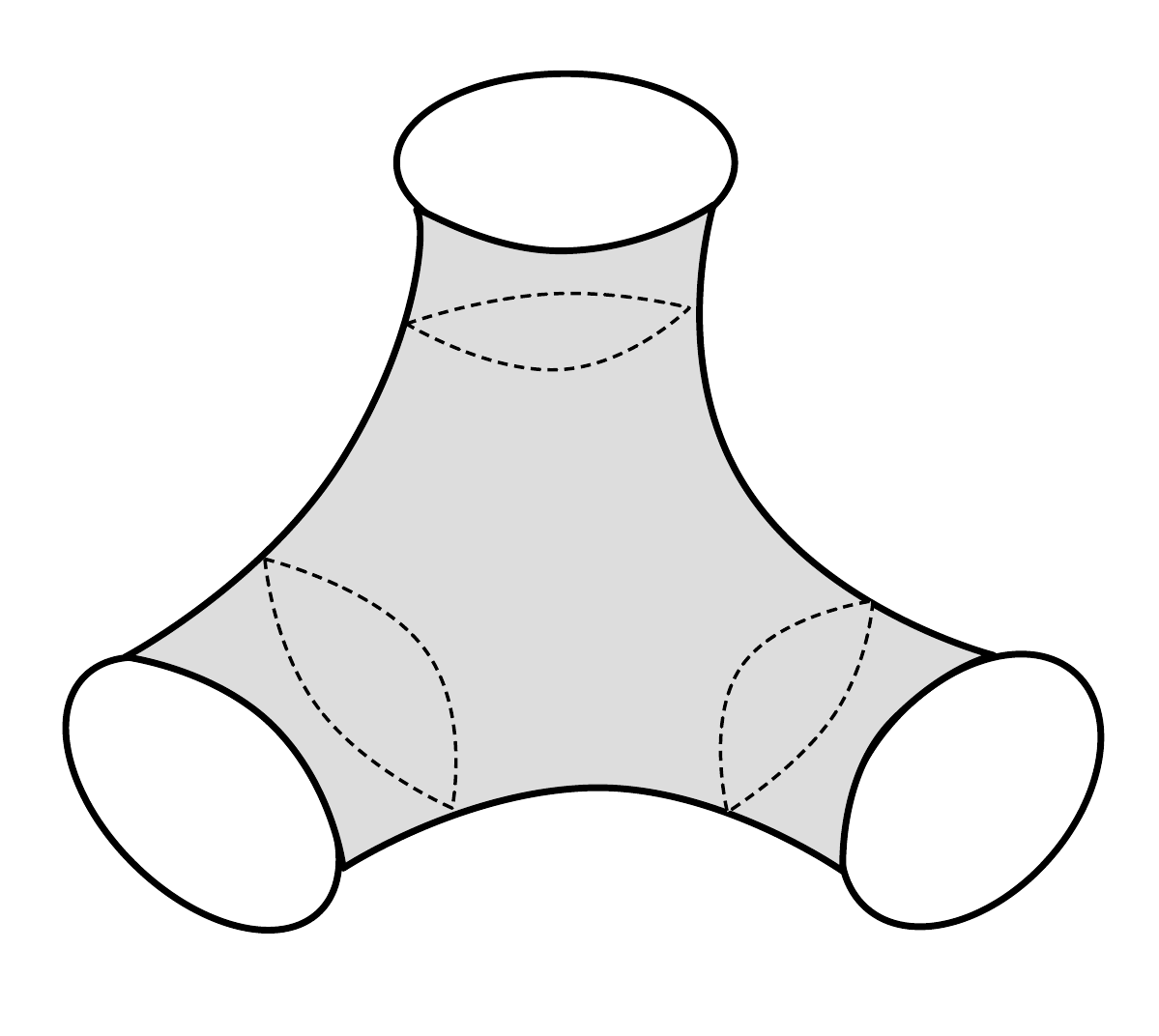}
\put (37,37) {$\Gamma_1$}
\put (50,51) {$\Gamma_2$}
\put (58,33) {$\Gamma_3$}
\end{overpic}
}}
\end{align}
We coarse grain the state by a simultaneous diagonal projection on each copy analogous to our coarse graining procedure for the PETS state. The Renyi entropies of this coarse grained state match with the partition functions of multi-boundary wormholes with genus-two boundaries and replica symmetric boundary conditions. The entropy of the coarse grained state is given by,
\begin{equation}
   S(\overline{\rho})=\frac{\sum_{i=1}^3 \text{Area}(\Gamma_i)}{4G_N}
\end{equation}
Physically, coarse graining away the interior bounded by the three geodesics in (\ref{prpants}) according to this prescription sets the mutual information between any copies of the CFT in the coarse grained state to zero,
\begin{equation}
   I(\text{CFT}_1: \text{CFT}_2)=I(\text{CFT}_2: \text{CFT}_3)=I(\text{CFT}_1: \text{CFT}_3)=0
\end{equation}
All the calculations are analogous to those described earlier for the PETS state so we skip the details here.

\section{Wormholes from decohering transition matrices} \label{secTM}

The coarse graining formalism that we described in the previous section involved completely decohering the pure state in the Virasoro representation space retaining the correlations between descendents. In this section, we shall generalise this idea to decohere transition matrices between pure states and provide a holographic interpretation for the Renyi entropies of the decohered transition matrices using the partition functions on appropriate replica-symmetric wormhole geometries. This helps interpret more general families of wormhole solutions in individual CFTs.  From the Renyi entropies, we calculate the pseudo entropy of the decohered transition matrices using the replica trick and provide a holographic interpretation in terms of the areas of minimal surfaces on appropriate wormhole or black hole geometries as we shall illustrate in sections \ref{Holopseudo} and \ref{Eucl}. The pseudo entropy is an information theoretic measure which is a natural generalisation of entanglement entropy to transition matrices which arise for instance via post-selection where an initial state is post-selected onto a final state. This measure was introduced in \cite{Nakata:2020luh} where they explored several properties of pseudo entropy including and provided some general arguments for the holographic interpretation of the pseudo entropy of CFT subregions.

\subsection{Pseudo entropy and minimal surfaces on wormholes} \label{Holopseudo}

In this subsection, we consider the situation when there is no gravitational saddle computing the overlap between the two pure states but when there exists a two-boundary wormhole solving the boundary condition corresponding to the square of the overlap between the states. We argue using an example that the area of a codimension-2 minimal surface on the wormhole matches with the pseudo entropy of a decohered transition matrix between the two pure states.
Consider the following pair of pure states obtained by exciting the CFT vacuum by two local scalar primary operator insertions,
\begin{equation}
    \begin{split}
        & \ket{\Psi}=\mathcal{O}_1(z_1)\mathcal{O}_2(z_2)\ket{0}=\sum_p c_{12p} \left | {\cal B}\left[  \vcenter{\hbox{\vspace{0.12in}
	\begin{tikzpicture}[scale=.75]
	\draw[thick,<-] (-1/2,0) -- (-3/2,0);
	\node[below,scale=0.75] at (-1,0) {$h_p$};
	\draw[thick] (-3/2,0) -- (-3/2,0.8);
	\node[above,scale=0.75] at (-3/2,0.8) {$h_2$};
        \draw[thick] (-3/2,0) -- (-5/2,0);
        \node[left,scale=0.75] at (-5/2,0) {$h_1$};
	\end{tikzpicture}
	}} \right] \right |^2\ket{p} \\
        & \ket{\Psi'}=\mathcal{O}_3(z_3)\mathcal{O}_4(z_4)\ket{0}=\sum_q c_{34q} \left | {\cal B}\left[  \vcenter{\hbox{\vspace{0.12in}
	\begin{tikzpicture}[scale=.75]
	\draw[thick,<-] (-1/2,0) -- (-3/2,0);
	\node[below,scale=0.75] at (-1,0) {$h_q$};
	\draw[thick] (-3/2,0) -- (-3/2,0.8);
	\node[above,scale=0.75] at (-3/2,0.8) {$h_3$};
        \draw[thick] (-3/2,0) -- (-5/2,0);
        \node[left,scale=0.75] at (-5/2,0) {$h_4$};
	\end{tikzpicture}
	}} \right] \right |^2\ket{q}
    \end{split}
\end{equation} 
For simplicity, we assume that the operators are inserted on the $S^2$ in such a way that the cross ratio between the 4 insertions is real. There is no gravitational saddle that computes the overlap between these states because we assume that the operators $\mathcal{O}_3$ and $\mathcal{O}_4$ are distinct from $\mathcal{O}_1$ and $\mathcal{O}_2$. However, there are $2k$-boundary wormholes solving the boundary conditions corresponding to $(\langle \Psi' \ket{\Psi})^{2k}$ with the defects dual to the operators going across the wormholes. The partition functions of these wormholes can be computed using the Virasoro TQFT prescription \cite{Collier:2023fwi} or using the large-$c$ ensemble \cite{Chandra:2022bqq} and read,
\begin{equation}
   Z_{2k} \approx \left | \int dh_p \rho_0(h_p) C_0(h_1,h_2,h_p)^k C_0(h_3,h_4,h_p)^k \left (\vcenter{\hbox{\vspace{0.12in}
	\begin{tikzpicture}[scale=.75]
        \draw[thick] (-1/2,0) -- (-1/2,0.8);
        \node[above, scale=0.75] at (-1/2,0.8) {$3$};
        \draw[thick] (-1/2,0) -- (1/2,0);
        \node[right, scale=0.75] at (1/2,0) {$4$};
	\draw[thick] (-1/2,0) -- (-3/2,0);
	\node[below,scale=0.75] at (-1,0) {$p$};
	\draw[thick] (-3/2,0) -- (-3/2,0.8);
	\node[above,scale=0.75] at (-3/2,0.8) {$2$};
        \draw[thick] (-3/2,0) -- (-5/2,0);
        \node[left,scale=0.75] at (-5/2,0) {$1$};
	\end{tikzpicture}
	}} \right )^{2k} \right |^2
\end{equation}
The $k=1$ wormhole i.e the two boundary wormhole when drawn in the hyperbolic metric looks like \cite{Chandra:2022bqq}\footnote{In Maldacena-Maoz coordinates, the hyperbolic metric reads $ds^2=d\rho^2+\cosh^2(\rho)d\Sigma^2$ where $d\Sigma^2=e^{\Phi(z,\overline{z})}|dz|^2$ is the Liouville metric on the $S^2$ punctured at the 4 operator insertions \cite{Chandra:2022bqq}.},
\begin{align}
    \vcenter{\hbox{
\begin{overpic}[width=2.3in,grid=false]{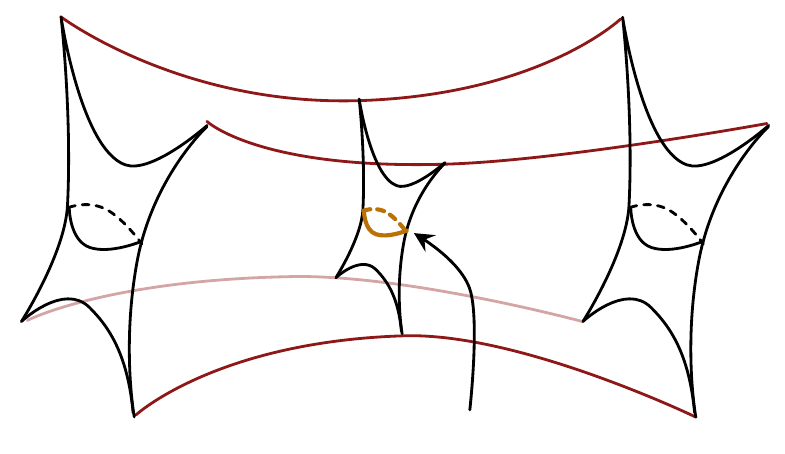}
\put (57, 0) {$\Gamma$}
\put (1,12)  {$1$}
\put (15,1) {$2$}
\put (25,43) {$3$}
\put (7,56) {$4$}
\put (72,12)  {$1$}
\put (86,1) {$2$}
\put (96,43) {$3$}
\put (78,56) {$4$}
\end{overpic}
}}
\end{align}
As observed in \cite{Chandra:2022bqq}, the area of the codimension-2 minimal surface $\Gamma$ shown in red above matches with the saddle point momentum $\gamma_L$ in the Liouville 4-point function $\langle \mathcal{O}_1 \mathcal{O}_2 \mathcal{O}_3 \mathcal{O}_4 \rangle_L$ when expanded in the $12 \to 34$ channel evaluated at the same cross ratio $x$ showing up in the boundary condition for the wormhole,
\begin{equation}
  \text{Area}(\Gamma)=2\pi \gamma_L(x)
\end{equation}

Now, we shall show that the area of the above surface matches with the pseudo-entropy of the transition matrix obtained by decohering a particular transition matrix between the pure states $\Psi$ and $\Psi'$,
\begin{multline}
    \overline{\rho}_{\Psi|\Psi'} = \overline{\rho}_{\Psi}\overline{\rho}_{\Psi'}=\sum_p |c_{12p}|^2 |c_{34p}|^2  \left |\vcenter{\hbox{\vspace{0.12in}
	\begin{tikzpicture}[scale=.75]
        \draw[thick] (-1/2,0) -- (-1/2,0.8);
        \node[above, scale=0.75] at (-1/2,0.8) {$3$};
        \draw[thick] (-1/2,0) -- (1/2,0);
        \node[right, scale=0.75] at (1/2,0) {$4$};
	\draw[thick] (-1/2,0) -- (-3/2,0);
	\node[below,scale=0.75] at (-1,0) {$p$};
	\draw[thick] (-3/2,0) -- (-3/2,0.8);
	\node[above,scale=0.75] at (-3/2,0.8) {$2$};
        \draw[thick] (-3/2,0) -- (-5/2,0);
        \node[left,scale=0.75] at (-5/2,0) {$1$};
	\end{tikzpicture}
	}} \right |^2 \\ \left | {\cal B}\left[  \vcenter{\hbox{\vspace{0.12in}
	\begin{tikzpicture}[scale=.75]
	\draw[thick,<-] (-1/2,0) -- (-3/2,0);
	\node[below,scale=0.75] at (-1,0) {$h_p$};
	\draw[thick] (-3/2,0) -- (-3/2,0.8);
	\node[above,scale=0.75] at (-3/2,0.8) {$h_2$};
        \draw[thick] (-3/2,0) -- (-5/2,0);
        \node[left,scale=0.75] at (-5/2,0) {$h_1$};
	\end{tikzpicture}
	}} \right] \right |^2\ket{p}\bra{p}  \left | {\cal B}\left[  \vcenter{\hbox{\vspace{0.12in}
	\begin{tikzpicture}[scale=.75]
	\draw[thick,<-] (-1/2,0) -- (-3/2,0);
	\node[below,scale=0.75] at (-1,0) {$h_p$};
	\draw[thick] (-3/2,0) -- (-3/2,0.8);
	\node[above,scale=0.75] at (-3/2,0.8) {$h_3$};
        \draw[thick] (-3/2,0) -- (-5/2,0);
        \node[left,scale=0.75] at (-5/2,0) {$h_4$};
	\end{tikzpicture}
	}} \right]^{\dagger} \right |^2
\end{multline}
where $ \overline{\rho}_{\Psi|\Psi'} $ is the decohered transition matrix between the two pure states defined as a product of the two coarse grained states $\overline{\rho}_{\Psi}$ and $\overline{\rho}_{\Psi'}$. We used the following identity involving OPE blocks to arrive at the second equality in the above equation,
\begin{equation}
    \bra{q}  \left | {\cal B}\left[  \vcenter{\hbox{\vspace{0.12in}
	\begin{tikzpicture}[scale=.75]
	\draw[thick,<-] (-1/2,0) -- (-3/2,0);
	\node[below,scale=0.75] at (-1,0) {$h_q$};
	\draw[thick] (-3/2,0) -- (-3/2,0.8);
	\node[above,scale=0.75] at (-3/2,0.8) {$h_3$};
        \draw[thick] (-3/2,0) -- (-5/2,0);
        \node[left,scale=0.75] at (-5/2,0) {$h_4$};
	\end{tikzpicture}
	}} \right]^{\dagger} \right |^2
 \left | {\cal B}\left[  \vcenter{\hbox{\vspace{0.12in}
	\begin{tikzpicture}[scale=.75]
	\draw[thick,<-] (-1/2,0) -- (-3/2,0);
	\node[below,scale=0.75] at (-1,0) {$h_p$};
	\draw[thick] (-3/2,0) -- (-3/2,0.8);
	\node[above,scale=0.75] at (-3/2,0.8) {$h_2$};
        \draw[thick] (-3/2,0) -- (-5/2,0);
        \node[left,scale=0.75] at (-5/2,0) {$h_1$};
	\end{tikzpicture}
	}} \right] \right |^2\ket{p}
=\delta_{pq} \left |\vcenter{\hbox{\vspace{0.12in}
	\begin{tikzpicture}[scale=.75]
        \draw[thick] (-1/2,0) -- (-1/2,0.8);
        \node[above, scale=0.75] at (-1/2,0.8) {$3$};
        \draw[thick] (-1/2,0) -- (1/2,0);
        \node[right, scale=0.75] at (1/2,0) {$4$};
	\draw[thick] (-1/2,0) -- (-3/2,0);
	\node[below,scale=0.75] at (-1,0) {$p$};
	\draw[thick] (-3/2,0) -- (-3/2,0.8);
	\node[above,scale=0.75] at (-3/2,0.8) {$2$};
        \draw[thick] (-3/2,0) -- (-5/2,0);
        \node[left,scale=0.75] at (-5/2,0) {$1$};
	\end{tikzpicture}
	}} \right |^2
\end{equation}
It is easy to see that the Renyis of the decohered transition matrix match with the wormhole partition functions,
\begin{equation}
    \text{Tr}( \overline{\rho}_{\Psi|\Psi'}^k) \approx Z_{2k}
\end{equation}
Thus, using the replica trick, we see that the pseudo-entropy of the decohered transition matrix is given by,
\begin{equation}
    S(\overline{\rho}_{\Psi|\Psi'}):= -\partial_k \log \left (\frac{ \text{Tr}( \overline{\rho}_{\Psi|\Psi'}^k)}{( \text{Tr}( \overline{\rho}_{\Psi|\Psi'}))^k}\right)\bigg |_{k=1}= -\partial_k \log \left (\frac{Z_{2k}}{Z_2^k}\right)\bigg |_{k=1} = \frac{c}{6}(2\pi \gamma_L(x))
\end{equation}
To verify the above result, note that $Z_{2k}\approx \left |\int dh_p \rho_0(h_p)( \dots )^k \right |^2$ where the dots involve terms that depend on the Renyi index implicitly through the dependence of the saddlepoint weight and hence drop out in the $k\to 1$ limit when we impose the saddle point equation for $Z_2$. Thus, we have shown that the area of the minimal surface on the two boundary wormhole matches with the pseudo-entropy of a decohered transition matrix,
\begin{equation}
   S(\overline{\rho}_{\Psi|\Psi'}) = \frac{\text{Area}(\Gamma)}{4G_N}
\end{equation}

\subsection{Pseudo entropy and sub-dominant Euclidean-HRT surfaces} \label{Eucl}

In the previous example, there was no bulk saddle computing the overlap between the two pure states. In this subsection, we consider the situation where the overlap between the two states can be computed holographically using a one-sided Euclidean black hole geometry which is not necessarily time-reversal symmetric. We illustrate using an example that the area of a subdominant HRT surface on the black hole matches with the pseudo entropy of a decohered transition matrix between the two pure states.
Consider the following pair of pure states
\begin{equation}
    \begin{split}
        & \ket{\Psi}=\mathcal{O}_1(z_1)\mathcal{O}_2(z_2)\ket{0} \\
        & \ket{\Psi'}=\mathcal{O}_1(z_1')\mathcal{O}_2(z_2')\ket{0}
    \end{split}
\end{equation}
where $\mathcal{O}_1$ and $\mathcal{O}_2$ are scalar primary operators above the multi-twist threshold inserted on the $S^2$ such that the cross ratio in the 4-point function computing the overlap between the two states is complex. In this case, the overlap between the states can be computed holographically from the on-shell action of a hyperbolic ball geometry with two propagating defects. However, the bulk geometry is not time reversal symmetric. In the semiclassical limit, the partition function matches with the Virasoro identity block evaluated at a complex cross ratio $(1-x)$ where $x=\frac{(z_1-z_2)(\overline{z}_1'-\overline{z}_2')}{(z_1\overline{z}_2'-1)(z_2\overline{z}_1'-1)}$,
\begin{equation}
    Z_{\text{grav}} \approx \left |\vcenter{\hbox{\vspace{0.12in}
	\begin{tikzpicture}[scale=.75]
        \draw[thick] (-1/2,0) -- (-1/2,0.8);
        \node[above, scale=0.75] at (-1/2,0.8) {$2$};
        \draw[thick] (-1/2,0) -- (1/2,0);
        \node[right, scale=0.75] at (1/2,0) {$2$};
	\draw[thick] (-1/2,0) -- (-3/2,0);
	\node[below,scale=0.75] at (-1,0) {$\id$};
	\draw[thick] (-3/2,0) -- (-3/2,0.8);
	\node[above,scale=0.75] at (-3/2,0.8) {$1$};
        \draw[thick] (-3/2,0) -- (-5/2,0);
        \node[left,scale=0.75] at (-5/2,0) {$1$};
	\end{tikzpicture}
	}}\right |^2
\end{equation}
Since the cross ratio is complex, the saddle point in the expansion of the identity block in the dual channel lands on complex weights. However, since the weights are complex conjugates of each other, the Cardy entropy evaluated at these weights would be real. 

Consider the transition matrix constructed from these states,
\begin{equation} \label{tranmat}
    \sigma_{\Psi | \Psi'}=\ket{\Psi}\bra{\Psi'}=\sum_{p,q}c_{12p}c_{12q}^* \left |{\cal B}\left[  \vcenter{\hbox{\vspace{0.12in}
	\begin{tikzpicture}[scale=.75]
	\draw[thick,<-] (-1/2,0) -- (-3/2,0);
	\node[below,scale=0.75] at (-1,0) {$h_p$};
	\draw[thick] (-3/2,0) -- (-3/2,0.8);
	\node[above,scale=0.75] at (-3/2,0.8) {$h_{2}$};
        \draw[thick] (-3/2,0) -- (-5/2,0);
        \node[left,scale=0.75] at (-5/2,0) {$h_{1}$};
	\end{tikzpicture}
	}} \right] \right |^2 \ket{p} \bra{q} \left |{\cal B'}\left[  \vcenter{\hbox{\vspace{0.12in}
	\begin{tikzpicture}[scale=.75]
	\draw[thick,<-] (-1/2,0) -- (-3/2,0);
	\node[below,scale=0.75] at (-1,0) {$h_q$};
	\draw[thick] (-3/2,0) -- (-3/2,0.8);
	\node[above,scale=0.75] at (-3/2,0.8) {$h_{2}$};
        \draw[thick] (-3/2,0) -- (-5/2,0);
        \node[left,scale=0.75] at (-5/2,0) {$h_{1}$};
	\end{tikzpicture}
	}} \right]^\dagger \right |^2
\end{equation}
Clearly, this `pure' transition matrix has a vanishing pseudo entropy,
\begin{equation}
  S( \sigma_{\Psi|\Psi'})=-\text{Tr}(\Hat{\sigma}_{\Psi|\Psi'}\log \Hat{\sigma}_{\Psi|\Psi'}):=-\partial_k \log (\text{Tr}\Hat{\sigma}_{\Psi|\Psi'}^k)  |_{k=1}=0
\end{equation}
where $\Hat{\sigma}_{\Psi|\Psi'}=\frac{\ket{\Psi}\bra{\Psi'}}{\langle \Psi' \ket{\Psi}}$ is the normalised transition matrix for which $\text{Tr}(\Hat{\sigma}_{\Psi|\Psi'})=1$. This is consistent with the fact that the dominant extremal surface in the dual geometry has vanishing area. Now, we shall that we can capture the area of the subdominant extremal surface by decohering the tranisition matrix $\sigma_{\Psi | \Psi'}$ in (\ref{tranmat}).
To this end, we define a decohered version of $\sigma_{\Psi | \Psi'}$ by a diagonal projection in the Virasoro representation space,
\begin{equation}
    \overline{\sigma}_{\Psi | \Psi'}=\sum_p |c_{12p}|^2 \left |{\cal B}\left[  \vcenter{\hbox{\vspace{0.12in}
	\begin{tikzpicture}[scale=.75]
	\draw[thick,<-] (-1/2,0) -- (-3/2,0);
	\node[below,scale=0.75] at (-1,0) {$h_p$};
	\draw[thick] (-3/2,0) -- (-3/2,0.8);
	\node[above,scale=0.75] at (-3/2,0.8) {$h_{2}$};
        \draw[thick] (-3/2,0) -- (-5/2,0);
        \node[left,scale=0.75] at (-5/2,0) {$h_{1}$};
	\end{tikzpicture}
	}} \right] \right |^2 \ket{p} \bra{p} \left |{\cal B'}\left[  \vcenter{\hbox{\vspace{0.12in}
	\begin{tikzpicture}[scale=.75]
	\draw[thick,<-] (-1/2,0) -- (-3/2,0);
	\node[below,scale=0.75] at (-1,0) {$h_p$};
	\draw[thick] (-3/2,0) -- (-3/2,0.8);
	\node[above,scale=0.75] at (-3/2,0.8) {$h_{2}$};
        \draw[thick] (-3/2,0) -- (-5/2,0);
        \node[left,scale=0.75] at (-5/2,0) {$h_{1}$};
	\end{tikzpicture}
	}} \right]^\dagger \right |^2
\end{equation}
where the prime on the OPE block is just to indicate that it is evaluated at the moduli corresponding to the primed locations of the operators.
The von Neumann entropy of the above decohered transition matrix can be computed in the semiclassical limit by a saddle point calculation giving,
\begin{equation}
    S( \overline{\sigma}_{\Psi | \Psi'})=\frac{c}{6}\pi(\gamma_*+\overline{\gamma}_*)
\end{equation}
where $(\gamma_*,\overline{\gamma}_*)$ are chiral saddlepoint momenta expressible in terms of the cross ratio which is complex in this setup. Using the relation to saddlepoint weights $(h_*,\overline{h}_*)=\left (\frac{c}{24}(1+\gamma_*^2),\frac{c}{24}(1+\overline{\gamma}_*^2)\right)$, we see that the coarse grained entropy matches with the Cardy entropy evaluated at these saddlepoint weights,
\begin{equation} \label{Pseudoentropy}
    S( \overline{\sigma}_{\Psi | \Psi'})=S_0(h_*,\overline{h}_*)=2\pi\sqrt{\frac{c}{6}(h_*-\frac{c}{24}})+2\pi\sqrt{\frac{c}{6}(\overline{h}_*-\frac{c}{24}})
\end{equation}
The Renyi entropies of the decohered transition matrix read,
\begin{equation}
    \text{Tr}( \overline{\sigma}_{\Psi | \Psi'}^k) \approx \left |\int dh_p \rho_0(h_p)C_0(h_1,h_2,h_p)^k \left (\vcenter{\hbox{\vspace{0.12in}
	\begin{tikzpicture}[scale=.75]
        \draw[thick] (-1/2,0) -- (-1/2,0.8);
        \node[above, scale=0.75] at (-1/2,0.8) {$2$};
        \draw[thick] (-1/2,0) -- (1/2,0);
        \node[right, scale=0.75] at (1/2,0) {$1$};
	\draw[thick] (-1/2,0) -- (-3/2,0);
	\node[below,scale=0.75] at (-1,0) {$p$};
	\draw[thick] (-3/2,0) -- (-3/2,0.8);
	\node[above,scale=0.75] at (-3/2,0.8) {$2$};
        \draw[thick] (-3/2,0) -- (-5/2,0);
        \node[left,scale=0.75] at (-5/2,0) {$1$};
	\end{tikzpicture}
	}}\right )^k \right |^{2} 
\end{equation}
This agrees with the semiclassical partition function of a $k$-boundary wormhole with the defects going across the wormhole between adjacent boundaries,
\begin{equation}
    Z_k \approx \text{Tr}( \overline{\sigma}_{\Psi | \Psi'}^k)
\end{equation}
The partition function for the 2-boundary wormhole was computed in \cite{Chandra:2022bqq} and for $k>2$, it can be computed using the Virasoro TQFT prescription \cite{Collier:2023fwi}.\footnote{As an aside, note that we can also express the wormhole amplitudes in terms of the overlap between two GHZ-like entangled states defined in this case as,
\begin{equation}
 \begin{split}
   & \ket{\Psi_k}=\sum_p c_{12p}^k \left(\left | {\cal B}\left[  \vcenter{\hbox{\vspace{0.12in}
	\begin{tikzpicture}[scale=.75]
	\draw[thick,<-] (-1/2,0) -- (-3/2,0);
	\node[below,scale=0.75] at (-1,0) {$h_p$};
	\draw[thick] (-3/2,0) -- (-3/2,0.8);
	\node[above,scale=0.75] at (-3/2,0.8) {$h_2$};
        \draw[thick] (-3/2,0) -- (-5/2,0);
        \node[left,scale=0.75] at (-5/2,0) {$h_1$};
	\end{tikzpicture}
	}} \right] \right |^2\ket{p} \right)^{\otimes k}\\
	&\ket{\Psi_k'}=\sum_p c_{12p}^k \left(\left | {\cal B}'\left[  \vcenter{\hbox{\vspace{0.12in}
	\begin{tikzpicture}[scale=.75]
	\draw[thick,<-] (-1/2,0) -- (-3/2,0);
	\node[below,scale=0.75] at (-1,0) {$h_p$};
	\draw[thick] (-3/2,0) -- (-3/2,0.8);
	\node[above,scale=0.75] at (-3/2,0.8) {$h_2$};
        \draw[thick] (-3/2,0) -- (-5/2,0);
        \node[left,scale=0.75] at (-5/2,0) {$h_1$};
	\end{tikzpicture}
	}} \right] \right |^2\ket{p} \right)^{\otimes k}
 \end{split}
\end{equation}
where the prime on the OPE block is just to emphasize that it is evaluated at the primed operator locations.
We thus have the following identity relating wormhole amplitudes with the overlap of the above GHZ-like entangled states,
\begin{equation}
    Z_k \approx \langle \Psi_k'\ket{\Psi_k}
\end{equation}}
These wormholes are replica-symmetric solutions solving the boundary conditions corresponding to $(\langle \Psi ' \ket{\Psi})^k$. Since there is a saddlepoint in the partition function as $k\to 1$, we can think of the wormhole topologies as being described by branching around an extremal surface on the $k=1$ geometry i.e the black hole and the area of the extremal surface can be read off from the partition functions as,
\begin{equation}
   \frac{A}{4G_N}=-\partial_k \log \left(\frac{Z_k}{Z_1^k}\right)\bigg |_{k=1}=S_0(h_*,\overline{h}_*)
\end{equation}
Therefore, by comparing with (\ref{Pseudoentropy}), we see that the pseudo entropy of the decohered transition matrix matches with the area of the sub-dominant Euclidean-HRT surface,
\begin{equation}
    S( \overline{\sigma}_{\Psi | \Psi'})= \frac{A}{4G_N}
\end{equation}
thereby providing a CFT interpretation for the area of a subdominant Euclidean-HRT surface.

\subsection{More general wormholes from decohering transition matrices}

We can generalise the previous discussion of decohering transition matrices to provide an interpretation for wormholes with non replica-symmetric boundary conditions in individual CFTs. Let us illustrate this idea by taking the example of the two-boundary wormhole with boundaries being 4-punctured spheres with different moduli (cross ratios). The wormhole satisfies the boundary conditions corresponding to the product of overlaps between CFT states ${}_L\langle \Psi' \ket{\Psi}_L {}_R\langle \Psi' \ket{\Psi}_R$ with the subscripts denoting the left and right boundaries of the wormhole. The unprimed states corresponding to either boundary are created by the action of scalar primary operators $\mathcal{O}_1$ and $\mathcal{O}_2$ on the vacuum possibly at different locations on the two boundary spheres whereas the primed states are created by the action of scalar primary operators $\mathcal{O}_3$ and $\mathcal{O}_4$ on the vacuum. It was shown in \cite{Chandra:2022bqq} that the gravitational partition function on the wormhole in the semiclassical limit is given by
\begin{equation}
   Z_{WH}\approx \left |\int dh_p \rho_0(h_p) C_0(h_1,h_2,h_p) C_0(h_3,h_4,h_p)\vcenter{\hbox{\vspace{0.12in}
	\begin{tikzpicture}[scale=.75]
        \draw[thick] (-1/2,0) -- (-1/2,0.8);
        \node[above, scale=0.75] at (-1/2,0.8) {$3$};
        \draw[thick] (-1/2,0) -- (1/2,0);
        \node[right, scale=0.75] at (1/2,0) {$4$};
	\draw[thick] (-1/2,0) -- (-3/2,0);
	\node[below,scale=0.75] at (-1,0) {$p$};
	\draw[thick] (-3/2,0) -- (-3/2,0.8);
	\node[above,scale=0.75] at (-3/2,0.8) {$2$};
        \draw[thick] (-3/2,0) -- (-5/2,0);
        \node[left,scale=0.75] at (-5/2,0) {$1$};
	\end{tikzpicture}
	}}(x_L) \vcenter{\hbox{\vspace{0.12in}
	\begin{tikzpicture}[scale=.75]
        \draw[thick] (-1/2,0) -- (-1/2,0.8);
        \node[above, scale=0.75] at (-1/2,0.8) {$3$};
        \draw[thick] (-1/2,0) -- (1/2,0);
        \node[right, scale=0.75] at (1/2,0) {$4$};
	\draw[thick] (-1/2,0) -- (-3/2,0);
	\node[below,scale=0.75] at (-1,0) {$p$};
	\draw[thick] (-3/2,0) -- (-3/2,0.8);
	\node[above,scale=0.75] at (-3/2,0.8) {$2$};
        \draw[thick] (-3/2,0) -- (-5/2,0);
        \node[left,scale=0.75] at (-5/2,0) {$1$};
	\end{tikzpicture}
	}} (x_R)  \right |^2
\end{equation}
where $x_L$ and $x_R$ are respectively the cross ratios between the operator insertions on the two boundaries.
The relevant transition matrices to capture the above wormhole amplitude are,
\begin{equation}
    \begin{split}
        & \rho_{\Psi_L| \Psi_R} = \ket{\Psi}_L \bra{\Psi}_R =\sum_{p,q} c_{12p} c^*_{12q}\left |{\cal B}_L\left[  \vcenter{\hbox{\vspace{0.12in}
	\begin{tikzpicture}[scale=.75]
	\draw[thick,<-] (-1/2,0) -- (-3/2,0);
	\node[below,scale=0.75] at (-1,0) {$h_p$};
	\draw[thick] (-3/2,0) -- (-3/2,0.8);
	\node[above,scale=0.75] at (-3/2,0.8) {$h_{2}$};
        \draw[thick] (-3/2,0) -- (-5/2,0);
        \node[left,scale=0.75] at (-5/2,0) {$h_{1}$};
	\end{tikzpicture}
	}} \right] \right |^2 \ket{p} \bra{q}\left |{\cal B}_R\left[  \vcenter{\hbox{\vspace{0.12in}
	\begin{tikzpicture}[scale=.75]
	\draw[thick,<-] (-1/2,0) -- (-3/2,0);
	\node[below,scale=0.75] at (-1,0) {$h_q$};
	\draw[thick] (-3/2,0) -- (-3/2,0.8);
	\node[above,scale=0.75] at (-3/2,0.8) {$h_{2}$};
        \draw[thick] (-3/2,0) -- (-5/2,0);
        \node[left,scale=0.75] at (-5/2,0) {$h_{1}$};
	\end{tikzpicture}
	}} \right]^\dagger \right |^2  \\
        & \rho_{\Psi'_R| \Psi'_L} = \ket{\Psi'}_R \bra{\Psi'}_L = \sum_{p,q} c_{34p} c^*_{34q}\left |{\cal B}_R\left[  \vcenter{\hbox{\vspace{0.12in}
	\begin{tikzpicture}[scale=.75]
	\draw[thick,<-] (-1/2,0) -- (-3/2,0);
	\node[below,scale=0.75] at (-1,0) {$h_p$};
	\draw[thick] (-3/2,0) -- (-3/2,0.8);
	\node[above,scale=0.75] at (-3/2,0.8) {$h_{3}$};
        \draw[thick] (-3/2,0) -- (-5/2,0);
        \node[left,scale=0.75] at (-5/2,0) {$h_{4}$};
	\end{tikzpicture}
	}} \right] \right |^2 \ket{p} \bra{q}\left |{\cal B}_L\left[  \vcenter{\hbox{\vspace{0.12in}
	\begin{tikzpicture}[scale=.75]
	\draw[thick,<-] (-1/2,0) -- (-3/2,0);
	\node[below,scale=0.75] at (-1,0) {$h_q$};
	\draw[thick] (-3/2,0) -- (-3/2,0.8);
	\node[above,scale=0.75] at (-3/2,0.8) {$h_{3}$};
        \draw[thick] (-3/2,0) -- (-5/2,0);
        \node[left,scale=0.75] at (-5/2,0) {$h_{4}$};
	\end{tikzpicture}
	}} \right]^\dagger \right |^2
    \end{split}
\end{equation}
where we have expanded the states using Virasoro OPE blocks. The subscripts on the OPE blocks appear because the insertion points of the operators are different on the two boundaries. We decohere the above transition matrices in the Virasoro representation space and denote the decohered transition matrices by  $\overline{\rho}_{\Psi_L| \Psi_R}$ and $\overline{\rho}_{\Psi'_L| \Psi'_R}$ respectively,
\begin{equation}
    \begin{split}
        & \overline{\rho}_{\Psi_L| \Psi_R}  =\sum_p |c_{12p}|^2 \left |{\cal B}_L\left[  \vcenter{\hbox{\vspace{0.12in}
	\begin{tikzpicture}[scale=.75]
	\draw[thick,<-] (-1/2,0) -- (-3/2,0);
	\node[below,scale=0.75] at (-1,0) {$h_p$};
	\draw[thick] (-3/2,0) -- (-3/2,0.8);
	\node[above,scale=0.75] at (-3/2,0.8) {$h_{2}$};
        \draw[thick] (-3/2,0) -- (-5/2,0);
        \node[left,scale=0.75] at (-5/2,0) {$h_{1}$};
	\end{tikzpicture}
	}} \right] \right |^2 \ket{p} \bra{p}\left |{\cal B}_R\left[  \vcenter{\hbox{\vspace{0.12in}
	\begin{tikzpicture}[scale=.75]
	\draw[thick,<-] (-1/2,0) -- (-3/2,0);
	\node[below,scale=0.75] at (-1,0) {$h_p$};
	\draw[thick] (-3/2,0) -- (-3/2,0.8);
	\node[above,scale=0.75] at (-3/2,0.8) {$h_{2}$};
        \draw[thick] (-3/2,0) -- (-5/2,0);
        \node[left,scale=0.75] at (-5/2,0) {$h_{1}$};
	\end{tikzpicture}
	}} \right]^\dagger \right |^2  \\
        & \overline{\rho}_{\Psi'_R| \Psi'_L}  = \sum_p |c_{34p}|^2 \left |{\cal B}_R\left[  \vcenter{\hbox{\vspace{0.12in}
	\begin{tikzpicture}[scale=.75]
	\draw[thick,<-] (-1/2,0) -- (-3/2,0);
	\node[below,scale=0.75] at (-1,0) {$h_p$};
	\draw[thick] (-3/2,0) -- (-3/2,0.8);
	\node[above,scale=0.75] at (-3/2,0.8) {$h_{3}$};
        \draw[thick] (-3/2,0) -- (-5/2,0);
        \node[left,scale=0.75] at (-5/2,0) {$h_{4}$};
	\end{tikzpicture}
	}} \right] \right |^2 \ket{p} \bra{p}\left |{\cal B}_L\left[  \vcenter{\hbox{\vspace{0.12in}
	\begin{tikzpicture}[scale=.75]
	\draw[thick,<-] (-1/2,0) -- (-3/2,0);
	\node[below,scale=0.75] at (-1,0) {$h_p$};
	\draw[thick] (-3/2,0) -- (-3/2,0.8);
	\node[above,scale=0.75] at (-3/2,0.8) {$h_{3}$};
        \draw[thick] (-3/2,0) -- (-5/2,0);
        \node[left,scale=0.75] at (-5/2,0) {$h_{4}$};
	\end{tikzpicture}
	}} \right]^\dagger \right |^2
    \end{split}
\end{equation}
The wormhole amplitude can be expressed as the overlap between the two decohered transition matrices,
\begin{equation}
      \vcenter{\hbox{\includegraphics[width=1.4in]{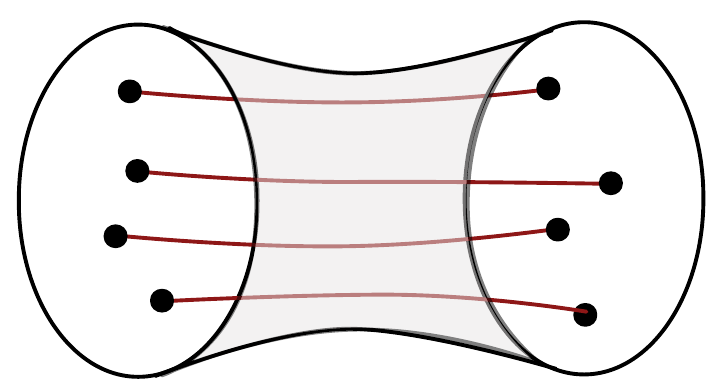}}}\approx \text{Tr}(\overline{\rho}_{\Psi_L| \Psi_R}\,\,\overline{\rho}_{\Psi'_R| \Psi'_L})
\end{equation}
Therefore, we have shown that the two-boundary wormhole solving the boundary condition corresponding to ${}_L\langle \Psi' \ket{\Psi}_L {}_R\langle \Psi' \ket{\Psi}_R$ computes the overlap between two decohered transition matrices $\overline{\rho}_{\Psi_L| \Psi_R}$ and $\overline{\rho}_{\Psi'_R| \Psi'_L}$ constructed from the 4 pure states appearing in the boundary condition.\footnote{
We can also express the above wormhole amplitude as the overlap between two GHZ-like entangled states defined as
\begin{equation}
 \begin{split}
   & \ket{\Psi_2}=\sum_p c_{12p}c_{34p} \left |{\cal B}_L\left[  \vcenter{\hbox{\vspace{0.12in}
	\begin{tikzpicture}[scale=.75]
	\draw[thick,<-] (-1/2,0) -- (-3/2,0);
	\node[below,scale=0.75] at (-1,0) {$h_p$};
	\draw[thick] (-3/2,0) -- (-3/2,0.8);
	\node[above,scale=0.75] at (-3/2,0.8) {$h_{2}$};
        \draw[thick] (-3/2,0) -- (-5/2,0);
        \node[left,scale=0.75] at (-5/2,0) {$h_{1}$};
	\end{tikzpicture}
	}} \right] \right |^2 \ket{p} \otimes \left |{\cal B}_R\left[  \vcenter{\hbox{\vspace{0.12in}
	\begin{tikzpicture}[scale=.75]
	\draw[thick,<-] (-1/2,0) -- (-3/2,0);
	\node[below,scale=0.75] at (-1,0) {$h_p$};
	\draw[thick] (-3/2,0) -- (-3/2,0.8);
	\node[above,scale=0.75] at (-3/2,0.8) {$h_{3}$};
        \draw[thick] (-3/2,0) -- (-5/2,0);
        \node[left,scale=0.75] at (-5/2,0) {$h_{4}$};
	\end{tikzpicture}
	}} \right] \right |^2 \ket{p}\\
	& \ket{\Psi_2'}=\sum_p c_{12p}c_{34p} \left |{\cal B}_L\left[  \vcenter{\hbox{\vspace{0.12in}
	\begin{tikzpicture}[scale=.75]
	\draw[thick,<-] (-1/2,0) -- (-3/2,0);
	\node[below,scale=0.75] at (-1,0) {$h_p$};
	\draw[thick] (-3/2,0) -- (-3/2,0.8);
	\node[above,scale=0.75] at (-3/2,0.8) {$h_{3}$};
        \draw[thick] (-3/2,0) -- (-5/2,0);
        \node[left,scale=0.75] at (-5/2,0) {$h_{4}$};
	\end{tikzpicture}
	}} \right] \right |^2 \ket{p} \otimes \left |{\cal B}_R\left[  \vcenter{\hbox{\vspace{0.12in}
	\begin{tikzpicture}[scale=.75]
	\draw[thick,<-] (-1/2,0) -- (-3/2,0);
	\node[below,scale=0.75] at (-1,0) {$h_p$};
	\draw[thick] (-3/2,0) -- (-3/2,0.8);
	\node[above,scale=0.75] at (-3/2,0.8) {$h_{2}$};
        \draw[thick] (-3/2,0) -- (-5/2,0);
        \node[left,scale=0.75] at (-5/2,0) {$h_{1}$};
	\end{tikzpicture}
	}} \right] \right |^2 \ket{p}
 \end{split}
\end{equation}
so that we have the following identity in the semiclassical limit,
\begin{equation}
    Z_{WH} \approx \langle \Psi_2'\ket{\Psi_2} \approx \text{Tr}(\overline{\rho}_{\Psi_L| \Psi_R}\overline{\rho}_{\Psi'_L| \Psi'_R})
\end{equation}
}

The observation made using the above example that asymmetric (Quasi-Fuchsian) two-boundary wormholes compute overlaps between decohered transition matrices is robust and can also be extended to wormholes with higher-genus boundaries. For example, consider the two-boundary wormhole with the topology of $\Sigma_{1,1} \times \mathbb{R}$ i.e the boundaries are once-punctured tori, solving the boundary conditions corresponding to the product of overlaps between the PETS and TFD states, ${}_L\langle \text{PETS} \ket{\text{TFD}}_L {}_R\langle \text{PETS} \ket{\text{TFD}}_R$. The partition function on this wormhole was computed in \cite{Chandra:2022bqq} in the semiclassical limit,
\begin{equation}
   \vcenter{\hbox{\includegraphics[width=1.4in]{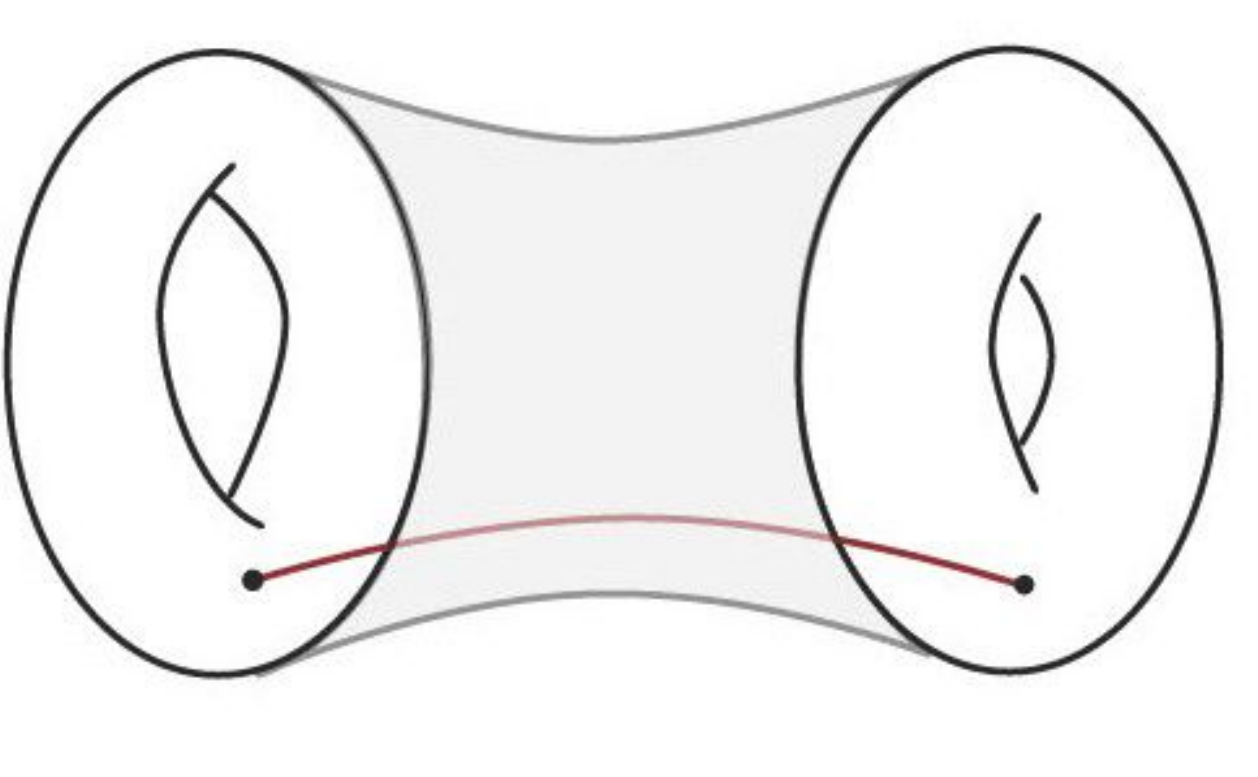}}} \approx \left | \int dh_p \rho_0(h_p) C_0(h_{\mathcal{O}},h_p,h_p) \vcenter{\hbox{
	\begin{tikzpicture}[scale=0.75]
	\draw[thick] (0,0) circle (1);
	\node[above] at (0,1) {$p$};
	\draw[thick] (1,0) -- (2,0);
	\node[right] at (2,0) {$\mathcal{O}$};
        \node[scale=0.75] at (0,0) {$\tau_L$};
	\end{tikzpicture}
	}} 
	\vcenter{\hbox{
	\begin{tikzpicture}[scale=0.75]
	\draw[thick] (0,0) circle (1);
	\node[above] at (0,1) {$p$};
	\draw[thick] (1,0) -- (2,0);
	\node[right] at (2,0) {$\mathcal{O}$};
        \node[scale=0.75] at (0,0) {$-\tau_R$};
	\end{tikzpicture}
	}} \right |^2
\end{equation}
We can express this amplitude as the overlap of the decohered transition matrix between the PETS states on either boundary with the decohered transition matrix between the TFD states on either boundary,
\begin{equation}
   \vcenter{\hbox{\includegraphics[width=1.4in]{more-figures/torusonepointwormhole.pdf}}} \approx \text{Tr}(\overline{\rho}_{\text{PETS}_L|\text{PETS}_R}\,\,\overline{\rho}_{\text{TFD}_R|\text{TFD}_L})
\end{equation}
Note that in the above expression $\overline{\rho}_{\text{TFD}_R|\text{TFD}_L}$ is the transition matrix between $\widetilde{\text{TMD}}$ states defined in (\ref{TMDtilde}) and not the TMD states in (\ref{TMD}).

\section{The West Coast model for 3d gravity} \label{secevap}

We discuss evaporation of the black hole solutions discussed in section \ref{secCG} using a setup generalizing that of the West Coast model \cite{Penington:2019kki} to 3d gravity. We shall first describe the setup more generally and illustrate by studying the evaporation of three types of black hole solutions. The idea of \cite{Penington:2019kki} is to discuss flavoured black hole microstates $\ket{\Psi^i}$  with $i= 1 \,\text {to}\, N_f$ where $N_f$ is the number of flavours. We shall assume that the number of flavours is parametrically large, i.e, $N_f \gg 1$. To discuss evaporation in this setup, we need to entangle these flavoured microstates with an auxiliary `radiation' reservoir whose Hilbert space is spanned by radiation states $\{\ket{i}_R\}$. Consider the following state of the black hole entangled with radiation,
\begin{equation}
    \ket{\psi}=\sum_{i=1}^{N_f} \ket{\Psi^i}_{BH}\ket{i}_R
\end{equation}
Now, we trace out the BH degrees of freedom to get a reduced state for the radiation,
\begin{equation} \label{Radstate}
   \rho_R=\sum_{i,j}\langle\Psi^j \ket{\Psi^i}\ket{i}_R\bra{j}_R
\end{equation}
In the remainder of the section, we shall compute the von Neumann entropy of the radiation using the replica trick by considering replica wormhole geometries appropriate to the evaporation of black hole geometries dual to pure states in the CFT discussed in section \ref{secCG}. In each case, we observe that the wormholes which showed up in our coarse graining formalism are also responsible for saturating the Page curve for the radiation at late times.

\subsection{Simplest pure state black hole} \label{simpl}

Here, we shall discuss the evaporation of the black hole dual to the state $\ket{\Psi}=\mathcal{O}_1\mathcal{O}_2\ket{0}$ where $\mathcal{O}_1$ and $\mathcal{O}_2$ are scalar primary operators above the multi-twist threshold using the setup described above.
Let us assume that the operator $\mathcal{O}_1$ is charged under a global symmetry and $\mathcal{O}_2$ is uncharged so that 
\begin{equation}
    \ket{\Psi^i}=\mathcal{O}_1^i \mathcal{O}_2\ket{0}
\end{equation}
denote flavoured microstates. The Renyis of the radiation state $\rho_R$ in (\ref{Radstate}) can be computed holographically by filling in the boundary conditions corresponding to $\text{Tr}(\rho_R^k)$ with bulk geometries,
\begin{equation}
   \text{Tr}(\rho_R^k)=N_f Z_1^k +\dots +N_f^k Z_k
\end{equation}
where $Z_1$ denotes the gravitational partition function on the hyperbolic ball with two defects and $Z_k$ denotes the partition function on the $k$-boundary wormhole with defects going cyclically across the wormhole. The prefactors in the above expression come from the sum over flavours with each `flavour loop' giving a factor of $N_f$. Thus, the prefactor reads $N_f$ if the charged defect starts and ends on the same boundary since there would only one flavour loop and reads $N_f^k$ if the charged defect goes across the wormhole since there would be $k$ flavour loops. The `$\dots$' denote non replica-symmetric contributions which we shall not consider in this work. For example, the fourth Renyi of the radiation state is computed using the following geometries,
\begin{equation}
   \text{Tr}(\rho_R^4)= N_f \left [\vcenter{\hbox{
\begin{overpic}[width=1in]{more-figures/singlebdry.pdf}
\end{overpic}
}} \right ]^4 +\dots+ N_f^4 \left [\vcenter{\hbox{
\begin{overpic}[width=1.2in,grid=false]{more-figures/fourpointfourbdry.pdf}
\end{overpic}
}}\right]
\end{equation}
Using the replica trick, we can compute the von Neumann entropy of the radiation state,
\begin{equation}
  S(\rho_R)=-\partial_k \log \left (\frac{\text{Tr} \rho_R^k}{(\text{Tr}\rho_R)^k}\right )\bigg |_{k=1}=\text{min}\{\log N_f, S_{BH}\}
\end{equation}
where the first term comes from the black hole saddle and the second one comes from the wormhole saddles. The wormhole contribution is expected to match with the black hole entropy $S_{BH}$ because we have observed that these are also the replica wormholes obtained by branching around the horizon of the black hole. Thus, we see that the entropy of the radiation is consistent with the expectation from the island rule for the Page curve with the Page transition occuring at ``Page time" $\log N_f=S_{BH}$. It would be interesting to fill in the `$\dots$' by computing the contribution from non replica-symmetric geometries and checking if they can smoothen the Page transition possibly using the resolvent method introduced in \cite{Penington:2019kki}.

\subsection{Black hole with multiple minimal surfaces}

Now, we discuss the more interesting example of evaporation of the black hole geometry dual to the pure state $ \ket{\Psi}=\mathcal{O}_1\mathcal{O}_2\mathcal{O}_3\ket{0}$. To this end, we introduce the following flavoured black hole microstates,
\begin{equation} \label{threedefBH}
    \ket{\Psi^i}=\mathcal{O}^i_1\mathcal{O}_2\mathcal{O}_3\ket{0}
\end{equation}
Here, we are assuming that each of the three scalar primary operators are above the multi-twist threshold so that on the time-symmetric slice of the black hole geometry, there are minimal geodesics around each pair of operators which we shall denote as $\Gamma_{12}$, $\Gamma_{23}$ and $\Gamma_{13}$, and a minimal geodesic surrounding all three of them which we shall call $\Gamma_{123}$, in the hyperbolic metric on the thrice-punctured unit disk. Furthermore, we assume that the operator $\mathcal{O}_1$ is charged under a global symmetry with $i$ denoting the flavour label whereas the other two operators are uncharged. As described in section (\ref{CGinner}), we can construct replica wormholes by branching around any of the 4 minimal geodesics. The wormholes obtained by branching around $\Gamma_{12}$ have defects dual to $\mathcal{O}_1$ and $\mathcal{O}_2$ going across the wormhole in a cyclic fashion while the defect dual to $\mathcal{O}_3$ starts and ends on the same boundary, and correspondingly for wormholes obtained by branching around $\Gamma_{23}$ and $\Gamma_{13}$. The wormholes obtained by branching around $\Gamma_{123}$ have all three defects going across.  Due to the competition between these wormhole saddles, the entangling of this black hole to the auxiliary system is expected to give an interesting Page curve. To this end, note that the Renyis of the radiation state computed holographically read
\begin{equation}
   \text{Tr}(\rho_R^k)=N_f Z_1^k +N_f Z^{23}_k+N_f^k Z^{12}_k+N_f^k Z^{13}_k+N_f^k Z^{123}_k
\end{equation}
where the first term is the partition function of the black hole and the remaining terms are contributions from the wormholes described above. The superscripts denote the labels of the defects going across the wormhole. For example, the various geometries contributing to the fourth Renyi entropy are given by
\begin{multline}
  \text{Tr}(\rho_R^4)= N_f  \left[\vcenter{\hbox{
\begin{overpic}[width=0.8in]{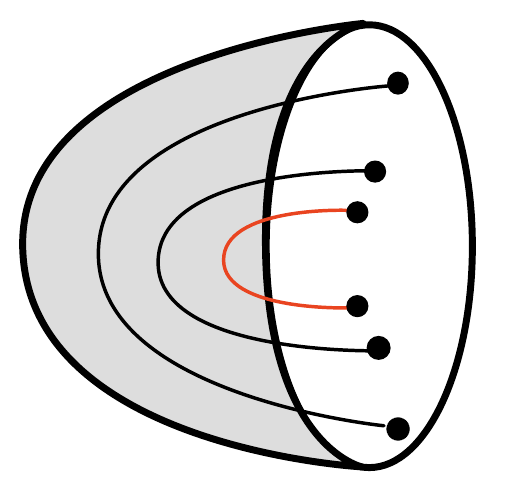}
\end{overpic}
}}\right ]^4+ N_f \left [\vcenter{\hbox{
\begin{overpic}[width=1.2in,grid=false]{more-figures/fourpointfourbdryprobeout.pdf}
\put (30,60) {$\tiny 1$}
\put (60,60) {$\tiny 1$}
\put (30,35) {$\tiny 1$}
\put (60,35) {$\tiny 1$}
\end{overpic}
}} \right ]+ N_f^4 \left [\vcenter{\hbox{
\begin{overpic}[width=1in,grid=false]{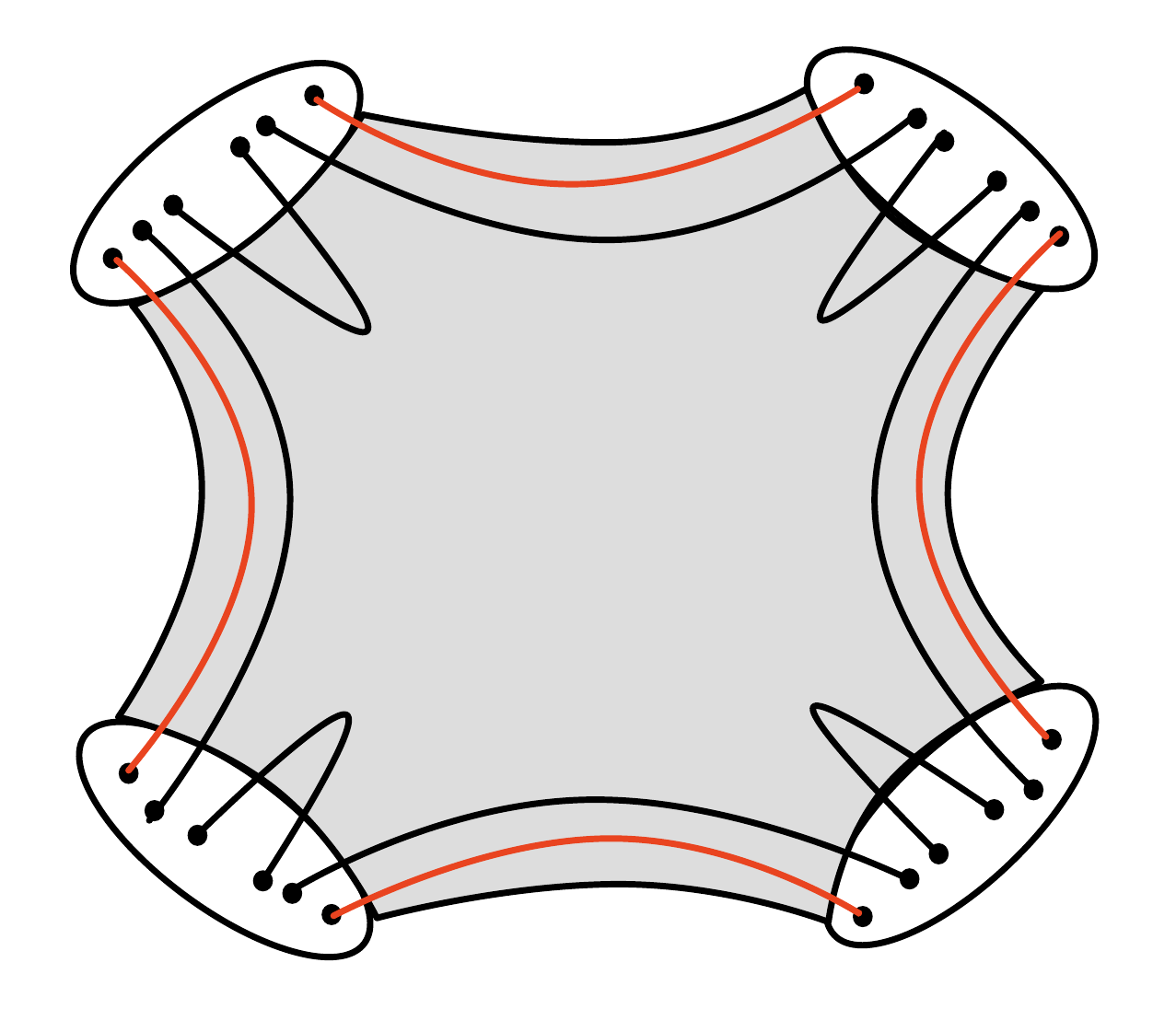}
\put (35,50) {$\tiny 2$}
\put (65,50) {$\tiny 2$}
\put (35,30) {$\tiny 2$}
\put (65,30) {$\tiny 2$}
\end{overpic}
}} \right ]\\+ N_f^4 \left [\vcenter{\hbox{
\begin{overpic}[width=1in,grid=false]{more-figures/chargedsixpoint.pdf}
\put (35,50) {$\tiny 3$}
\put (65,50) {$\tiny 3$}
\put (35,30) {$\tiny 3$}
\put (65,30) {$\tiny 3$}
\end{overpic}
}} \right ] + N_f^4 \left [\vcenter{\hbox{
\begin{overpic}[width=1in,grid=false]{more-figures/fourpointfourbdryprobein.pdf}
\end{overpic}
}} \right ]
\end{multline}
where the charged defect has been marked in red and the defect starting and ending on the same boundary has been labelled.
Using the replica trick, we compute the entropy of the radiation state,
\begin{equation} \label{Pagethreedef}
   S(\rho_R)=\text{min} \left \{ \log N_f, \log N_f + \frac{\text{Area}(\Gamma_{23})}{4G_N}, \frac{\text{Area}(\Gamma_{12})}{4G_N}, \frac{\text{Area}(\Gamma_{13})}{4G_N}, \frac{\text{Area}(\Gamma_{123})}{4G_N}\right \}
\end{equation}
Observe that the second term is always subdominant for any number of flavours so does not affect the Page curve. At ``early times" i.e when the number of flavours is small, the first term dominates and thus the entropy of the radiation increases with the number of flavours following which the Page transition occurs and the entropy saturates to the minimum of the last three terms. Note that interestingly, this calculation is showing that only the minimal surfaces surrounding the charged defect affect the Page transition. In addition, we see that the entropy of the radiation in this setup saturates to the black hole entropy only if the outermost minimal surface has an area smaller than that of the inner (non-trivial) minimal surfaces. This effect was not observed in the West Coast model for JT gravity \cite{Penington:2019kki} because in that setup, the black hole with an EOW brane has a single minimal surface corresponding to the apparent horizon. A similar effect can also be seen in a higher dimensional setup by constructing Euclidean black hole solutions with multiple minimal surfaces on the time-symmetric slice for instance by the backreaction of multiple heavy spherically symmetric thin shells considered for instance in \cite{Sasieta:2022ksu,Balasubramanian:2022gmo}. 

The fact that the entropy of radiation need not saturate to the black hole entropy seems to suggest that this model does not describe an evaporating black hole at late times but rather describes an `old' black hole past the Page time perturbed by the backreaction of infalling matter\footnote{I thank Tom Hartman for suggesting this interpretation.}. If the infalling matter was light consisting say of a few probe particles, then it would not affect the Page transition for the black hole. However, if the infalling matter is heavy enough to backreact and create a new minimal surface, then it could make the `old' black hole `young' again. In our setup, this would mean that if the operator $\mathcal{O}_3$ was a probe operator, then the Page curve for the black hole in (\ref{threedefBH}) would be similar to that for the black hole in section \ref{simpl}. However, in this section, we are assuming that $\mathcal{O}_3$ is heavy enough to create a new minimal surface thereby pushing the horizon outward so the Page transition could be affected by the competition between the minimal surfaces as we have shown in (\ref{Pagethreedef}). Even for this case, we can still ensure that the entropy of radiation saturates to the area of the outermost minimal surface (i.e the horizon) by making all the three defects charged and entangling them to the radiation reservoir. This follows from our observation in (\ref{Pagethreedef}) that it is entropically favourable for a charged defect to go through the replica wormholes so making all three defects charged would mean that the replica wormholes with all three defects going through them dominate the Page curve calculation at late times ensuring that the entropy of radiation matches with the coarse grained entropy of the horizon.

\subsection{Multi-boundary black holes}

We can readily generalise the idea to discuss the Page curve for multi-boundary black holes in this setup. We illustrate here using the example of the PETS black hole discussed in section \ref{CGmulti}. To this end, we consider the operator exciting the TFD to be charged under a global symmetry thereby defining flavoured PETS states,
\begin{equation}
   \ket{\text{PETS}^i}=\mathcal{O}^i \ket{\text{TFD}}
\end{equation}
where $\mathcal{O}$ is a scalar primary operator heavy enough so that there are two minimal surfaces on the spatial slice of the PETS black hole which is a punctured cylinder.
By entangling the flavoured PETS state with the radiation reservoir and tracing out the black hole degrees of freedom, we observe that the matrix elements of the radiation state are given by overlaps between the flavoured PETS states,
\begin{equation}
  (\rho_R)_{ij}=\langle \text{PETS}^i \ket{\text{PETS}^j}
\end{equation}
The Renyis of the radiation state are computed holographically using the PETS black hole and the replica-symmetric torus two-point wormholes. 
\begin{equation}
    \text{Tr}(\rho_R^k)=N_f Z_1^k + N_f^k Z_k
\end{equation}
where the first term is the contribution from the black hole while the second term is the replica wormhole contribution. For example,
\begin{equation} \label{radRen}
   \text{Tr}(\rho_R^4)=N_f\left [\vcenter{\hbox{
\begin{overpic}[width=1in,grid=false]{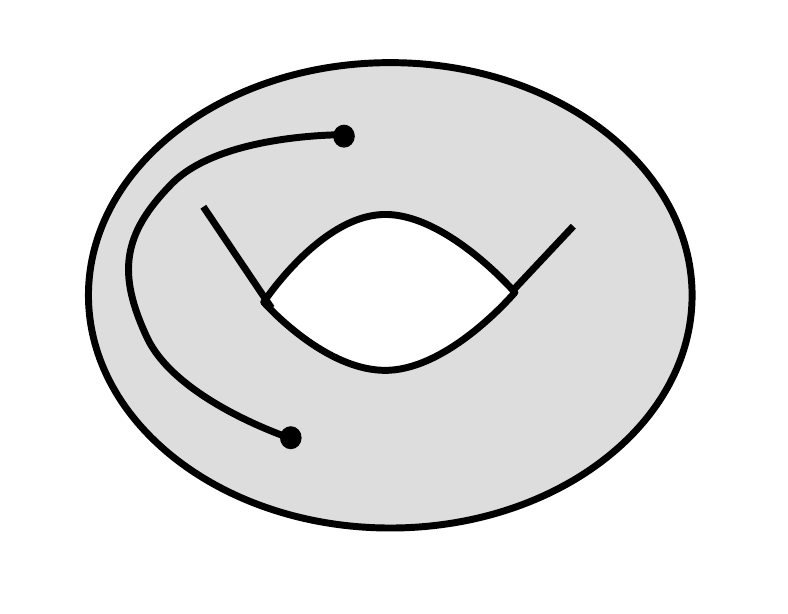}
\end{overpic}
}} \right]^4+ N_f^4 \left [\vcenter{\hbox{
\begin{overpic}[width=1in,grid=false]{more-figures/torustwopointfourbdry.pdf}
\end{overpic}
}} \right]
\end{equation}
The entropy of the radiation computed using the replica trick reads
\begin{equation}
   S(\rho_R)=\text{min}\left \{ \log N_f, \frac{\text{Area}(\Gamma_1)+\text{Area}(\Gamma_2)}{4G_N} \right \}
\end{equation}
Thus, at late times, the entropy saturates to the coarse grained entropy of the PETS black hole which matches with the sum of the areas of the two minimal surfaces bounding the interior. This is consistent with the prediction from the island rule because at early times, the dominant extremal surface is the trivial surface so the entanglement wedge of radiation covers the full spacetime while at late times, the domain of dependence of the interior region on the time-symmetric slice (which constitutes the island) is in the entanglement wedge of the radiation.

\section{Discussion} \label{secDisc}

In this paper, we have provided an interpretation for several families of Euclidean wormhole solutions of 3d gravity in individual 2d CFTs. The replica symmetric multi-boundary wormhole solutions help generalise the idea in \cite{Chandra:2022fwi} of coarse graining pure states in CFT$_d$ dual to spherically symmetric one-sided black hole geometries in AdS$_{d+1}$ using Euclidean wormhole solutions in AdS$_{d+1}$, to coarse grain more general pure states in 2d CFT which are dual to non-spherically symmetric or even multi-boundary black hole solutions in AdS$_3$. These geometries are time-symmetric and have one or more minimal surfaces on the time-symmetric slice the outermost of which is to be interpreted as an apparent horizon on the corresponding Lorentzian black hole geometries. We observed that the coarse grained state defined by decohering the contribution of Virasoro primaries behind an apparent horizon retaining the correlations between Virasoro descendents has the following two robust features:
\begin{enumerate}
  \item The Renyi entropies of the coarse grained state match with the partition functions of replica wormhole geometries described by branching around the time-symmetric apparent horizon of the black hole.
\item The coarse grained state preserves the correlations of boundary gravitons and of probe particles outside the horizon.
\end{enumerate}
We used these two features to more generally define a coarse grained state outside the apparent horizon on the time-symmetric slice of the black hole geometry. While coarse graining multi-copy CFT states, we observed that coarse graining away the interior region of the multi-boundary black hole dual to the CFT state by decohering the Virasoro primary contribution in each copy sets the mutual information between any two copies of the CFT to zero. 

We then generalized the idea of decohering pure states to decohere transition matrices between pure states. This helps interpret more general families of Euclidean wormhole solutions in individual CFTs including those with non-replica symmetric boundary conditions. We showed using examples that the pseudo entropy of appropriate decohered transition matrices can be interpreted holographically in terms of the area of a codimension-2 minimal surface on a wormhole or in terms of the area of a sub-dominant (in the HRT sense) extremal surface on a non time-symmetric Euclidean black hole geometry. Finally, we generalized the setup of the West Coast model \cite{Penington:2019kki} to 3d gravity to discuss evaporation of various black hole geometries in 3d gravity and computed the Page curve for the entropy of radiation in each case using the same replica wormholes which showed up in the coarse graining formalism. 

In the following, we comment on a future direction which is another interesting application involving some the families of Euclidean wormhole goemetries discussed in this paper.

\subsection{Relation to global charge violating amplitudes}

We briefly comment on the relation of the wormholes discussed in section \ref{secTM} to global charge violating amplitudes. Let us assume that the CFT has a global symmetry. In the setup described in section \ref{Holopseudo}, consider the 4 operators to be charged under the global symmetry and assume that $\mathcal{O}_3$ and $\mathcal{O}_4$ have the same conformal dimensions as $\mathcal{O}_1$ and $\mathcal{O}_2$ respectively but have different global charges. This would mean that in the dual gravitational theory, there is no saddle that computes the overlap between these states. However, as described in section \ref{Holopseudo}, there are $2k$-boundary wormholes such that the charged particles travel across the wormholes and are annihilated on different boundaries. These wormholes describe a semiclassical contribution to $(\langle \Psi \ket{\Psi'})^{2k}$ and therefore quantify the amount of global charge violation in semiclassical gravity due to the existence of Euclidean wormhole solutions. This idea of using Euclidean wormholes to quantify global charge violation was illustrated in \cite{Bah:2022uyz} where they constructed two-boundary wormholes in AdS$_{d+1}$ sourced by spherically symmetric shells which are light but boosted using `negative Euclidean evolution'. It would be interesting to construct analogous wormhole solutions in AdS$_3$ which are sourced by light but boosted point particles and understand their interpretation in terms of the dual CFT.

\bigskip

\noindent \textbf{Acknowledgements} I thank Tom Hartman for several helpful discussions and comments on the draft. I also thank Ven Chandrasekaran, Scott Collier, Alex Maloney and Baur Mukhametzhanov for useful discussions. I thank Scott Collier for mentioning to me about the wormhole amplitudes to be computed in the upcoming paper \cite{CollierInProgress}. This work was supported by NSF grant PHY-2014071.

\renewcommand{\baselinestretch}{1}\small
\bibliographystyle{ourbst}
\bibliography{biblio3}

\end{document}